\documentclass[aps,prd,twocolumn,nofootinbib,eqsecnum,superscriptaddress,longbibliography]{revtex4-2}

\usepackage[centertags]{amsmath}
\usepackage{amssymb}
\usepackage{latexsym}
\usepackage{enumerate}
\usepackage{physics}
\usepackage{graphicx}
\usepackage{mathrsfs}
\usepackage[hidelinks]{hyperref}
\usepackage{stmaryrd}
\usepackage{import}
\usepackage{tensor}
\usepackage{xcolor}
\usepackage{bm}
\usepackage{multirow}
\usepackage{breakurl}
\usepackage{float}
\usepackage{lipsum}
\usepackage{siunitx}
\usepackage{comment}
\usepackage{mathtools}
\usepackage{autobreak}
\usepackage{subfig}
\usepackage{orcidlink}
\mathtoolsset{centercolon}
\usepackage{comment}

\newcommand{\be}{\begin{equation}}
\newcommand{\ee}{\end{equation}}
\newcommand{\bse}{\begin{subequations}}
\newcommand{\ese}{\end{subequations}}
\newcommand{\ba}{\begin{align}}
\newcommand{\ea}{\end{align}}
\newcommand{\nn}{\nonumber}
\newcommand{\p}{\partial}

\renewcommand{\dd}{\mathrm{d}}

\newcommand{\dI}{\mathrm{I}}
\newcommand{\dJ}{\mathrm{J}}

\newcommand{\dm}{\mathrm{m}}

\newcommand{\ubar}{\bar{u}}
\newcommand{\vbar}{\bar{v}}

\include{preamble}

\allowdisplaybreaks

\begin{document}

\title{Quasi-Keplerian parametrization for compact binaries on hyperbolic orbits \\in scalar-tensor theories at second post-Newtonian order}

\begin{abstract}

We obtain the generalized quasi-Keplerian parametrization for compact binaries on quasihyperbolic orbits   at second post-Newtonian (2PN) order in a class of massless scalar-tensor theories, extending the analogous results for quasielliptic systems \href{https://doi.org/10.1103/PhysRevD.109.104003}{\mbox{[Phys.Rev.D 109, 104003 (2024)]}}.
In particular, we compute the conservative scattering angle and impact parameter at 2PN.
Our results are consistent with the 2PN conservative scatter-to-bound map in these theories between the scattering angle and the periastron advance.
We then compute the total energy and angular momentum lost by the system and study the limiting cases of parabolic orbits and bremsstrahlung, including a re-expansion of our results at seventh post-Minkowskian order. Flux-balance arguments then allow us to compute the dissipative contributions to the scattering angle at 1.5PN and 2.5PN, completing the full scattering angle at 2.5PN in these theories. Finally, we obtain, \emph{in general relativity}, an expression for the 3PN impact parameter in the conservative sector, correcting previous literature.

\end{abstract}

\author{Davide Usseglio\,\orcidlink{0000-0003-2427-9547}} \email{davide.usseglio-ssm@unina.it}
\affiliation{Scuola Superiore Meridionale, Largo San Marcellino 10, 80138, Naples, Italy}
\affiliation{INFN, Sezione di Napoli, Complesso Universitario di Monte S. Angelo, Via Cintia Edificio 6, 80126, Naples, Italy}

\author{David Trestini\,\orcidlink{0000-0002-4140-0591}}
\email{david.trestini@southampton.ac.uk}
\affiliation{School of Mathematical Sciences and STAG Research Centre, University of Southampton, Southampton, United Kingdom, SO17 1BJ}

\author{Abhishek Chowdhuri\,\orcidlink{0000-0003-4474-790X}}
\email{chowdhuria@tsinghua.edu.cn}
\affiliation{The Institute of Mathematical Sciences, CIT Campus, Tharamani, Chennai 600113, India}
\affiliation{Homi Bhabha National Institute, Training School Complex, Anushakti Nagar, Mumbai 400085, India}
\affiliation{Department of Astronomy, Tsinghua University, Haidian District, Beijing 100084, China}

\maketitle

\section{Introduction}

 In order to make full use of the scientific potential of  gravitational wave detectors, there has been a push toward improving waveform models. Experimentally, the current and future gravitational detectors  are mostly sensitive to gravitational waves generated by compact binaries on bound orbits. One of the most relevant approaches to model these systems is the post-Newtonian (PN) approximation, which yields fully analytical waveforms in the regime of slow velocities and weak fields. This approximation is well adapted  to the inspiral phase of a compact binary on a bound orbit: indeed, the virial theorem guarantees that a largely separated system is automatically slowly orbiting.

The systems detected by LIGO, Virgo, and \mbox{KAGRA} (LVK)  collaboration are typically circularized, but there has recently been a push to extending the parameter space of waveform models to include, for example, eccentricity and spin. This is because future detectors (ET, CE, LISA, TianQin, Taiji, etc.) will detect  a broader range of the parameter space, but a more complete model is also useful to  avoid modeling systematics even in current detectors. Theoretically, it is interesting to study an even broader region of parameter space, such as to include unbound binary systems on hyperbolic orbits, even though gravitational waves generated by these systems are expected to not be detectable. Indeed, the two body problem is very clean and thus useful to study properties of the underlying theory.

However, the growing interest for the unbound problem is not only theoretical. Recent work succeeded in establishing maps between the bound and unbound problems in limiting cases, and there is hope to make these maps more general (see~\cite{Kalin:2019rwq,Kalin:2019inp,Cho:2021arx,Gonzo:2024xjk}). Using such maps, one could use our understanding of the unbound problem to obtain novel results about the bound problem.

The natural framework to study hyperbolic encounters is the post-Minkowskian (PM) approximation, which is a weak field expansion $(G\ll1)$. The scattering trajectory can be seen as small deviation from a straight line, but notably there is no assumption on the values for the relative velocity. State-of-the-art results in this approximation have been obtained using techniques based on quantum field theory and modern scattering amplitude methods \cite{Cheung:2018wkq,Badger:2023eqz}, which have allowed the community to push the analytical knowledge of observables for a scattering scenario at very high orders in perturbation theory (see \cite{Driesse:2024feo,Driesse:2024xad} for state-of-the-art results up to 5PM).

However, the PN expansion is also relevant for the unbound, scattering case only, the absence of a virial theorem in this case means that the slow velocity (post-Newtonian) expansion is orthogonal to the large separation (post-Minkowskian) expansion. Indeed, although post-Minkowksian produce a complete resummation of the PN series at each order in $G$, they are typically harder to obtain, and run into special functions like elliptic integrals, harmonic polylogarithms , and even Calabi-Yau manifolds entering at 5PM~\cite{Driesse:2024xad,Frellesvig:2023bbf,Driesse:2024feo}. Thus, PN methods can be  precious both for cross-checking PM results and completing analytical knowledge of coefficients inaccessible with PM methods.

In PN theory, the hyperbolic problem has mostly been studied as a doppelgänger of the elliptic problem. Indeed, the quasielliptic motion was studied using an elegant generalization of the Kepler solution to first post-Newtonian (1PN) order, dubbed the quasi-Keplerian (QK) parametrization, which was introduced by Damour and Deruelle~\cite{DD85}, and generalized up to 3PN order~\cite{DS88,SW93,MGS04} and partially to 4PN order~\cite{ChoTanay2022}. At 1PN order, Damour and Deruelle showed that the quasihyperbolic motion could be obtained through analytical continuation of the elliptic motion~\cite{DD85}, and Blanchet and Schäfer \cite{BS89} used this result to obtain the energy and angular momentum losses at 1PN order beyond Einstein's quadrupole formula. This statement seems to break down after 3PN order, and the QK parametrization for the quasihyperbolic case~\cite{Cho:2018upo, Bini:2020hmy} was developed independently from the quasielliptic case. Using this parametrization, it was then possible to compute the fluxes at 3PN order~\cite{Cho:2021onr, Cho:2022pqy} and obtain dissipative corrections to, e.g., the scattering angle at 3PN order~\cite{Bini:2017wfr, bini2021radiative}.

All these waveform models have mainly been developed assuming that general relativity (GR) is the underlying theory of gravity, but future detectors might unveil very small deviations to GR, which need to be modeled. One approach to modeling alternative theories is to take a theory-agnostic, parametrized approach~\cite{Will:2018bme,Yunes:2009ke,BSat94,Mishra:2010tp}. Here, we instead take a theory-dependent approach, and derive predictions for the gravitational waveform within the class of massless scalar-tensor (ST) theories of gravity, which were first introduced by Jordan~\cite{jordan1955schwerkraft}, Fierz~\cite{Fierz:1956zz}, and Brans and Dicke~\cite{BransDicke}, and later generalized in~\cite{Nordtvedt:1970uv,Wagoner:1970vr}. At relatively low PN orders, other popular theories, such as scalar Gauss-Bonnet, reduce to the theory under scrutiny in this work~\cite{Shiralilou:2021mfl}. See also Refs.~\cite{Diedrichs:2023foj, Schon:2021pcv, Taherasghari:2023rwn, Higashino:2022izi,Wu:2023jwd} for studies of compact binaries in extensions of these theories and~\cite{AbhishekChowdhuri:2023rfv} for dark matter effects in GR.

Previous results in massless ST~theories using the PN expansion were derived using (i) effective field theory methods~\cite{Dgef92,Dgef93,Dgef94,Dgef96,Dgef98}, (ii)  the direct integration of the relaxed field equations (DIRE)~\cite{Wiseman:1992dv,WW96,Mirshekari:2013vb,Lang:2013fna,Lang:2014osa}, and (iii) the post-Newtonian, multipolar post-Minkowskian (PN-MPM) formalism~\cite{B18_i,B18_ii,BBT22,Bernard:2019yfz,Bernard:2023eul}. It is the latter  that we will use hereafter, as it has been successfully used in GR to compute the energy flux and GW phasing at very high 4.5PN order~\cite{bfhlt_letter}. 

Gravitational radiation in ST theories was first studied by~Refs.~\cite{Dgef92,Dgef93,Damour:1996ke,Dgef96,Dgef98} in an alternative (but equivalent) formulation of the action. The equations of motion and conserved energy and angular momentum were first obtained at 2.5PN order in~\cite{Mirshekari:2013vb}, then at 3PN order in~\cite{B18_i,B18_ii}, including the contribution of the (dipolar) tail term. The waveform and flux were obtained in terms of positions and velocities at 1.5PN beyond the leading quadrupolar radiation, i.e., at 2.5PN beyond the leading dipolar radiation \cite{Lang:2013fna,Lang:2014osa,BBT22}. The fluxes, the phase, as well as the gravitational and scalar waveform amplitudes were then obtained in the case of  circular orbits at 1.5PN~\cite{Sennett:2016klh,BBT22}. In order to extend these results to the case of quasielliptic orbits, the QK parametrization was established at 2PN order by one of us in these theories~\cite{T24_QK}, and used to compute the fluxes and secular evolution of the orbital parameters at 1.5PN beyond the quadrupolar radiation~\cite{Trestini:2024mfs}.

This work aims at extending the QK parametrization for ST theories of Ref.~\cite{T24_QK} to the case of hyperbolic orbits. The main application will be completion of the full scattering angle at 2.5PN order in scalar-tensor theory, including both conservative and dissipative contributions. The conservative sector is in agreement with Ref.~\cite{Jain:2023vlf}, and the dissipative is new. We hope that such results will be useful in particular to benchmark future PM scattering calculations in this class of ST theories.
Indeed, there are currently few PM results for alternative theories of gravity; see, however, \cite{Bini:2024icd, Bhattacharyya:2024aeq,Bhattacharyya:2024kxj}. 

The paper is organized as follows. After a brief notational reminder, we define in Sec.~\ref{sec:STtheory} the exact theories we are considering, and recall the general setup for PN-MPM computation in these theories. In Sec.~\ref{sec:motion}, we derive the QK parametrization for hyperbolic orbits at 2PN order, obtain the gauge invariant scattering angle and impact parameter, and perform a check of the scatter-to-bound map between the scattering angle and periastron advance. In Sec.~\ref{sec:fluxes}, we compute the total radiated energy and angular momentum of the system at Newtonian order beyond Einstein's quadrupole formula in GR, or equivalently, at 1PN order beyond the $-1$PN order dominant dipolar radiation of the scalar field. In Sec.~\ref{sec:dissipative}, we compute the dissipative contributions to a few orbitals elements: the scattering angle, the impact parameter, the time eccentricity and the asymptotic velocity. In Sec.~\ref{sec:parabolic_bremsstrahlung}, we study the parabolic and bremsstrahlung limits of the energy and angular momentum losses, and relegate their expansion at 7PM order to Supplemental Material~\cite{Suppl_Mat}. We finally conclude in Sec.~\ref{sec:conclusion}. In Appendix~\ref{app:QK_details}, we detail the derivation of the QK parametrization, and Appendix~\ref{app:QK_parameters} compiles lengthy results for the conservative QK parameters in scalar-tensor theories. Then, Appendix~\ref{app:diss_quantities} compiles some lengthy radiative corrections to these QK parameters. Finally, in Appendix~\ref{app:b_GR}, we switch back to general relativity and give the 3PN expression of the conservative impact parameter $b$, correcting previous literature~\cite{DeVittori:2014psa, Cho:2018upo, Cho:2021onr}.

Finally, most the results of this paper are given in machine-readable form (Wolfram language) in the Supplemental Material~\cite{Suppl_Mat}. 

\subsection*{Notation conventions and parameter overview} \label{sec:not}

Throughout this work, PN orders are defined relative to the Newtonian limit and the standard quadrupolar gravitational radiation of GR. Accordingly, leading-order dipolar radiation appears at~$-0.5$PN order in the waveform and at~$-1$PN order in the energy flux. The 2PN QK parametrization developed here thus corresponds to a next-to-next-to-leading-order correction, while Newtonian fluxes and waveforms are classified as next-to-leading-order contributions.

We adopt a 3+1 decomposition of spacetime and represent three-dimensional Euclidean vectors using bold symbols. The position of the field point in the center-of-mass (CM) frame is written as $\bm{X} = R\, \bm{N}$, where $\bm{N}$ is a unit vector. In spherical coordinates, this becomes $(R,\Theta,\Phi)$. The coordinate time is denoted by $t$, and the retarded time is given by \mbox{$U = t - R/c$}, and this is distinguished from the QK motion's eccentric anomaly, $u$. Both scalar and spin-weighted (with weight $s = -2$) spherical harmonics are used, denoted by $Y^{\ell \dm}(\Theta,\Phi)$ and $Y^{\ell \dm}_{-2}(\Theta,\Phi)$, respectively. The index $\dm$ should not be mistaken for the total mass $m$.

The individual positions of the two compact objects are denoted $\bm{y}_1$ and $\bm{y}_2$, and their relative separation is given by \mbox{$\bm{x} = \bm{y}_1 - \bm{y}_2$}, with magnitude \mbox{$r = |\bm{x}|$} and direction \mbox{$\bm{n} = \bm{x}/r$}. The relative velocity is defined as \mbox{$\bm{v} = \dd\bm{x}/\dd t$}. For a nonprecessing eccentric binary in the CM frame, we define the orthonormal basis \mbox{$(\bm{n}, \bm{\lambda}, \bm{\ell})$}, where $\bm{\lambda}$ lies within the orbital plane, satisfies \mbox{$\bm{\lambda} \cdot \bm{v} > 0$}, and $\bm{\ell} = \bm{n} \times \bm{\lambda}$. Introducing a fixed inertial triad \mbox{$(\bm{n}_0, \bm{\lambda}_0, \bm{\ell})$}, the orbital motion can be expressed in polar coordinates $(r, \phi)$, such that $\bm{n} = \cos\phi\,\bm{n}_0 + \sin\phi\,\bm{\lambda}_0$. The velocity then becomes \mbox{$\bm{v} = \dot{r}\,\bm{n} + r\dot{\phi}\,\bm{\lambda}$}. This implies the relations \mbox{$\bm{n} \cdot \bm{v} = \dot{r}$}, \mbox{$v^2 = \dot{r}^2 + r^2\dot{\phi}^2$}, and \mbox{$\bm{n} \times \bm{v} = r \dot{\phi} \bm{\ell}$}.

We use multi-index notation, where $L = i_1 \cdots i_\ell$ indicates $\ell$ spatial indices (with $K$ indicating $k$ indices similarly). For derivatives and vector components, we write $\partial_L = \partial_{i_1} \cdots \partial_{i_\ell}$, $\partial_{aL-1} = \partial_a \partial_{i_1} \cdots \partial_{i_{\ell-1}}$, and likewise for products: $n_L = n_{i_1} \cdots n_{i_\ell}$, $n_{aL-1} = n_a n_{i_1} \cdots n_{i_{\ell-1}}$. The symmetric tracefree (STF) part of a tensor is denoted with either a hat or angle brackets: for instance, $\mathrm{STF}_L [\partial_L] =\hat{\partial}L = \partial{\langle i_1} \cdots \partial_{i_\ell \rangle}$, $\mathrm{STF}_L [n_L]=\hat{n}L = n{\langle i_1} \cdots n_{i_\ell \rangle}$, and $\mathrm{STF}_L [x_L] =\hat{x}L = r^\ell n{\langle i_1} \cdots n_{i_\ell \rangle}$. Higher-order time derivatives of a function $F(t)$ are denoted as $F^{(n)}(t) = \dd^n F / \dd t^n$.

The scalar field asymptotically approaches the constant value $\phi_0$ at spatial infinity, and we define the normalized scalar field as $\varphi \equiv \phi / \phi_0$. The scalar coupling function $\omega(\phi)$ is expanded around $\phi_0$, with $\omega_0 \equiv \omega(\phi_0)$. Similarly, the scalar-dependent masses $m_A(\phi)$ (defined in Sec.~\ref{subsec:fieldEquations}) are expanded about $\phi_0$, yielding $m_A \equiv m_A(\phi_0)$ for $A \in {1,2}$. In the CM frame, we define the total mass $m = m_1 + m_2$, the reduced mass $\mu = m_1 m_2 / m$, the symmetric mass ratio $\nu = \mu / m \in \ ]0,1/4]$, and the fractional mass difference $\delta = (m_1 - m_2) / m \in \ [0,1[$. These two mass parameters are related via $\delta^2 = 1 - 4\nu$.

Following~\cite{B18_i, BBT22}, we summarize in Table~\ref{table} a list of ST and PN parameters derived from the expansions of $\omega(\phi)$ and $m_A(\phi)$. The ST parameters arise directly from these expansions, while the PN parameters are specific combinations of the former that extend the standard parametrized post-Newtonian (PPN) formalism to a general ST framework~\cite{will1972conservation,Will:2018bme}.

\begin{widetext}
\begin{small}
\begin{center}\begin{table}[h]
\begin{tabular}{|c||cc|}
	\hline
	& \multicolumn{2}{|c|}{\textbf{ST parameters}} \\[2pt]
	\hline &&\\[-10pt]
	General & \multicolumn{2}{c|}{$\omega_0=\omega(\phi_0),\qquad\omega_0'=\eval{\frac{\dd\omega}{\dd\phi}}_{\phi=\phi_0}, \qquad\omega_0''=\eval{\frac{\dd^2\omega}{\dd\phi^2}}_{\phi=\phi_0},\qquad\varphi = \frac{\phi}{\phi_{0}},\qquad\tilde{g}_{\mu\nu}=\varphi\,g_{\mu\nu},$} \\[12pt]
	& \multicolumn{2}{|c|}{$\tilde{G} = \frac{G(4+2\omega_{0})}{\phi_{0}(3+2\omega_{0})},\qquad \zeta = \frac{1}{4+2\omega_{0}},$} \\[8pt]
	& \multicolumn{2}{|c|}{$\lambda_{1} = \frac{\zeta^{2}}{(1-\zeta)}\left.\frac{\dd\omega}{\dd\varphi}\right\vert_{\varphi=1},\qquad \lambda_{2} = \frac{\zeta^{3}}{(1-\zeta)}\left.\frac{\dd^{2}\omega}{\dd\varphi^{2}}\right\vert_{\varphi=1}, \qquad \lambda_{3} = \frac{\zeta^{4}}{(1-\zeta)}\left.\frac{\dd^{3}\omega}{\dd\varphi^{3}}\right\vert_{\varphi=1}.$} \\[7pt]
	\hline &&\\[-7pt]
	~Sensitivities~ & \multicolumn{2}{|c|}{$s_A = \eval{\frac{\dd \ln{m_A(\phi)}}{\dd\ln{\phi}}}_{\phi=\phi_0},\qquad s_A^{(k)} = \eval{\frac{\dd^{k+1}\ln{m_A(\phi)}}{\dd(\ln{\phi})^{k+1}}}_{\phi=\phi_0},\qquad(A=1,2)$} \\[9pt]
    & \multicolumn{2}{|c|}{$s'_A = s_A^{(1)},\qquad s''_A = s_A^{(2)},\qquad s'''_A = s_A^{(3)},$} \\[5pt]
	& \multicolumn{2}{|c|}{$\mathcal{S}_+ = \frac{1-s_1 - s_2}{\sqrt{\alpha}}\,,\qquad \mathcal{S}_- = \frac{s_2 - s_1}{\sqrt{\alpha}}.$} \\[7pt]	\hline\hline 
	Order & \multicolumn{2}{|c|}{\textbf{PN parameters}} \\[2pt]
	\hline &&\\[-10pt]
	N & \multicolumn{2}{|c|}{$\alpha= 1-\zeta+\zeta\left(1-2s_{1}\right)\left(1-2s_{2}\right)$}   \\[5pt]
	\hline &&\\[-10pt]
	1PN & $\overline{\gamma} = -\frac{2\zeta}{\alpha}\left(1-2s_{1}\right)\left(1-2s_{2}\right),$ & Degeneracy \\[5pt]
	&~~$\overline{\beta}_{1} = \frac{\zeta}{\alpha^{2}}\left(1-2s_{2}\right)^{2}\left(\lambda_{1}\left(1-2s_{1}\right)+2\zeta s'_{1}\right),$~~~~&  $\alpha(2+\overline{\gamma})=2(1-\zeta)$ \\[5pt]
	& $\overline{\beta}_{2} = \frac{\zeta}{\alpha^{2}}\left(1-2s_{1}\right)^{2}\left(\lambda_{1}\left(1-2s_{2}\right)+2\zeta s'_{2}\right),$~~~~& \\[5pt]
	&  $\overline{\beta}_+ = \frac{\overline{\beta}_1+\overline{\beta}_2}{2}, \qquad \overline{\beta}_- = \frac{\overline{\beta}_1-\overline{\beta}_2}{2}.$ &  \\[5pt]
	\hline &\\[-10pt]
	2PN & $\overline{\delta}_{1} = \frac{\zeta\left(1-\zeta\right)}{\alpha^{2}}\left(1-2s_{1}\right)^{2}\,,\qquad \overline{\delta}_{2} = \frac{\zeta\left(1-\zeta\right)}{\alpha^{2}}\left(1-2s_{2}\right)^{2},$ & Degeneracy \\[5pt]
	&  $\overline{\delta}_+ = \frac{\overline{\delta}_1+\overline{\delta}_2}{2}, \qquad \overline{\delta}_- = \frac{\overline{\delta}_1-\overline{\delta}_2}{2},$ &  $16\overline{\delta}_{1}\overline{\delta}_{2} = \overline{\gamma}^{2}(2+\overline{\gamma})^{2}$\\[5pt]
	& $~~\overline{\chi}_{1} = \frac{\zeta}{\alpha^{3}}\left(1-2s_{2}\right)^{3}\left[\left(\lambda_{2}-4\lambda_{1}^{2}+\zeta\lambda_{1}\right)\left(1-2s_{1}\right)-6\zeta\lambda_{1}s'_{1}+2\zeta^{2}s''_{1}\right],~~$  &  \\[5pt]
	& $\overline{\chi}_{2} = \frac{\zeta}{\alpha^{3}}\left(1-2s_{1}\right)^{3}\left[\left(\lambda_{2}-4\lambda_{1}^{2}+\zeta\lambda_{1}\right)\left(1-2s_{2}\right)-6\zeta\lambda_{1}s'_{2}+2\zeta^{2}s''_{2}\right],$ &  \\[5pt]
	&  $\overline{\chi}_+ = \frac{\overline{\chi}_1+\overline{\chi}_2}{2}, \qquad \overline{\chi}_- = \frac{\overline{\chi}_1-\overline{\chi}_2}{2}.$ &  \\[5pt]
\hline
\end{tabular}
\caption{Summary of parameters for the general ST theory and notations for PN parameters. \label{table}}\end{table}
\end{center}
\end{small}
\end{widetext}

\section{Massless scalar-tensor theories}\label{sec:STtheory} 

\label{subsec:fieldEquations}

Following Refs.\cite{MW13, Lang:2013fna, Lang:2014osa, B18_i, B18_ii, BBT22, T24_QK, Trestini:2024mfs}, we consider a general class of ST theories involving a single massless scalar field $\phi$ minimally coupled to the metric~$g_{\mu\nu}$. The dynamics are governed by the action  \begin{align}\label{eq:STactionJF}
S_{\mathrm{ST}} &= \frac{c^{3}}{16\pi G} \int\dd^{4}x\,\sqrt{-g}\left[\phi R - \frac{\omega(\phi)}{\phi}g^{\alpha\beta}\p_{\alpha}\phi\p_{\beta}\phi\right] \nn\\*
&\qquad\quad+S_{\mathrm{m}}\left(\mathfrak{m},g_{\alpha\beta}\right)\,,
\end{align}
where $R$ is the Ricci scalar, $g$ is the determinant of the metric, $\omega$ is a function of the scalar field, and $\mathfrak{m}$ collectively denotes the matter fields.

This is the so-called Jordan frame formulation, in which matter is coupled only to the ``physical" metric $g_{\alpha\beta}$.

In this context, the matter action $S_{\mathrm{m}}$ depends solely on the matter fields and the metric. However, since the strong equivalence principle is violated in ST theories, internal gravitational binding energy must be taken into account. Specifically, the scalar field controls the effective gravitational constant, thereby influencing the balance between gravitational and nongravitational forces within a body. Consequently, the scalar field indirectly affects the body's structure and internal gravity.

To incorporate these effects, we adopt the approach introduced by Eardley~\cite{Eardley1975} (see also~\cite{Nordtvedt:1990zz}), modeling $S_{\mathrm{m}}$ as the effective action of $N$ nonspinning point particles, whose masses $m_A(\phi)$ depend on the local value of the scalar field:  
\be\label{eq:matteract}
S_{\mathrm{m}} = - c \sum_{A} \int\,m_{A}(\phi) \sqrt{-\left(g_{\alpha\beta}\right)_{A}\dd y_{A}^{\alpha}\,\dd y_{A}^{\beta}}\,,
\ee
where $y_A^\alpha$ are the spacetime coordinates of particle $A$, and $\left(g_{\alpha\beta}\right)_{A}$ denotes the metric evaluated\footnote{Divergences are regularized via Hadamard regularization~\cite{BFreg}, which is equivalent at this order to dimensional regularization~\cite{BDE04,BDEI05dr}.} at that location. Through the mass dependence $m_A(\phi)$, the matter action acquires an implicit dependence on the scalar field. One then introduces the sensitivities of each particle to variations in the scalar field via  \be\label{sAk} s_A^{(k)} \equiv \eval{\frac{\dd^{k+1}\ln{m_A(\phi)}}{\dd(\ln{\phi})^{k+1}}}_{\phi=\phi_0}\,, \ee  with $s_A \equiv s_A^{(1)}$, and $\phi_0$ is the asymptotic value of the scalar field, assumed constant in time to neglect cosmological evolution.

For stationary, ordinary black holes, all information
regarding the matter that formed the black hole has disappeared behind the horizon, hence the mass can depend
only on the Planck scale: $m_A \propto M_\text{Planck} \propto G^{-1/2} \propto \phi^{1/2}$. This uniquely fixes the sensitivities to $s^\text{BH}_A={1}/{2}$.

To analyze perturbations in these theories, it is convenient to define a rescaled scalar field and a conformally related metric,  \be\label{eq:def_gt} \varphi\equiv \frac{\phi}{\phi_{0}}\qquad\mathrm{and}\qquad\tilde{g}_{\alpha\beta}\equiv \varphi\, g_{\alpha\beta}\,, \ee  such that both $\tilde{g}_{\alpha\beta}$ and $g_{\alpha\beta}$ share the same asymptotic behavior at spatial infinity. Quantities expressed in terms of $(\varphi, \tilde{g}_{\alpha\beta})$ belong to the so-called Einstein frame.

Perturbations are defined via $\psi\equiv\varphi-1$ and $h^{\mu\nu}\equiv \sqrt{-\tilde{g}}\tilde{g}^{\mu\nu}-\eta^{\mu\nu}$, where $\eta^{\mu\nu}=\text{diag}(-1,1,1,1)$ denotes the Minkowski metric. The field equations in this frame become  
\begin{subequations}\label{eq:rEFE}
\begin{align}
& \Box_{\eta}\,h^{\mu\nu} = \frac{16\pi G}{c^{4}}\tau^{\mu\nu}\,,\\
& \Box_{\eta}\,\psi = -\frac{8\pi G}{c^{4}}\tau_{s}\,,
\end{align}
\end{subequations}
where $\Box_{\eta}$ is the flat-space d'Alembertian operator. The source terms are  
\begin{subequations}
\begin{align}\label{eq:taumunu}
& \tau^{\mu\nu} = \frac{\varphi}{\phi_{0}} (- g) T^{\mu\nu} +\frac{c^{4}}{16\pi G}\Lambda^{\mu\nu}[h,\psi]\,,\\
\label{taus} & \tau_{s} = -\frac{\varphi}{\phi_{0}(3+2\omega)}\sqrt{-g}\left(T-2\varphi\frac{\p T}{\p \varphi}\right) -\frac{c^{4}}{8\pi G}\Lambda_s[h,\psi]\,.
\end{align}
\end{subequations}

Here, $T^{\mu\nu}= 2 (-g)^{-1/2}\delta S_{\mathrm{m}}/\delta g_{\mu\nu}$ is the stress-energy tensor, $T = g_{\mu\nu}T^{\mu\nu}$ its trace, and $\p T/\p \varphi$ is the partial derivative of $T(g_{\mu\nu}, \varphi)$ with respect to $\varphi$ at fixed $g_{\mu\nu}$. The nonlinear source terms $\Lambda^{\mu\nu}$ and $\Lambda^s$ are functionals of $h^{\mu\nu}$ and $\psi$, defined in Eqs.(2.8) and (2.9) of \cite{BBT22}, and start at quadratic order in the fields.

The system \eqref{eq:rEFE} is solved using the PN-MPM algorithm \cite{BD86,Blanchet:2013haa}. For a complete treatment in scalar-tensor theories, see~\cite{BBT22}. In this work, we only require the expressions of the source multipole moments $\dI_L$, $\dJ_L$, and $\dI_L^s$, which are STF functions of retarded time $U$ and fully determine the linearized fields $h_1^{\mu\nu}$ and $\psi_1$. For example, the scalar field reads \be \psi_1 = - \frac{2}{c^2}\sum_{\ell=0}^{+\infty}\frac{(-)^\ell}{\ell!}\partial_L\left[\frac{\dI_L^s(U)}{R}\right].\label{psi1} \ee In scalar-tensor theories, the scalar monopole $\dI^s$ is not constant. However, its variation in time is suppressed by two PN orders: $\dd \dI^s/\dd t = \mathcal{O}(c^{-2})$, implying that dipole radiation dominates at leading order. For convenience, we define \bse\be \dI^s(U) = \frac{1}{\phi_0}\left[ m^s+\frac{E^s(U)}{c^2} \right], \ee with \be m^s = - \frac{1}{3+2\omega_0}\sum_A m_A (1-s_A) \ee\ese where $m^s$ is constant, and $E^s(U)$ encapsulates the PN time dependence.

Far from the source, where $G m/(c^2 R_\mathrm{obs}) \ll 1$, gravitational and scalar radiation can also be expanded in multipoles. Introducing radiative coordinates $(T,R,\bm{N})$ [or equivalently $(T,R,\Theta,\Phi)$], the asymptotic waveform can be expressed in terms of three sets of radiative moments~\cite{Blanchet:2013haa, BBT22}: $\mathcal{U}_L$, $\mathcal{V}_L$, and $\mathcal{U}^s_L$. In transverse-traceless gauge, the waveforms are  \begin{subequations}\label{eq:radiative_expansion}
\begin{align}
 	h_{ij}^\text{TT} &= - \frac{4G}{c^2 R} \perp^\text{TT}_{ijab} \sum_{\ell=2}^{+\infty} \frac{1}{c^\ell \ell!}\Bigg[ N_{L-2}\,\mathcal{U}_{ab L-2}(U) \nn\\*
 	&\qquad- \frac{2\ell}{c(\ell+1)} N_{c L-2} \epsilon_{cd(a}\mathcal{V}_{b)dL-2}(U)\Bigg] + \mathcal{O}\Big(\frac{1}{R^2}\Big)\,,\label{seq:tensor_radiative_expansion}\\
 	\psi &= - \frac{2G}{c^2 R}\sum_{\ell=0}^{+\infty} \frac{1}{c^\ell \ell!} N_L \mathcal{U}_L^s(U) + \mathcal{O}\Big(\frac{1}{R^2}\Big)\,, \label{seq:scalar_radiative_expansion}
\end{align}
\end{subequations}
where \mbox{$\perp_{ijab}^\mathrm{TT} = \perp_{a(i}\perp_{j)b}-\frac{1}{2}\!\perp_{ij}\perp_{ab}$} with \mbox{$\perp_{ij} =\delta_{ij}-N_i N_j$}.
The radiative moments are related to the source moments by the relations
\begin{align}
 \mathcal{U}_L(U) &=  \overset{(\ell)}{\dI}_{\!\!L}(U)+\mathcal{O}\left(\frac{1}{c^3}\right)\,,\\
 \mathcal{V}_L(U) &=  \overset{(\ell)}{\dJ}_{\!\!L}(U) +\mathcal{O}\left(\frac{1}{c^3}\right)\,,\\
\mathcal{U}^s_L(U) &= \overset{(\ell)}{\dI^s }_{\!\!L}(U)+\mathcal{O}\left(\frac{1}{c^3}\right)\,,
 \end{align}
 where the $\mathcal{O}(1/c^3)$ corrections include nonlinear effects from the MPM scheme, such as tails and memory terms (see \cite{BBT22} for details). Since this work focuses only on linear radiation, these corrections will not be considered.

\section{Motion for hyperbolic orbits}
\label{sec:motion}

We now proceed with the computation of the QK parametrization for hyperbolic orbits. We will follow essentially the same steps as Ref.~\cite{T24_QK}, and obtain a closed form, 2PN-accurate, unbound solution to the equations of motion for nonspinning particles. This solution will be parametrized by only two independent variables, the conserved energy $E$ and angular momentum $J = |\bm{J}|$. In practice, we will actually use the \textit{reduced} energy $\bar{\varepsilon}$ and angular momentum $\overline{\jmath}$, which are defined as
\begin{equation}
    \bar{\varepsilon} = \dfrac{2E}{\mu c^2} \hspace{1cm} \text{and} \hspace{1cm}  \overline{\jmath} = \dfrac{2J^2E}{\mu^3 (\tilde{G}\alpha m)}\,.
\end{equation}
Note that these reduced variables are decorated with a bar, to distinguish them from their counterparts in the bound case (see~\cite{T24_QK}). The barred and unbarred energy and angular momentum differ only by a sign, namely, \mbox{$\bar{\varepsilon} = - \varepsilon$} and \mbox{$\bar{\jmath} = - j$}, such that we have \mbox{$\varepsilon>0$} and \mbox{$j>0$} for bound systems and \mbox{$\bar{\varepsilon}>0$} and \mbox{$\bar{\jmath}>0$} for unbound systems.

In Sec.~\ref{subsec:QK_hyperbolic}, we will work out the expressions for the 2PN QK parameters for a hyperbolic encounter in a generic theory of gravity, expressed in terms of the coefficients  (denoted $A$, $B$, $C$, etc.)
parametrizing the polynomial relation linking
the energy and angular momentum of the binary system
to the relative trajectories and velocities (the details of this derivation are relegated to Appendix~\ref{app:QK_details}). In Sec.~\ref{subsec:QK_gauge_invariant}, we discuss the derivation of two special, gauge invariant QK parameters: the scattering angle and impact parameter. The scattering angle will be then related to the periastron advance \textit{via} a scatter-to-bound map in Sec.~\ref{subsec:scatter_to_bound}. 

These coefficients $(A, B, C, \ldots)$ can then be specialized to the case of ST theory by replacing them by their expressions in terms of $(\bar{\varepsilon},\bar{\jmath})$, which were given in Ref.~\cite{T24_QK}. Thanks to the values of these coefficients, we obtain explicit expressions of the various QK parameters in terms of $(\bar{\varepsilon},\bar{\jmath})$. These lengthy expressions are relegated to Appendix~\ref{app:QK_parameters} (see also the Supplemental Material~\cite{Suppl_Mat}).

\subsection{The quasi-Keplerian parametrization for hyperbolic orbits}
\label{subsec:QK_hyperbolic}

Consider the conservative dynamics of a spinless binary system, characterized by $E$ and $J$. The dynamics are thus planar, and can be described by the two equations of motion for, respectively, the radial and angular motion, namely 
\begin{subequations}\label{eq:2PN_motion_radial_angular}\begin{align}\label{eq:2PN_motion_radial}
    \dot{r}^2 &= s^{-4}\dot{s}^2 = \mathcal{R}(s,E,J), \\*
    \dot{\phi} & = \mathcal{S}(s,E,J), \label{eq:2PN_motion_angular}
\end{align}\end{subequations}
where $s\equiv 1/r$ is the inverse of the radial distance between the two compact objects. At this stage, we are working in a generic theory of gravity, that we can  study perturbatively as a PN expansion. In this sense, we can then write both $\mathcal{R}$  and $\mathcal{S}$ for Eq.~\eqref{eq:2PN_motion_radial_angular}  as fifth-order polynomials in $s$, with the following structure
\begin{subequations}\label{eq:defRS}\begin{align}\label{seq:defR}
    \mathcal{R}(s,E,J) &= A + 2 B s + C s^2 + D_1 s^3 + D_2 s^4 + D_3 s^5 , \\*
    \label{seq:defS} \mathcal{S}(s,E,J) &= F s^2 + I_1 s^3 + I_2 s^4 + I_3 s^5 ,
\end{align}\end{subequations}
where $A$, $B$, $C$ and $F$ are of order $\mathcal{O}(1)$, $D_1$ and $I_1$ are of order $\mathcal{O}(c^{-2})$ and $D_2$, $D_3$, $I_2$ and $I_3$ are of order $\mathcal{O}(c^{-4})$. This structure for the equations of motion is satisfied in GR, in ST theories, and potentially in other alternative theories of gravity. Notice the absence of logarithmic terms, which appear at 3PN in GR, and of nonlocal tails, which appear at 4PN in GR but only at 3PN in ST theories.

Both in GR and ST theories, we have:
\begin{itemize}
    \item $A<0$, $B>0$, and $C<0$ for quasielliptic orbits,
    \item $A>0$, $B>0$, and $C<0$ for quasihyperbolic orbits,
    \item $A=0$, $B>0$, and $C<0$ for quasiparabolic orbits.
\end{itemize}
The polynomial $\mathcal{R}(s)$ has exactly two roots that are nonvanishing in the \mbox{$c\rightarrow \infty$} limit, which we denote \mbox{$s_+ \ge s_-$}. In the quasielliptic case studied in Ref.~\cite{T24_QK}, the two roots were positive, namely, \mbox{$0 < s_- < s_+$}, such that the accessible radii were \mbox{$r_p < r < r_a$}, where    the periastron and  apastron were defined as \mbox{$r_p = 1/s_+$}  and \mbox{$r_a = 1/s_-$}, respectively. Here, in the quasihyperbolic case, $s_- < 0 < s_+$, so the negative root does not correspond to anything physical, and the accessible radii correspond to $r > 1/s_+$. Moreover, in the quasihyperbolic case, it is useful to introduce the excess velocity \mbox{$v_\infty \equiv \lim_{s \rightarrow 0} \sqrt{\mathcal{R}(s)} = \sqrt{A}$}.

The QK representation of the motion has already been derived in the case of bound orbits  in terms of $A$, $B$, etc. in Ref.~\cite{T24_QK}.  There, the starting point of the parametrization was the computation of the two physically meaningful gauge invariant periods of the system, namely, the radial period $P$ (associated to the mean motion $n=2\pi / P$) and the increase in true anomaly per radial period, denoted $\Phi$ [associated to the periastron advance $K = \Phi/(2\pi)$]. In this work, we are interested in obtaining the QK parametrization in the unbound case, which is not periodic, so physical periods cannot be naturally defined anymore. Thus, our definition of the analogous unbound quantities $\bar{n}$ and $\bar{K}$ will exhibit a high degree of arbitrariness, and we will exploit this to (i) simplify the expression of the QK parametrization (see Appendix~\ref{app_A} for the detailed derivation) and (ii) immediately recover results from the literature in the GR case \cite{Cho:2018upo,Bini:2020hmy}.

Following the same derivation as in GR~\cite{Cho:2018upo,Bini:2020hmy}, which we describe in detail in Appendix~\ref{app_A}, we obtain the following parametrization: 
\begin{subequations}
\begin{align}
    r &= \bar{a}_r(\bar{e}_r \cosh(\ubar) - 1) \,, \label{eq:radial_unbound_fin} \\
   \bar{n}(t-t_0) &=  \bar{e}_t \sinh(\ubar) -  \ubar + 
    \bar{g}_{t} \, \vbar +  \bar{f}_{t} \, \sin(\vbar)\,, \label{eq:Kepler_unbound_fin} \\
    \frac{\phi-\phi_0}{\bar{K}} &=  \vbar +  \bar{f}_{\phi} \,\sin(2\vbar) +  \bar{g}_{\phi} \,\sin(3\vbar) \,, \label{eq:angular_unbound_fin}
\end{align}
\end{subequations}
where we decorate all QK parameters with a bar to clearly distinguish them from their quasielliptic counterparts. Here, $\ubar$ is a parameter ranging from $-\infty$ to $+\infty$, which we call the \textit{eccentric anomaly}. We have also introduced the \textit{true anomaly} (for unbound orbits), which is defined by 
\begin{equation}\label{def:vbar}
    \vbar = 2 \arctan\left( \sqrt{\dfrac{\bar{e}_\phi+1}{\bar{e}_\phi -1}} \tanh{\dfrac{\ubar}{2}} \right) \,.
\end{equation}
Contrary to the bound case, all three eccentricities, $\bar{e}_r$, $\bar{e}_t$, and $\bar{e}_\phi$, are greater than $1$. Importantly, note that here, we have used a naming convention for $\bar{f}_t$ and $\bar{g}_t$, which is in close analogy to the usual notations in the bound case~\cite{SW93, MGS04, ChoTanay2022}. Some previous papers \cite{Cho:2018upo,Bini:2020hmy} have swapped the two, which can lead to confusion.

The various QK parameters are given in terms of $A$,~$B$,~etc., as follows (see also the Supplemental Material~\cite{Suppl_Mat}):
\begin{widetext}
\begin{subequations}\label{eq:QK_ABC}
\begin{align}
    \bar{n} &= \dfrac{A^{3/2}}{B} \,,\\*
    \bar{K} &= \frac{F}{\sqrt{-C}} +\frac{B \left(3 D_1 F-2 C I_1\right)}{2 (-C)^{5/2}} \nonumber\\*
     &\qquad\qquad+ \frac{1}{16 (-C)^{9/2}}\bigg[ 105 B^2 D_1^2 F-15 A C D_1^2 F-60 B^2 C D_2 F+12 A C^2 D_2 F+140 B^3 D_3 F\nonumber\\*
  &\qquad\qquad -60 A B
   C D_3 F-60 B^2 C D_1 I_1+12 A C^2 D_1 I_1 +24 B^2 C^2 I_2-8 A C^3 I_2-40 B^3 C
   I_3+24 A B C^2 I_3 
       \bigg]\,,\\
     \bar{a}_r & = \frac{B}{A}-\frac{D_1}{2 C} - \frac{\left(2 B D_1^2-2 B C D_2+4 B^2 D_3-A C D_3\right)}{2 C^3}  ,\\ \nonumber
    \bar{e}_r & = \frac{\sqrt{B^2-A C}}{B} +\frac{A \left(2 B^2-A C\right) D_1}{2 B^2 C \sqrt{B^2-A C}}  + \frac{A}{8 B^3 C^3 \left(B^2-A C\right)^{3/2}}\bigg[ 16 B^6 D_1^2-24 A B^4 C D_1^2+5 A^2 B^2 C^2 D_1^2+2 A^3 C^3 D_1^2 \\ 
    & -16 B^6 C D_2+28
   A B^4 C^2 D_2-12 A^2 B^2 C^3 D_2+32 B^7 D_3-64 A B^5 C D_3+36 A^2 B^3 C^2
   D_3-4 A^3 B C^3 D_3 \bigg], \\ \nonumber
   \bar{e}_t & = \frac{\sqrt{B^2-A C}}{B} +\frac{A D_1}{2 C \sqrt{B^2-A C}}  + \frac{A}{8 B C^3 \left(B^2-A C\right)^{3/2}}\bigg[ 8 B^4 D_1^2-12 A B^2 C D_1^2+3 A^2 C^2 D_1^2-8 B^4 C D_2 \\ 
    & +12 A B^2 C^2 D_2-4 A^2
   C^3 D_2+16 B^5 D_3-28 A B^3 C D_3+12 A^2 B C^2 D_3  \bigg],\\ \nonumber
    \bar{e}_\phi & = \frac{\sqrt{B^2-A C}}{B} +\frac{A}{2 B^2 C
   \sqrt{B^2-A C} F} \left[ 3 B^2 D_1 F-2 A C D_1 F-2 B^2 C I_1+2 A C^2 I_1 \right] \\ \nonumber
   & + \frac{A}{8 B^3 C^3 \left(B^2-A C\right)^{3/2} F^2}\bigg[ 26 B^6 D_1^2 F^2-36 A B^4 C D_1^2 F^2+A^2 B^2 C^2 D_1^2 F^2+8 A^3 C^3 D_1^2
   F^2-32 B^6 C D_2 F^2\\ \nonumber
   & +60 A B^4 C^2 D_2 F^2-28 A^2 B^2 C^3 D_2 F^2+79 B^7 D_3
   F^2-165 A B^5 C D_3 F^2+97 A^2 B^3 C^2 D_3 F^2-11 A^3 B C^3 D_3 F^2\\ \nonumber
   &+8 B^6 C
   D_1 F I_1-36 A B^4 C^2 D_1 F I_1+44 A^2 B^2 C^3 D_1 F I_1-16 A^3 C^4 D_1 F
   I_1-16 B^6 C^2 I_1^2+40 A B^4 C^3 I_1^2\\ \nonumber
   &-32 A^2 B^2 C^4 I_1^2+8 A^3 C^5
   I_1^2+16 B^6 C^2 F I_2-32 A B^4 C^3 F I_2+16 A^2 B^2 C^4 F I_2-30 B^7 C F
   I_2+66 A B^5 C^2 F I_3\\ 
   &-42 A^2 B^3 C^3 F I_3+6 A^3 B C^4 F I_3 \bigg],\\
     \bar{f}_{4t} & = -\frac{A^{3/2} \sqrt{B^2-A C} D_3}{2 B (-C)^{5/2}} , \\
     \bar{g}_{4t} & = \frac{A^{3/2} \left(3 D_1^2-4 C D_2+12 B D_3\right)}{8 B (-C)^{5/2}}  , \\
     \bar{f}_{4\phi} & = \frac{B^2-A C}{32 C^4 F^2} \left( D_1^2 F^2-4 C D_2 F^2+20 B D_3 F^2+4 C D_1 F I_1-8 C^2 I_1^2+8 C^2 F I_2-24 B C F
   I_3 \right), \\
     \bar{g}_{4\phi} & = \frac{\left(B^2-A C\right)^{3/2} \left(D_3 F-2 C I_3\right)}{24 C^4 F} \,.
    \end{align}\end{subequations}
\end{widetext}
We have checked explicitly that replacing $A$, $B$, etc., by their GR values yields exactly the QK parametrization of \cite{Cho:2018upo, Bini:2020hmy} (see Table VIII of Ref.~\cite{Bini:2020hmy}) .
In the case of ST theories, the explicit expressions of $A$, $B$, etc., in terms of the reduced energy $\varepsilon = - \bar{\varepsilon}$ and angular momentum $j = - \overline{\jmath}$ are given in (3.10) of \cite{T24_QK}. Replacing these coefficients by their explicit values in Eq.~\eqref{eq:QK_ABC}, we obtain explicit expressions of the QK parameters in terms of $(\bar{\varepsilon}, \bar{\jmath})$ for ST theories, which we relegate to Appendix~\ref{app:QK_parameters} (see also the Supplemental Material~\cite{Suppl_Mat}).  As a consistency test, we verified that we could recover the QK parametrization of GR directly from Eq.~\eqref{eq:QK_param_ST} 
by sending all ST parameters to their GR limits.
\vspace{1cm}
\subsection{Gauge invariant quantities: Conservative scattering angle and impact parameter}
\label{subsec:QK_gauge_invariant}

From this parametrization, we can obtain two physical quantities of interest for the scattering problem:  the scattering angle $\chi$ and the impact parameter $b$. The full 2PN expressions of the scattering angle $\chi$ and the impact parameter $b$ in the case of ST theories, expressed explicitly in terms of the energy and angular momentum, are relegated to Appendix~\ref{app:QK_parameters}.  

\subsubsection{The scattering angle $\chi$}

The scattering angle is defined as $\chi = \Delta \phi - \pi$, where the accumulated azimuthal angle is defined as \mbox{$\Delta \phi \equiv \lim_{t\rightarrow +\infty} \phi(t) - \lim_{t\rightarrow -\infty} \phi(t)$}, as shown in \mbox{\cite{Bini:2012ji,Bini:2017wfr}}. It reads at 2PN order (see also the Supplemental Material~\cite{Suppl_Mat}),
\begin{widetext}
\begin{align} \nonumber
        \chi &= -\pi  + \frac{\sqrt{A}}{C^2}\left(\frac{\left(3 B^2-2 A C\right) D_1 F}{B^2-A C}-2 C I_1\right) + \frac{\sqrt{A}}{24 C^4 \left(B^2-A C\right)^2} \bigg[ 315 B^5 D_1^2 F-570 A B^3 C D_1^2 F+243 A^2 B C^2 D_1^2 F\\ \nonumber
        &\qquad\qquad -180 B^5 C D_2 F+336 A B^3 C^2 D_2 F-156 A^2 B C^3 D_2 F+420 B^6 D_3 F-880 A
        B^4 C D_3 F  +524 A^2 B^2 C^2 D_3 F \\ \nonumber
        &\qquad\qquad -64 A^3 C^3 D_3 F-180 B^5 C D_1 I_1+336 A B^3 C^2 D_1 I_1-156 A^2 B C^3 D_1 I_1  +72 B^5 C^2 I_2 -144 A
        B^3 C^3 I_2 \\ \nonumber
        &\qquad\qquad +72 A^2 B C^4 I_2 -120 B^6 C I_3+272 A B^4 C^2 I_3-184 A^2 B^2 C^3 I_3+32 A^3 C^4 I_3 \bigg] \nonumber \\
        &\qquad +  \dfrac{1}{\sqrt{-C}}\arctan  \left(\frac{B+\sqrt{B^2-A C}}{\sqrt{-A C}}\right) \Bigg[ 4 F  +\frac{B}{C^2}  \left(6 D_1 F-4 C
        I_1\right) \nonumber \\ 
        &\qquad\qquad + \frac{1}{4 C^{4}} \bigg( 105 B^2 D_1^2 F-15 A C D_1^2 F -60 B^2 C D_2 F+12 A C^2 D_2 F+140 B^3 D_3 F-60 A B C D_3 F\\ \nonumber
        &\qquad\qquad \qquad\qquad -60 B^2 C D_1 I_1+12 A C^2 D_1 I_1+24 B^2 C^2
   I_2-8 A C^3 I_2 -40 B^3 C I_3 +24 A B C^2 I_3 \bigg)  \Bigg] .
\end{align}
\end{widetext}
Our scattering angle agrees\footnote{\label{footnote:Ebar_vs_E} The comparison with 
\cite{Bini:2017wfr}
is nontrivial because of different definitions for the energy.
In that paper, $\bar{E}$ is defined in an EOB setting, and we can express our $E$  in terms of their $\bar{E}$ as follows: 
$$E = mc^2 \left( \sqrt{1+2\nu(\sqrt{2 \bar{E} + 1 }-1)} -1\right)\,.$$
When performing the PN expansion of this relation and the required order, we find perfect agreement.}
at 2PN, in the GR case,  with (5.49) of \cite{Bini:2012ji} and with  (45) of \cite{Bini:2017wfr}.
We also find agreement\footnote{\label{footnote:typoJain}The agreement is obtained after a typo is corrected in Ref. \cite{Jain:2023vlf}: the global prefactor of the $C_\text{2PN}^0$ coefficient in Eq. (3.18) of the published version of Ref. \cite{Jain:2023vlf} should be $\frac{1}{4\alpha^2 (1+\alpha^2)^2}$. We also had to account for the different definition of energy described in Footnote~\ref{footnote:Ebar_vs_E}.} at 2PN, in the case of ST theories, with Eqs.~(3.14)-(3.16) of~\cite{Jain:2023vlf}. We relegate our explicit 2PN expression for the scattering angle in ST theories to Appendix~\ref{app:QK_parameters}.

\subsubsection{The impact parameter $b$}

The impact parameter was defined in Refs.~\cite{DeVittori:2014psa, Cho:2018upo} as
\begin{equation}\label{eq:b_def_rv}
    b = \lim_{r \rightarrow \infty} \frac{|\mathbf{r}\times\mathbf{v}|}{v_{\infty}} \,,
\end{equation}
which is not a manifestly gauge invariant definition. 
The structure of Eqs.~\eqref{eq:2PN_motion_radial} and \eqref{seq:defR}, together with Eq.~\eqref{eq:b_def_rv}, means that, at 2PN, the impact parameter then simply reads
\begin{align} \label{eq:b_def_gauge_dependent}
        b &= \frac{F}{\sqrt{A}}\,.
\end{align}
In fact, from the structure of the QK parametrization in~GR~\cite{MGS04,Cho:2018upo}, it is clear that Eq.~\eqref{eq:b_def_gauge_dependent} even holds at 3PN in GR (in ST theories, this structure is spoiled by the tails appearing at 3PN in the equations of motion).

In GR, we have checked explicitly that Eq.~\eqref{eq:b_def_gauge_dependent} yields the same results in both ADM or modified harmonic coordinates at 3PN, suggesting that Eq.~\eqref{eq:b_def_gauge_dependent} is indeed gauge invariant, even if this is not manifest. 
The 3PN impact parameter we found  in this case disagrees with previous literature~\cite{DeVittori:2014psa, Cho:2018upo, Cho:2021onr}. We provide more details about this discrepancy in Appendix~\ref{app:b_GR}, and the corrected expression is given by Eq.~\eqref{eq:b_GR}.

In ST theories, the expression for the impact parameter had never been derived. 
Here, we obtained it for the first time using the expression of $A$ and $F$ in harmonic coordinates~\cite{T24_QK} (the corresponding expressions in ADM coordinates are not known in ST theories).
We relegate the explicit expression for $b$ to Appendix~\ref{app:QK_parameters}, see Eq.~\eqref{subeq:b_ST}. We also provide it in the Supplemental Material~\cite{Suppl_Mat}.

As further confirmation our of result, we also computed the impact parameter in both GR and ST theories using the manifestly gauge invariant definition of Refs.~\cite{Dscatt16,Bini:2018ywr,bini2021radiative}, which reads
\begin{align} \label{eq:b_def_Jpcm}
b &= \frac{J}{p_{\text{cm}} }\,,
\end{align}
where $p_{\text{cm}}$ is defined\footnote{The exact definition is complicated to relate to our variables and does not naturally appear in our formalism, so we refer the reader to Refs.~\cite{Dscatt16,Bini:2018ywr,bini2021radiative} for more details.} in an effective one body (EOB) context as the incoming 3-momentum of either particle in the center-of-mass frame before the scattering event. We found results identical to the ones obtained using Eq.~\eqref{eq:b_def_rv}. Since $p_\text{cm}$ can be related to $E$ in a ``special relativistic", theory-independent way, the definition \eqref{eq:b_def_Jpcm} clarifies why the expression of $b$ in terms of $(E,J)$ is identical in GR and ST theories [compare Eqs.~\eqref{eq:b_GR} and \eqref{subeq:b_ST}].

\subsection{Explicit check of the scatter-to-bound map}
\label{subsec:scatter_to_bound}

It has been proven \cite{Kalin:2019rwq,Kalin:2019inp} on very general grounds that for the conservative dynamics, in the absence of hereditary terms, there is a very simple formula mapping the scattering angle in the unbound configuration to the pericenter advance in the bound configuration,
\begin{align}
    \Delta \Phi(E,J) = \chi(E, J) + \chi(E, -J) \,,
\end{align}
where $\Delta \Phi = 2\pi(K-1)$.
This result is independent of the particular theory under consideration or the perturbation scheme. However, it requires that the motion derives from a Hamiltonian that is local in the positions and velocities. In particular, at 3PN in ST theories or at 4PN in GR, this map has not yet been fully proven due to the presence of hereditary tails in the equations of motion~\cite{BD92,B18_i}, except by adding some additional hypotheses, such as a large eccentricity expansion, to include nonlocal-in-time effects.

In GR, this map was also verified at 3PN order in Sec. 5.5 of ~\cite{Cho:2021arx} for the dissipative dynamics. In this work, we found that our results are in agreement with the predictions of the scatter-to-bound map at 2PN in the class ST theories under scrutiny, where the scattering angle is given by Eq.~\eqref{scattering_angle_ST_cons} and the periastron advance is given by Eq.~(C1b) of \cite{T24_QK}
.

\section{Radiated energy and angular momentum}
\label{sec:fluxes}

In this section, we will compute the instantaneous fluxes of energy and angular momentum for quasihyperbolic motion at Newtonian order, i.e., at 1PN relative order with respect to the leading $-1$PN dipolar contribution \cite{BBT22}. We will then integrate over the entire orbit to obtain the total energy and angular momentum lost by the system after the scattering process.

It was shown in \cite{T24_QK} that the Newtonian fluxes can be expressed in terms of the source moments as follows [neglecting $\mathcal{O}(c^{-3})$ terms]:
\begin{subequations}
\label{eq:fluxes_N}\begin{align}
\mathcal{F} &= 	\frac{G\phi_0}{5c^{5}}\overset{(3)}{\mathrm{I}}_{\!\!ij}\overset{(3)}{\mathrm{I}}_{\!\!ij}  \label{seq:tensor_energy_flux_N} \,,\\
\mathcal{G}_i &=  \frac{2G\phi_0}{5c^{5}}\epsilon_{iab}\overset{(2)}{\mathrm{I}}_{\!\!ak}\overset{(3)}{\mathrm{I}}_{\!\!bk}  \label{seq:tensor_angular_momentum_flux_N}\,,\\
\mathcal{F}^s &=  \frac{G \phi_0 (3+2\omega_0)}{c^3}\Bigg[ \frac{\overset{(2)}{\mathrm{I}}{}^{\!\!s}_{\!\!a}\overset{(2)}{\mathrm{I}}{}^{\!\!s}_{\!\!a} }{3}+ \frac{\overset{(1)}{E}{}^{\!s}\overset{(1)}{E}{}^{\!s}}{\phi_0^2 c^2} + \frac{\overset{(3)}{\mathrm{I}}{}^{\!\!s}_{\!\!ab}\overset{(3)}{\mathrm{I}}{}^{\!\!s}_{\!\!ab} }{30c^2} \Bigg]  \label{seq:scalar_energy_flux_N} \,,\\
\mathcal{G}^s_i &= \frac{G \phi_0 (3+2\omega_0)}{c^3}\epsilon_{iab}\Bigg[\frac{1}{3}\overset{(1)}{\mathrm{I}}{}^{\!\!s}_{\!\!a}\overset{(2)}{\mathrm{I}}{}^{\!\!s}_{\!\!b} +  \frac{1}{15c^2}\overset{(2)}{\mathrm{I}}{}^{\!\!s}_{\!\!ak}\overset{(3)}{\mathrm{I}}{}^{\!\!s}_{\!\!bk}  \Bigg] \,.
\end{align}    
\end{subequations} 

The source moments are, in turn, expressed in terms of the relative positions and velocities of the two particles in (4.6) of \cite{BBT22}, whereas the acceleration needed to compute the time derivatives is given by  (3.10) of~\cite{B18_ii} (but see also~\cite{MW13}). Since  only  nonspinning particles are considered, there is no precession of the orbital plane, so it is possible to write $\mathcal{G}_i = \mathcal{G} \ell_i$ and $\mathcal{G}^s_i = \mathcal{G}^s \ell_i$, where $\bm{\ell}$ is the constant unit vector orthogonal to the orbital plane. 
Finally, only the 1PN truncation of the QK parametrization derived in Sec.~\ref{subsec:QK_hyperbolic} is required, and it allows us to express the fluxes as pure functions of the eccentric anomaly $\bar{u}$. We relegate their explicit expressions in terms of $\bar{u}$ in the Supplemental Material~\cite{Suppl_Mat}.

In this work, we will be interested in the total energy and angular momentum lost by the binary after the scattering event, namely, 
\begin{subequations}
\begin{align}
    \Delta E = - \int_{-\infty}^{+\infty} \mathrm{d} t \left[\mathcal{F}(t)+\mathcal{F}_s(t)\right] \,, \\
     \Delta J = - \int_{-\infty}^{+\infty} \mathrm{d} t \left[\mathcal{G}(t)+\mathcal{G}_s(t)\right] \,.
\end{align}
\end{subequations}
Note that in these conventions, \mbox{$\Delta E <0$} and \mbox{$\Delta J<0$}. In practice, these are computed by performing the following change of variables in the integral: \mbox{$\bar{n}\, \mathrm{d}t = \bar{e}_t \sinh{(\bar{u})}\,\mathrm{d}u + \mathcal{O}(c^{-4})$}, where we have used Eq.~\eqref{eq:radial_unbound_fin}. We then reexpress the fluxes in terms of the pair or variables $(\bar{x},\bar{e}_t)$, where we have defined
\begin{align}
    \bar{x}=\Big(\frac{\tilde{G}\alpha m \bar{K}\bar{n}}{c^3}\Big)^{2/3}
\end{align}
and used the following 1PN-accurate relations, which can be directly obtained from Eq.~\eqref{eq:QK_param_ST}:
\begin{subequations}
    \begin{align}\nonumber
    \overline{\jmath} &= \bar{e}_t^2-1 + \dfrac{1}{4}\big[9+\nu -8(\beta_+ - \beta_-\delta - \bar{\gamma}) \\*
    &\qquad - \bar{e}_t^2 (17-7\nu +8\bar{\gamma})\big]\bar{x} + O(\bar{x}^2)\,,\\ \nonumber
    \bar{\varepsilon} & = \bar{x} - \dfrac{1}{12(\bar{e}_t^2-1)}\big[ 9 +\nu - 8 (\beta_+ - \beta_-\delta - \bar{\gamma}) \\*
    &\qquad + \bar{e}_t^2 (15-\nu + 8\bar{\gamma}) \big]\bar{x}^2 + O(\bar{x}^3)\,.
\end{align}
\end{subequations}
Note that, contrary to the bound case, $\bar{x}$ is not related to any physical frequency, and it is simply a convenient parametrization. 

The expressions of the integrands can then be expressed, up to neglected 2PN terms, as linear combinations of terms \mbox{$\propto  (\bar{e}_t \cosh{\bar{u}} -1)^{-N}$} and \mbox{$\propto  \sinh(\bar{u})(\bar{e}_t \cosh{\bar{u}} -1)^{-N}$}. The integrals are finally performed thanks to the formulas \cite{BS89} 
\begin{subequations}
\begin{align}
\int_{-\infty}^{\infty} \! \frac{\mathrm{d} \bar{u}}{(\bar{e}_t \cosh{\bar{u}} -1)^{N+1}} &=\! \left.  \frac{2}{N!} \frac{\mathrm{d}^N}{\mathrm{d} x^N} \!\!\left(\frac{\arccos(-x/\bar{e_t})}{\sqrt{\bar{e}_t^2 - x^2}}\right) \!  \right|_{x= 1}\!\!\!\! ,\nonumber\\*
&\\
\int_{-\infty}^{\infty} \frac{\mathrm{d} \bar{u} \sinh(\bar{u})}{(\bar{e}_t \cosh{\bar{u}} -1)^{N+1}} &= 0 \,.
\end{align}
\end{subequations}

Finally, we reexpress the losses of energy in terms of $(\bar{\varepsilon},\overline{\jmath})$ using Eq.~\eqref{eq:QK_param_ST}. The total losses of energy and angular momentum are decomposed into scalar and gravitational contributions, namely, \mbox{$\Delta E =  \left[\Delta E\right]_{\text{grav}} + \left[\Delta E\right]_{\text{scalar}} $} and  \mbox{$\Delta J =  \left[\Delta J\right]_{\text{grav}} + \left[\Delta J\right]_{\text{scalar}}$}. We find that these contributions read (see also the Supplemental Material~\cite{Suppl_Mat}):
\begin{widetext}
\begin{subequations}\label{eq:DeltaE_DeltaJ_expressions}
\begin{align}
    \left[\Delta E\right]_{\text{grav}} &= - \frac{170 c^2 (1 + \bar{\gamma}/2) m \nu^2 \overline{\varepsilon}^{7/2} } {3 \overline{\jmath}^{7/2}}\Bigl[\overline{\jmath}^{1/2}\left(1 + \frac{673}{1275} \overline{\jmath}\right) + \left(1 + \frac{366}{425} \overline{\jmath} + \frac{37}{425} \overline{\jmath}^2\right) \bigl(\pi -  \arctan\sqrt{\overline{\jmath}}\bigr)\Bigr]  \,, \label{seq:DeltaE_grav_expression}\\
    \left[\Delta J\right]_{\text{grav}} &= -\frac{24 (1 +  \bar{\gamma}/2) \tilde{G} \alpha m^2 \nu^2 \overline{\varepsilon}^2 }{c \overline{\jmath}^2}\left[\overline{\jmath}^{1/2}\left( 1+ \frac{2}{15} \overline{\jmath}\right) + \left(1 + \frac{7}{15} \overline{\jmath}\right) \bigl(\pi -  \arctan\sqrt{\overline{\jmath}}\bigr)\right] \,, \label{seq:DeltaJ_grav_expression}\\
    \left[\Delta E\right]_{\text{scalar}} &=  -\frac{c^2 m \nu^2 \overline{\varepsilon}^{5/2} }{\overline{\jmath}^2} \Bigg(4 \zeta \mathcal{S}_{-}^2 \bigg\{1 + \frac{1}{\sqrt{\overline{\jmath}}}\left(1 + \frac{1}{3} \overline{\jmath}\right) \bigl(\pi -  \arctan\sqrt{\overline{\jmath}}\bigr)\bigg\} \nonumber \\ 
    &\hspace{0.5cm}+ \overline{\varepsilon} \Biggl\{\frac{208}{9} \bar{\beta}_{+} -  \frac{571}{90} \bar{\gamma} -  \frac{104}{3} \bar{\beta}_{+}^2 \bar{\gamma}^{-1} + \frac{481}{18} \zeta \mathcal{S}_{-}^2 -  \frac{64}{9} \bar{\beta}_{+} \zeta \mathcal{S}_{-}^2 +  \frac{112}{9} \zeta \bar{\gamma} \mathcal{S}_{-}^2 -  \frac{196}{9} \bar{\beta}_{+} \zeta \bar{\gamma}^{-1} \mathcal{S}_{-}^2   \nonumber \\
    & \hspace{1.5cm} + \frac{208}{3} \bar{\beta}_{-}^2 \zeta \bar{\gamma}^{-2} \mathcal{S}_{-}^2 + \frac{208}{3} \bar{\beta}_{+}^2 \zeta \bar{\gamma}^{-2} \mathcal{S}_{-}^2  -  \frac{196}{9} \bar{\beta}_{-} \zeta \bar{\gamma}^{-1} \mathcal{S}_{-} \mathcal{S}_{+} + \frac{416}{3} \bar{\beta}_{-} \bar{\beta}_{+} \zeta \bar{\gamma}^{-2} \mathcal{S}_{-} \mathcal{S}_{+}\nonumber \\
    & \hspace{1.5cm} + \delta\bigg(\frac{64}{9} \bar{\beta}_{-} \zeta \mathcal{S}_{-}^2 -  \frac{116}{9} \bar{\beta}_{-} \zeta \bar{\gamma}^{-1} \mathcal{S}_{-}^2 -  \frac{116}{9} \bar{\beta}_{+} \zeta \bar{\gamma}^{-1} \mathcal{S}_{-} \mathcal{S}_{+}\bigg)  -  \frac{239}{18} \zeta \mathcal{S}_{-}^2 \nu  \nonumber \\
    & \hspace{1.5cm}+ \frac{1}{\overline{\jmath}}\Bigg[\frac{80}{3} \bar{\beta}_{+} -  \frac{55}{6} \bar{\gamma} - 40 \bar{\beta}_{+}^2 \bar{\gamma}^{-1} + \frac{202}{3} \zeta \mathcal{S}_{-}^2 -  \frac{128}{3} \bar{\beta}_{+} \zeta \mathcal{S}_{-}^2 + \frac{140}{3} \zeta \bar{\gamma} \mathcal{S}_{-}^2 -  \frac{20}{3} \bar{\beta}_{+} \zeta \bar{\gamma}^{-1} \mathcal{S}_{-}^2 \nonumber \\
    & \hspace{2.5cm} + 80 \bar{\beta}_{-}^2 \zeta \bar{\gamma}^{-2} \mathcal{S}_{-}^2  + 80 \bar{\beta}_{+}^2 \zeta \bar{\gamma}^{-2} \mathcal{S}_{-}^2 -  \frac{20}{3} \bar{\beta}_{-} \zeta \bar{\gamma}^{-1} \mathcal{S}_{-} \mathcal{S}_{+} + 160 \bar{\beta}_{-} \bar{\beta}_{+} \zeta \bar{\gamma}^{-2} \mathcal{S}_{-} \mathcal{S}_{+}   \nonumber \\
    & \hspace{2.5cm}+ \delta\left(\frac{128}{3} \bar{\beta}_{-} \zeta \mathcal{S}_{-}^2 -  \frac{100}{3} \bar{\beta}_{-} \zeta \bar{\gamma}^{-1} \mathcal{S}_{-}^2 -  \frac{100}{3} \bar{\beta}_{+} \zeta \bar{\gamma}^{-1} \mathcal{S}_{-} \mathcal{S}_{+}\right) -  \frac{62}{3} \zeta \mathcal{S}_{-}^2 \nu   \Bigg] \nonumber \\
    & \hspace{1.5cm} -  \frac{1}{1 + \overline{\jmath}}\Bigg[\frac{59}{6} \zeta \mathcal{S}_{-}^2 + \frac{4}{3} \bar{\beta}_{+} \zeta \mathcal{S}_{-}^2 + 4 \zeta \bar{\gamma} \mathcal{S}_{-}^2 -  \frac{4}{3} \bar{\beta}_{-} \zeta \mathcal{S}_{-}^2 \delta -  \frac{29}{6} \zeta \mathcal{S}_{-}^2 \nu  \nonumber \\
    & \hspace{2.7cm} +  \overline{\jmath} \left( \frac{17}{6} \zeta \mathcal{S}_{-}^2 +  \frac{4}{3} \zeta \bar{\gamma} \mathcal{S}_{-}^2 - \frac{7}{6} \zeta \mathcal{S}_{-}^2 \nu\right)  + \frac{4 \zeta \mathcal{S}_{-}^2}{\overline{\jmath}} \left( 1 +  \bar{\beta}_{+}  -  \bar{\beta}_{-}  \delta -   \nu\right)\Bigg]\nonumber \\
    & \hspace{0.5cm} + \bigl(\pi -  \arctan\sqrt{\overline{\jmath}}\bigr) \Bigg[\frac{1}{\sqrt{\overline{\jmath}}}\Bigg(32 \bar{\beta}_{+} -  \frac{47}{5} \bar{\gamma} - 48 \bar{\beta}_{+}^2 \bar{\gamma}^{-1} + 42 \zeta \mathcal{S}_{-}^2 - 20 \bar{\beta}_{+} \zeta \mathcal{S}_{-}^2 + 24 \zeta \bar{\gamma} \mathcal{S}_{-}^2 - 24 \bar{\beta}_{+} \zeta \bar{\gamma}^{-1} \mathcal{S}_{-}^2\nonumber \\
    & \hspace{2.5cm}  + 96 \bar{\beta}_{-}^2 \zeta \bar{\gamma}^{-2} \mathcal{S}_{-}^2 + 96 \bar{\beta}_{+}^2 \zeta \bar{\gamma}^{-2} \mathcal{S}_{-}^2 - 24 \bar{\beta}_{-} \zeta \bar{\gamma}^{-1} \mathcal{S}_{-} \mathcal{S}_{+} + 192 \bar{\beta}_{-} \bar{\beta}_{+} \zeta \bar{\gamma}^{-2} \mathcal{S}_{-} \mathcal{S}_{+}\nonumber \\
    & \hspace{2.5cm} + \delta\Big(20 \bar{\beta}_{-} \zeta \mathcal{S}_{-}^2 - 24 \bar{\beta}_{-} \zeta \bar{\gamma}^{-1} \mathcal{S}_{-}^2 - 24 \bar{\beta}_{+} \zeta \bar{\gamma}^{-1} \mathcal{S}_{-} \mathcal{S}_{+}\Big)  - 18 \zeta \mathcal{S}_{-}^2 \nu \Bigg)\nonumber \\
    & \hspace{1.5cm} + \frac{1}{\overline{\jmath}^{3/2}}\Bigg(\frac{80}{3} \bar{\beta}_{+} -  \frac{55}{6} \bar{\gamma} - 40 \bar{\beta}_{+}^2 \bar{\gamma}^{-1} + \frac{190}{3} \zeta \mathcal{S}_{-}^2 -  \frac{140}{3} \bar{\beta}_{+} \zeta \mathcal{S}_{-}^2 + \frac{140}{3} \zeta \bar{\gamma} \mathcal{S}_{-}^2 -  \frac{20}{3} \bar{\beta}_{+} \zeta \bar{\gamma}^{-1} \mathcal{S}_{-}^2\nonumber \\
    & \hspace{2.5cm} + 80 \bar{\beta}_{-}^2 \zeta \bar{\gamma}^{-2} \mathcal{S}_{-}^2 + 80 \bar{\beta}_{+}^2 \zeta \bar{\gamma}^{-2} \mathcal{S}_{-}^2 -  \frac{20}{3} \bar{\beta}_{-} \zeta \bar{\gamma}^{-1} \mathcal{S}_{-} \mathcal{S}_{+} + 160 \bar{\beta}_{-} \bar{\beta}_{+} \zeta \bar{\gamma}^{-2} \mathcal{S}_{-} \mathcal{S}_{+}\nonumber \\
    & \hspace{2.5cm} + \delta\Big(\frac{140}{3} \bar{\beta}_{-} \zeta \mathcal{S}_{-}^2 -  \frac{100}{3} \bar{\beta}_{-} \zeta \bar{\gamma}^{-1} \mathcal{S}_{-}^2 -  \frac{100}{3} \bar{\beta}_{+} \zeta \bar{\gamma}^{-1} \mathcal{S}_{-} \mathcal{S}_{+}\Big) -  \frac{50}{3} \zeta \mathcal{S}_{-}^2 \nu \Bigg)\nonumber \\
    & \hspace{1.5cm} + \sqrt{\overline{\jmath}} \Bigg(\frac{16}{3} \bar{\beta}_{+} -  \frac{13}{10} \bar{\gamma} - 8 \bar{\beta}_{+}^2 \bar{\gamma}^{-1} + \frac{13}{3} \zeta \mathcal{S}_{-}^2 + \frac{4}{3} \zeta \bar{\gamma} \mathcal{S}_{-}^2 -  \frac{20}{3} \bar{\beta}_{+} \zeta \bar{\gamma}^{-1} \mathcal{S}_{-}^2 \nonumber \\
    & \hspace{2.5cm} + 16 \bar{\beta}_{-}^2 \zeta \bar{\gamma}^{-2} \mathcal{S}_{-}^2  + 16 \bar{\beta}_{+}^2 \zeta \bar{\gamma}^{-2} \mathcal{S}_{-}^2 -  \frac{20}{3} \bar{\beta}_{-} \zeta \bar{\gamma}^{-1} \mathcal{S}_{-} \mathcal{S}_{+} + 32 \bar{\beta}_{-} \bar{\beta}_{+} \zeta \bar{\gamma}^{-2} \mathcal{S}_{-} \mathcal{S}_{+} \nonumber \\
    & \hspace{2.5cm} -\delta \Big( \frac{4}{3} \bar{\beta}_{-} \zeta \bar{\gamma}^{-1} \mathcal{S}_{-}^2 +  \frac{4}{3} \bar{\beta}_{+} \zeta \bar{\gamma}^{-1} \mathcal{S}_{-} \mathcal{S}_{+}\Big) -  \frac{7}{3} \zeta \mathcal{S}_{-}^2 \nu  \Bigg) \Bigg]\Biggr\}\Bigg) \,, \label{seq:DeltaE_scalar_expression}\\
    \left[\Delta J\right]_{\text{scalar}} &= -\frac{2  \tilde{G} \alpha m^2 \nu^2\overline{\varepsilon} }{3 c \overline{\jmath}} \Bigg(4 \zeta \mathcal{S}_{-}^2 \bigg\{\overline{\jmath}^{1/2} + \bigl(\pi -  \arctan\sqrt{\overline{\jmath}}\bigr)\bigg\}\nonumber \\
    &  - \overline{\varepsilon} \Bigg\{ \frac{1}{\sqrt{\overline{\jmath}}}\Bigg[  16 \bar{\beta}_{+} \zeta \mathcal{S}_{-}^2+3\bar{\gamma} - 38 \zeta \mathcal{S}_{-}^2  - 24 \zeta \bar{\gamma} \mathcal{S}_{-}^2 - 24 \bar{\beta}_{+} \zeta \bar{\gamma}^{-1} \mathcal{S}_{-}^2 - 24 \bar{\beta}_{-} \zeta \bar{\gamma}^{-1} \mathcal{S}_{-} \mathcal{S}_{+} - 20 \zeta \mathcal{S}_{-}^2 \nu\nonumber \\
    & \hspace{1.7cm} - \delta \Big(16 \bar{\beta}_{-} \zeta \mathcal{S}_{-}^2- 24 \bar{\beta}_{-} \zeta \bar{\gamma}^{-1} \mathcal{S}_{-}^2 - 24 \bar{\beta}_{+} \zeta \bar{\gamma}^{-1} \mathcal{S}_{-} \mathcal{S}_{+}\Big)  \Bigg] \nonumber \\
    & \hspace{0.5cm} + \overline{\jmath}^{1/2} \Bigg[ \frac{2}{5} \bar{\gamma} - 17 \zeta \mathcal{S}_{-}^2 - 8 \zeta \bar{\gamma} \mathcal{S}_{-}^2 + 7 \zeta \mathcal{S}_{-}^2 \nu\Bigg]\nonumber \\
    & \hspace{0.5cm} + \frac{1}{1 + \overline{\jmath}}\Bigg[\overline{\jmath}^{3/2} \Big( \frac{17}{2} \zeta \mathcal{S}_{-}^2 + 4 \zeta \bar{\gamma} \mathcal{S}_{-}^2 - \frac{7}{2} \zeta \mathcal{S}_{-}^2 \nu\Big) +  \frac{8\zeta \mathcal{S}_{-}^2}{\overline{\jmath}^{1/2}}\Big( 1 +  \bar{\beta}_{+} -  \bar{\beta}_{-}  \delta -   \nu\Big)\nonumber \\
    & \hspace{1.9cm} + \overline{\jmath}^{1/2} \Bigg(21 \zeta \mathcal{S}_{-}^2 + 4 \bar{\beta}_{+} \zeta \mathcal{S}_{-}^2 + 8 \zeta \bar{\gamma} \mathcal{S}_{-}^2 - 4 \bar{\beta}_{-} \zeta \mathcal{S}_{-}^2 \delta - 11 \zeta \mathcal{S}_{-}^2 \nu \Bigg)\Bigg] \nonumber \\
    & \hspace{0.5cm} - \bigl(\pi -\arctan\sqrt{\overline{\jmath}}\bigr)\Bigg[ 31 \zeta \mathcal{S}_{-}^2 - \frac{7}{5} \bar{\gamma}  - 4 \bar{\beta}_{+} \zeta \mathcal{S}_{-}^2 + 16 \zeta \bar{\gamma} \mathcal{S}_{-}^2 + 8 \bar{\beta}_{+} \zeta \bar{\gamma}^{-1} \mathcal{S}_{-}^2 + 8 \bar{\beta}_{-} \zeta \bar{\gamma}^{-1} \mathcal{S}_{-} \mathcal{S}_{+}\nonumber \\
    & \hspace{3.5cm} + \delta \Big(4 \bar{\beta}_{-} \zeta \mathcal{S}_{-}^2 - 8 \bar{\beta}_{-} \zeta \bar{\gamma}^{-1} \mathcal{S}_{-}^2 - 8 \bar{\beta}_{+} \zeta \bar{\gamma}^{-1} \mathcal{S}_{-} \mathcal{S}_{+} \Big)  - 15 \zeta \mathcal{S}_{-}^2 \nu \nonumber\\
    & \hspace{2cm} -  \frac{1}{\overline{\jmath}}\bigg(3 \bar{\gamma} - 38 \zeta \mathcal{S}_{-}^2 + 16 \bar{\beta}_{+} \zeta \mathcal{S}_{-}^2 - 24 \zeta \bar{\gamma} \mathcal{S}_{-}^2   - 24 \bar{\beta}_{+} \zeta \bar{\gamma}^{-1} \mathcal{S}_{-}^2 - 24 \bar{\beta}_{-} \zeta \bar{\gamma}^{-1} \mathcal{S}_{-} \mathcal{S}_{+}  \nonumber \\
    & \hspace{3cm} -  \delta \big(16 \bar{\beta}_{-} \zeta \mathcal{S}_{-}^2 - 24 \bar{\beta}_{-} \zeta \bar{\gamma}^{-1} \mathcal{S}_{-}^2 - 24 \bar{\beta}_{+} \zeta \bar{\gamma}^{-1} \mathcal{S}_{-} \mathcal{S}_{+}\big)  + 20 \zeta \mathcal{S}_{-}^2 \nu \bigg) \nonumber\\
    & \hspace{2cm} -  \frac{1}{1 + \overline{\jmath}}\bigg(25 \zeta \mathcal{S}_{-}^2 + 8 \bar{\beta}_{+} \zeta \mathcal{S}_{-}^2 + 8 \zeta \bar{\gamma} \mathcal{S}_{-}^2 - 8 \bar{\beta}_{-} \zeta \mathcal{S}_{-}^2 \delta - 15 \zeta \mathcal{S}_{-}^2 \nu\nonumber \\
    & \hspace{3cm} + \overline{\jmath} \left(17+ 8  \bar{\gamma}  - 7  \nu\right)\zeta \mathcal{S}_{-}^2 + \frac{8\zeta \mathcal{S}_{-}^2}{\overline{\jmath}} \left(1 +  \bar{\beta}_{+}  -  \bar{\beta}_{-}  \delta -   \nu\right)\bigg)\Bigg] \Bigg\}\Bigg) \,. \label{seq:DeltaJ_scalar_expression}
\end{align}
\end{subequations}
\end{widetext}

\section{Dissipative contributions to  orbital elements}
\label{sec:dissipative}

In the full dissipative problem, the radiation reaction makes the QK parameters vary throughout the motion by a small 1.5PN correction. For any QK parameter $\xi$, we denote its initial value (before the scattering event) by $\xi^-$, and its final value (after the scattering event) by $\xi^+$. We denote the difference by $\Delta \xi = \xi^+ - \xi^-$. Since a system now has evolving energy and angular momentum, one needs to choose a convention in order to uniquely characterize a scattering configuration. We choose to systematically describe a system using the initial energy and angular momentum $(E^-,J^-)$. 
Note, however, that $\Delta E$ and $\Delta J$  in \eqref{eq:DeltaE_DeltaJ_expressions} are only given to relative 1PN accuracy, so it does not matter whether they are computed with the initial or final QK parameters. 

The evolution of the QK parameters can be obtained at low orders using the linear response formula~\cite{Bini:2012ji,bini2021radiative,Bini:2022enm}
\begin{align}\label{eq:linear_response}
    \Delta \xi = \frac{\partial\xi^\text{cons}}{\partial E} \Delta E + \frac{\partial\xi^\text{cons}}{\partial J} \Delta J .
\end{align}
Applying this formula, we compute $\Delta b$, $\Delta v_\infty$ and $\Delta \bar{e}_t$, whose explicit expressions are relegated to Appendix~\ref{app:diss_quantities} (see also the Supplemental Material~\cite{Suppl_Mat}). 

The case of the scattering angle requires particular attention. Although one can formally define a scattering angle before and after the process (denoted $\chi^-$ and $\chi^+$, respectively), it makes physically more sense to treat this quantity globally. The total scattering angle $\chi$, including both conservative and dissipative contributions, is in fact the average of the two aforementioned quantities, and reads
\begin{align}
    \chi &= \frac{\chi^+ + \chi^-}{2} = \chi^- + \frac{\Delta \chi}{2} \,.
\end{align}
Identifying the conservative contribution as the one before radiation, \mbox{$\chi^{\text{cons}}(E^-,J^-) = \chi^-$}, it follows that the dissipative contribution is given by 
\begin{align}\label{eq:linear_response_chi}
\chi^\text{diss} = \frac{1}{2}\left[\frac{\partial \chi^\text{cons}}{\partial E} \Delta E+\frac{\partial \chi^\text{cons}}{\partial J} \Delta J\right] \,.
\end{align}
The formula \eqref{eq:linear_response_chi} was initially obtained in GR in Eq. (5.99) of \cite{Bini:2012ji}, but only at leading order in the radiation reaction, namely 2.5PN. It was later shown to hold even at 4.5PN thanks to the time-antisymmetric (or time-odd) character of the radiation reaction force, see Eq.~(3.25)~of~\cite{bini2021radiative}. It was then shown that time-\emph{asymmetric} contributions to the radiation reaction force would induce an extra term, denoted~$\Delta c_\phi$ in Eq.~(12.17) of~\cite{Bini:2022enm}, which would only start contributing at the order of radiation reaction squared, namely, 5PN in GR (see also Appendix~C of~\cite{Bini:2022enm} for details). These arguments immediately translate to ST theories, such that the radiation reaction terms are time-antisymmetric at this order, and the extra $\Delta c_\phi$ contribution can safely be ignored (it will only enter at the order of radiation reaction squared, namely, 3PN).
Thus, the complete 2.5PN scattering angle for ST theories finally reads 
\begin{align}
\chi &= \chi^\text{cons} + \chi^\text{diss} \,,
\end{align}
where $\chi^\text{cons}$ is given in Eq.~\eqref{scattering_angle_ST_cons} and where the dissipative contribution reads (see also the Supplemental Material~\cite{Suppl_Mat}):
\begin{widetext}
\begin{align}
    \chi_\text{diss} &= m \overline{\varepsilon}^{3/2} \nu^2 \Bigg\{\dfrac{4 \zeta \mathcal{S}_{-}^2}{\overline{\jmath}^2} \left[\dfrac{1}{\overline{\jmath}^{1/2}(1 + \overline{\jmath})}\left( 1 + \dfrac{2}{3}\overline{\jmath}\right) + \pi -  \arctan\sqrt{\overline{\jmath}}\right] \nonumber \\
    & + \frac{\overline{\varepsilon}}{(1 + \overline{\jmath})^2} \Bigg[\frac{1}{\overline{\jmath}^{3/2}}\Bigg(\frac{4976}{45} + \frac{448}{9} \bar{\beta}_{+} + \frac{340}{9} \bar{\gamma} -  \frac{224}{3} \bar{\beta}_{+}^2 \bar{\gamma}^{-1} + \frac{2191}{18} \zeta \mathcal{S}_{-}^2 -  \frac{724}{9} \bar{\beta}_{+} \zeta \mathcal{S}_{-}^2 + \frac{772}{9} \zeta \bar{\gamma} \mathcal{S}_{-}^2  \nonumber \\
    & \hspace{2.7cm}-  \frac{112}{9} \bar{\beta}_{+} \zeta \bar{\gamma}^{-1} \mathcal{S}_{-}^2+ \frac{448}{3} \bar{\beta}_{-}^2 \zeta \bar{\gamma}^{-2} \mathcal{S}_{-}^2 + \frac{448}{3} \bar{\beta}_{+}^2 \zeta \bar{\gamma}^{-2} \mathcal{S}_{-}^2 -  \frac{112}{9} \bar{\beta}_{-} \zeta \bar{\gamma}^{-1} \mathcal{S}_{-} \mathcal{S}_{+}  -  \frac{641}{18} \zeta \mathcal{S}_{-}^2 \nu \nonumber \\
    & \hspace{2.7cm}+ \frac{896}{3} \bar{\beta}_{-} \bar{\beta}_{+} \zeta \bar{\gamma}^{-2} \mathcal{S}_{-} \mathcal{S}_{+}+ \delta  \left(\frac{724}{9} \bar{\beta}_{-} \zeta \mathcal{S}_{-}^2 -  \frac{560}{9} \bar{\beta}_{-} \zeta \bar{\gamma}^{-1} \mathcal{S}_{-}^2 -  \frac{560}{9} \bar{\beta}_{+} \zeta \bar{\gamma}^{-1} \mathcal{S}_{-} \mathcal{S}_{+}\right) \Bigg) \nonumber \\
    & \hspace{1.7cm} + \frac{1}{\overline{\jmath}^{1/2}}\Bigg(\frac{514}{9} + \frac{208}{9} \bar{\beta}_{+} + \frac{359}{18} \bar{\gamma} -  \frac{104}{3} \bar{\beta}_{+}^2 \bar{\gamma}^{-1} + \frac{1189}{18} \zeta \mathcal{S}_{-}^2 -  \frac{280}{9} \bar{\beta}_{+} \zeta \mathcal{S}_{-}^2 + \frac{376}{9} \zeta \bar{\gamma} \mathcal{S}_{-}^2  \nonumber \\
    & \hspace{2.7cm} -  \frac{52}{9} \bar{\beta}_{+} \zeta \bar{\gamma}^{-1} \mathcal{S}_{-}^2 + \frac{208}{3} \bar{\beta}_{-}^2 \zeta \bar{\gamma}^{-2} \mathcal{S}_{-}^2 + \frac{208}{3} \bar{\beta}_{+}^2 \zeta \bar{\gamma}^{-2} \mathcal{S}_{-}^2 -  \frac{52}{9} \bar{\beta}_{-} \zeta \bar{\gamma}^{-1} \mathcal{S}_{-} \mathcal{S}_{+}  -  \frac{395}{18} \zeta \mathcal{S}_{-}^2 \nu \nonumber \\
    & \hspace{2.7cm} + \frac{416}{3} \bar{\beta}_{-} \bar{\beta}_{+} \zeta \bar{\gamma}^{-2} \mathcal{S}_{-} \mathcal{S}_{+} + \delta \left(\frac{280}{9} \bar{\beta}_{-} \zeta \mathcal{S}_{-}^2 -  \frac{260}{9} \bar{\beta}_{-} \zeta \bar{\gamma}^{-1} \mathcal{S}_{-}^2 -  \frac{260}{9} \bar{\beta}_{+} \zeta \bar{\gamma}^{-1} \mathcal{S}_{-} \mathcal{S}_{+}\right)  \Bigg) \nonumber \\
    & \hspace{1.7cm} + \frac{1}{\overline{\jmath}^{5/2}}\Bigg(\frac{170}{3} + \frac{80}{3} \bar{\beta}_{+} + \frac{115}{6} \bar{\gamma} - 40 \bar{\beta}_{+}^2 \bar{\gamma}^{-1} + \frac{190}{3} \zeta \mathcal{S}_{-}^2 -  \frac{140}{3} \bar{\beta}_{+} \zeta \mathcal{S}_{-}^2 + \frac{140}{3} \zeta \bar{\gamma} \mathcal{S}_{-}^2  \nonumber \\
    & \hspace{2.7cm} -  \frac{20}{3} \bar{\beta}_{+} \zeta \bar{\gamma}^{-1} \mathcal{S}_{-}^2 + 80 \bar{\beta}_{-}^2 \zeta \bar{\gamma}^{-2} \mathcal{S}_{-}^2 + 80 \bar{\beta}_{+}^2 \zeta \bar{\gamma}^{-2} \mathcal{S}_{-}^2 -  \frac{20}{3} \bar{\beta}_{-} \zeta \bar{\gamma}^{-1} \mathcal{S}_{-} \mathcal{S}_{+}  -  \frac{50}{3} \zeta \mathcal{S}_{-}^2 \nu \nonumber \\
    & \hspace{2.7cm} + 160 \bar{\beta}_{-} \bar{\beta}_{+} \zeta \bar{\gamma}^{-2} \mathcal{S}_{-} \mathcal{S}_{+} + \delta\left(\frac{140}{3} \bar{\beta}_{-} \zeta \mathcal{S}_{-}^2 -  \frac{100}{3} \bar{\beta}_{-} \zeta \bar{\gamma}^{-1} \mathcal{S}_{-}^2 -  \frac{100}{3} \bar{\beta}_{+} \zeta \bar{\gamma}^{-1} \mathcal{S}_{-} \mathcal{S}_{+}\right)  \Bigg) \nonumber \\
    & \hspace{1.7cm} + \overline{\jmath}^{1/2} \left(\frac{16}{5} + \frac{4}{3} \bar{\gamma} + \frac{32}{3} \zeta \mathcal{S}_{-}^2 + \frac{16}{3} \zeta \bar{\gamma} \mathcal{S}_{-}^2 -  \frac{8}{3} \zeta \mathcal{S}_{-}^2 \nu\right) \Bigg]  \nonumber \\
    & + \frac{\overline{\varepsilon}}{(1 + \overline{\jmath})^2} \bigl(\pi -  \arctan\sqrt{\overline{\jmath}}\bigr)\Bigg[ \frac{242}{15} + \frac{16}{3} \bar{\beta}_{+} + \frac{35}{6} \bar{\gamma} - 8 \bar{\beta}_{+}^2 \bar{\gamma}^{-1} + \frac{193}{6} \zeta \mathcal{S}_{-}^2 \nonumber \\
    & \hspace{1.7cm}  - 8 \bar{\beta}_{+} \zeta \mathcal{S}_{-}^2 + \frac{56}{3} \zeta \bar{\gamma} \mathcal{S}_{-}^2 -  \frac{4}{3} \bar{\beta}_{+} \zeta \bar{\gamma}^{-1} \mathcal{S}_{-}^2 + 16 \bar{\beta}_{-}^2 \zeta \bar{\gamma}^{-2} \mathcal{S}_{-}^2 + 16 \bar{\beta}_{+}^2 \zeta \bar{\gamma}^{-2} \mathcal{S}_{-}^2 -  \frac{4}{3} \bar{\beta}_{-} \zeta \bar{\gamma}^{-1} \mathcal{S}_{-} \mathcal{S}_{+} \nonumber \\
    & \hspace{1.7cm} + 32 \bar{\beta}_{-} \bar{\beta}_{+} \zeta \bar{\gamma}^{-2} \mathcal{S}_{-} \mathcal{S}_{+} + \delta\left(8 \bar{\beta}_{-} \zeta \mathcal{S}_{-}^2 -  \frac{20}{3} \bar{\beta}_{-} \zeta \bar{\gamma}^{-1} \mathcal{S}_{-}^2 -  \frac{20}{3} \bar{\beta}_{+} \zeta \bar{\gamma}^{-1} \mathcal{S}_{-} \mathcal{S}_{+}\right)  -  \frac{47}{6} \zeta \mathcal{S}_{-}^2 \nu \nonumber \\
    & \hspace{1.7cm} + \frac{1}{\overline{\jmath}^2}\Bigg(\frac{1942}{15} + \frac{176}{3} \bar{\beta}_{+} + \frac{265}{6} \bar{\gamma} - 88 \bar{\beta}_{+}^2 \bar{\gamma}^{-1} + \frac{953}{6} \zeta \mathcal{S}_{-}^2 -  \frac{304}{3} \bar{\beta}_{+} \zeta \mathcal{S}_{-}^2 + 112 \zeta \bar{\gamma} \mathcal{S}_{-}^2  \nonumber \\
    & \hspace{2.7cm} -  \frac{44}{3} \bar{\beta}_{+} \zeta \bar{\gamma}^{-1} \mathcal{S}_{-}^2 + 176 \bar{\beta}_{-}^2 \zeta \bar{\gamma}^{-2} \mathcal{S}_{-}^2 + 176 \bar{\beta}_{+}^2 \zeta \bar{\gamma}^{-2} \mathcal{S}_{-}^2 -  \frac{44}{3} \bar{\beta}_{-} \zeta \bar{\gamma}^{-1} \mathcal{S}_{-} \mathcal{S}_{+}  -  \frac{247}{6} \zeta \mathcal{S}_{-}^2 \nu \nonumber \\
    & \hspace{2.7cm} + 352 \bar{\beta}_{-} \bar{\beta}_{+} \zeta \bar{\gamma}^{-2} \mathcal{S}_{-} \mathcal{S}_{+} + \delta\left(\frac{304}{3} \bar{\beta}_{-} \zeta \mathcal{S}_{-}^2 -  \frac{220}{3} \bar{\beta}_{-} \zeta \bar{\gamma}^{-1} \mathcal{S}_{-}^2 -  \frac{220}{3} \bar{\beta}_{+} \zeta \bar{\gamma}^{-1} \mathcal{S}_{-} \mathcal{S}_{+}\right)  \Bigg) \nonumber \\
    & \hspace{1.7cm} + \frac{1}{\overline{\jmath}}\Bigg(\frac{1334}{15} + \frac{112}{3} \bar{\beta}_{+} + \frac{185}{6} \bar{\gamma} - 56 \bar{\beta}_{+}^2 \bar{\gamma}^{-1} + \frac{383}{3} \zeta \mathcal{S}_{-}^2 -  \frac{188}{3} \bar{\beta}_{+} \zeta \mathcal{S}_{-}^2 + 84 \zeta \bar{\gamma} \mathcal{S}_{-}^2 \nonumber \\
    & \hspace{2.7cm} -  \frac{28}{3} \bar{\beta}_{+} \zeta \bar{\gamma}^{-1} \mathcal{S}_{-}^2 + 112 \bar{\beta}_{-}^2 \zeta \bar{\gamma}^{-2} \mathcal{S}_{-}^2 + 112 \bar{\beta}_{+}^2 \zeta \bar{\gamma}^{-2} \mathcal{S}_{-}^2 -  \frac{28}{3} \bar{\beta}_{-} \zeta \bar{\gamma}^{-1} \mathcal{S}_{-} \mathcal{S}_{+} -  \frac{97}{3} \zeta \mathcal{S}_{-}^2 \nu  \nonumber \\
    & \hspace{2.7cm} + 224 \bar{\beta}_{-} \bar{\beta}_{+} \zeta \bar{\gamma}^{-2} \mathcal{S}_{-} \mathcal{S}_{+} + \delta\left(\frac{188}{3} \bar{\beta}_{-} \zeta \mathcal{S}_{-}^2 -  \frac{140}{3} \bar{\beta}_{-} \zeta \bar{\gamma}^{-1} \mathcal{S}_{-}^2 -  \frac{140}{3} \bar{\beta}_{+} \zeta \bar{\gamma}^{-1} \mathcal{S}_{-} \mathcal{S}_{+}\right)  \Bigg) \nonumber \\
    & \hspace{1.7cm} + \frac{1}{\overline{\jmath}^3}\Bigg(\frac{170}{3} + \frac{80}{3} \bar{\beta}_{+} + \frac{115}{6} \bar{\gamma} - 40 \bar{\beta}_{+}^2 \bar{\gamma}^{-1} + \frac{190}{3} \zeta \mathcal{S}_{-}^2 -  \frac{140}{3} \bar{\beta}_{+} \zeta \mathcal{S}_{-}^2 + \frac{140}{3} \zeta \bar{\gamma} \mathcal{S}_{-}^2  \nonumber \\
    & \hspace{2.7cm} -  \frac{20}{3} \bar{\beta}_{+} \zeta \bar{\gamma}^{-1} \mathcal{S}_{-}^2 + 80 \bar{\beta}_{-}^2 \zeta \bar{\gamma}^{-2} \mathcal{S}_{-}^2 + 80 \bar{\beta}_{+}^2 \zeta \bar{\gamma}^{-2} \mathcal{S}_{-}^2 -  \frac{20}{3} \bar{\beta}_{-} \zeta \bar{\gamma}^{-1} \mathcal{S}_{-} \mathcal{S}_{+} -  \frac{50}{3} \zeta \mathcal{S}_{-}^2 \nu  \nonumber \\
    & \hspace{2.7cm} + 160 \bar{\beta}_{-} \bar{\beta}_{+} \zeta \bar{\gamma}^{-2} \mathcal{S}_{-} \mathcal{S}_{+} + \delta\left(\frac{140}{3} \bar{\beta}_{-} \zeta \mathcal{S}_{-}^2 -  \frac{100}{3} \bar{\beta}_{-} \zeta \bar{\gamma}^{-1} \mathcal{S}_{-}^2 -  \frac{100}{3} \bar{\beta}_{+} \zeta \bar{\gamma}^{-1} \mathcal{S}_{-} \mathcal{S}_{+}\right)  \Bigg) \nonumber \\
    & \hspace{1.7cm} + \frac{16 \zeta \mathcal{S}_{-}^2\bigl(\pi -  \arctan\sqrt{\overline{\jmath}}\bigr)}{\overline{\jmath}^{5/2}}\left( \overline{\jmath} + 1 \right)^2\left( 1 -  \frac{1}{3} \bar{\beta}_{+} + \frac{2}{3} \bar{\gamma} + \frac{1}{3} \bar{\beta}_{-} \delta \right)  \Bigg]\Bigg\} \,.
\end{align}
\end{widetext}

We point out that, in this case as well, the GR limits of $\Delta \bar{e}_t$, $\Delta b$, and $\Delta v_\infty$ are in agreement with Eq.~(5.110) of \cite{Bini:2012ji} and Eqs.~(49) and (50) of \cite{Junker:1992kle}, respectively.
Taking the GR limit of the various ST parameters, this expression for the scattering angle is in agreement with Eq.~(5.115) of \cite{Bini:2012ji}. Although Ref.~\cite{Jain:2023vlf} obtained the 3PN-accurate result  for the scattering angle in the \emph{conservative} sector, we cannot claim with this result the completion of the full 3PN scattering angle. The first reason is that at 3PN, the hereditary tail appearing in the equations of motion also has a dissipative contribution, which we did not compute. The second reason is that 3PN is also the order of radiation reaction squared, where the linear response formula breaks down. The third reason is that the energy and angular momentum in the linear response formula was only needed at 1PN, so we did not need to include dissipative Schott terms~\cite{IW95, BFT24} : these enter the energy at 1.5PN in ST theories~\cite{Trestini:2024mfs}, are nonvanishing for hyperbolic orbits~\cite{Bini:2012ji}, and would be thus expected to contribute to the 3PN scattering angle.

\vspace{1cm}
\section{Parabolic and bremsstrahlung limits}
\label{sec:parabolic_bremsstrahlung}

\subsection{Parabolic limit}
\label{subsec:parabolic}

A quasiparabolic orbit, in the full dissipative problem, is defined as the orbit that is marginally bound, namely, whose final energy after the scattering event vanishes ($E^+ =0$), or equivalently, whose final eccentricity  goes to 1 ($e_t^+ = e_r^+ = e_\phi^+ = 1$). We first want to determine the corresponding initial energy $E^-$ in terms of our only degree of freedom, the initial angular momentum $J^-$. Since $\Delta E \sim \mathcal{O}(c^{-3})$, we can obtain $E^- = - \Delta E$ by setting $E=0$ and $J = J^-$ in \eqref{eq:DeltaE_DeltaJ_expressions} (its expression in terms of $J^+$ would be identical). We then obtain $J^+ = J^- + \Delta J$, where $\Delta J$ is obtained using \eqref{eq:DeltaE_DeltaJ_expressions} by setting again $E=0$ and $J = J^-$.  We finally find that in the parabolic limit, we have (see also the Supplemental Material~\cite{Suppl_Mat}):
\begin{widetext}
 \begin{subequations}\begin{align}
    \left. \Delta E \right|_{\mathrm{parabolic}}  &=  - \frac{c^2 m \pi \overline{\varepsilon}^{5/2} \nu^2}{\overline{\jmath}^{5/2}} \Bigg\{4 \zeta \mathcal{S}_{-}^2  + \frac{\overline{\varepsilon}}{\overline{\jmath}}\Bigg[\frac{170}{3} + \frac{80}{3} \bar{\beta}_{+} + \frac{115}{6} \bar{\gamma} - 40 \bar{\beta}_{+}^2 \bar{\gamma}^{-1} + \frac{190}{3} \zeta \mathcal{S}_{-}^2 -  \frac{140}{3} \bar{\beta}_{+} \zeta \mathcal{S}_{-}^2  + \frac{140}{3} \zeta \bar{\gamma} \mathcal{S}_{-}^2  \nonumber \\
    & \hspace{5cm}-  \frac{20}{3} \bar{\beta}_{+} \zeta \bar{\gamma}^{-1} \mathcal{S}_{-}^2  + 80 \bar{\beta}_{-}^2 \zeta \bar{\gamma}^{-2} \mathcal{S}_{-}^2 + 80 \bar{\beta}_{+}^2 \zeta \bar{\gamma}^{-2} \mathcal{S}_{-}^2  \nonumber \\
    & \hspace{5cm} -  \frac{20}{3} \bar{\beta}_{-} \zeta \bar{\gamma}^{-1} \mathcal{S}_{-} \mathcal{S}_{+}  + 160 \bar{\beta}_{-} \bar{\beta}_{+} \zeta \bar{\gamma}^{-2} \mathcal{S}_{-} \mathcal{S}_{+}  -  \frac{50}{3} \zeta \mathcal{S}_{-}^2 \nu \nonumber \\
    & \hspace{5cm} + \delta\left(\frac{140}{3} \bar{\beta}_{-} \zeta \mathcal{S}_{-}^2 -  \frac{100}{3} \bar{\beta}_{-} \zeta \bar{\gamma}^{-1} \mathcal{S}_{-}^2 -  \frac{100}{3} \bar{\beta}_{+} \zeta \bar{\gamma}^{-1} \mathcal{S}_{-} \mathcal{S}_{+}\right) \Bigg]\Bigg\}\,, \\
    \left. \Delta J\right|_{\mathrm{parabolic}}  &= - \frac{2 \alpha \tilde{G} m^2 \pi \overline{\varepsilon} \nu^2}{3 c \overline{\jmath}} \Bigg\{\! 4 \zeta \mathcal{S}_{-}^2 \! +   \frac{\overline{\varepsilon}}{\overline{\jmath}} \Bigg[36 + 15 \bar{\gamma} + 30 \zeta \mathcal{S}_{-}^2 \! - 24 \bar{\beta}_{+} \zeta \mathcal{S}_{-}^2 \! + 24 \zeta \bar{\gamma} \mathcal{S}_{-}^2  \!  + 24 \bar{\beta}_{+} \zeta \bar{\gamma}^{-1} \mathcal{S}_{-}^2  \!  + 24 \bar{\beta}_{-} \zeta \bar{\gamma}^{-1} \mathcal{S}_{-} \mathcal{S}_{+} \nonumber \\*
    & \hspace{5cm} + \delta\Big(24 \bar{\beta}_{-} \zeta \mathcal{S}_{-}^2 - 24 \bar{\beta}_{-} \zeta \bar{\gamma}^{-1} \mathcal{S}_{-}^2 - 24 \bar{\beta}_{+} \zeta \bar{\gamma}^{-1} \mathcal{S}_{-} \mathcal{S}_{+}\Big)  - 12 \zeta \mathcal{S}_{-}^2 \nu\Bigg]\Bigg\} \,.
\end{align}\end{subequations}
\end{widetext}
We have checked that these expressions are in full agreement with the parabolic limit ($e_t \rightarrow 1^-$) of the radiated energy and angular momentum during one radial period of an elliptic binary given in Eqs.~(4.11) and (4.12) of \cite{T24_QK}. Note that in the hyperbolic case, we have computed the energy loss, whereas in the elliptic case, it is the (orbit-averaged) energy \textit{flux} that is obtained, so we have to account for a factor of the radial period $P$ when comparing the two expressions.

Of particular interest will be the initial eccentricity of parabolic orbit, which is found under the constraint that  $\left.e_t^+\right|_\text{parabolic} = 1$, and reads (see also the Supplemental Material~\cite{Suppl_Mat})
\begin{widetext}
\begin{align}
    \left. \bar{e}_t^- \right|_{\mathrm{parabolic}} = 1 +  \frac{m \pi \overline{\varepsilon}^{3/2} \nu^2}{\overline{\jmath}^{3/2}} &\Bigg\{4 \zeta \mathcal{S}_{-}^2   + \frac{\overline{\varepsilon}}{\overline{\jmath}} \Bigg[\frac{170}{3} + \frac{80}{3} \bar{\beta}_{+} + \frac{115}{6} \bar{\gamma} - 40 \bar{\beta}_{+}^2 \bar{\gamma}^{-1} + \frac{214}{3} \zeta \mathcal{S}_{-}^2 -  \frac{116}{3} \bar{\beta}_{+} \zeta \mathcal{S}_{-}^2 + \frac{140}{3} \zeta \bar{\gamma} \mathcal{S}_{-}^2 \nonumber \\
    & \hspace{0.7cm} -  \frac{20}{3} \bar{\beta}_{+} \zeta \bar{\gamma}^{-1} \mathcal{S}_{-}^2 + 80 \bar{\beta}_{-}^2 \zeta \bar{\gamma}^{-2} \mathcal{S}_{-}^2 + 80 \bar{\beta}_{+}^2 \zeta \bar{\gamma}^{-2} \mathcal{S}_{-}^2 -  \frac{20}{3} \bar{\beta}_{-} \zeta \bar{\gamma}^{-1} \mathcal{S}_{-} \mathcal{S}_{+} -  \frac{74}{3} \zeta \mathcal{S}_{-}^2 \nu \nonumber \\
    & \hspace{0.7cm} + 160 \bar{\beta}_{-} \bar{\beta}_{+} \zeta \bar{\gamma}^{-2} \mathcal{S}_{-} \mathcal{S}_{+} + \delta\left(\frac{116}{3} \bar{\beta}_{-} \zeta \mathcal{S}_{-}^2 -  \frac{100}{3} \bar{\beta}_{-} \zeta \bar{\gamma}^{-1} \mathcal{S}_{-}^2 -  \frac{100}{3} \bar{\beta}_{+} \zeta \bar{\gamma}^{-1} \mathcal{S}_{-} \mathcal{S}_{+}\right)  \Bigg]\Bigg\} \,.
\end{align}
\end{widetext}
In particular, this result can be useful when integrating hereditary effects over bound orbits. See Sec. VI.E.2 of Ref.~\cite{Trestini:2024mfs} for more details.

\subsection{Bremsstrahlung limit}
\label{subsec:bremsstrahlung}

We consider the bremsstrahlung limit of a hyperbolic encounter, which corresponds to the large eccentricity expansion or, alternatively, to the limit for $J^-\rightarrow\infty$, in GR (see \cite{BS89, Cho:2021onr}). It is worth noticing that this limit corresponds to the PM expansion around straight-line motion, which is precisely the limiting case of a scattering event when the two bodies are far from each other. 

In this case,  we can directly compute the energy and angular momentum loss, which read in the bremsstrahlung limit (see also the Supplemental Material~\cite{Suppl_Mat})
\begin{widetext}\begin{subequations}\begin{align}
    \left. \Delta E \right|_{\mathrm{bremsstrahlung}}  = - \frac{c^2 m \pi \overline{\varepsilon}^{5/2} \nu^2}{\overline{\jmath}^{3/2}} & \Bigg\{\frac{2}{3} \zeta \mathcal{S}_{-}^2 \nonumber   + \overline{\varepsilon} \Bigg[\frac{37}{15} + \frac{8}{3} \bar{\beta}_{+} + \frac{7}{12} \bar{\gamma} - 4 \bar{\beta}_{+}^2 \bar{\gamma}^{-1} + \frac{13}{6} \zeta \mathcal{S}_{-}^2 + \frac{2}{3} \zeta \bar{\gamma} \mathcal{S}_{-}^2 \nonumber  \\* 
    & \hspace{2cm} -  \frac{10}{3} \bar{\beta}_{+} \zeta \bar{\gamma}^{-1} \mathcal{S}_{-}^2  + 8 \bar{\beta}_{-}^2 \zeta \bar{\gamma}^{-2} \mathcal{S}_{-}^2 + 8 \bar{\beta}_{+}^2 \zeta \bar{\gamma}^{-2} \mathcal{S}_{-}^2 \nonumber  \\ 
    & \hspace{2cm} -  \frac{10}{3} \bar{\beta}_{-} \zeta \bar{\gamma}^{-1} \mathcal{S}_{-} \mathcal{S}_{+} + 16 \bar{\beta}_{-} \bar{\beta}_{+} \zeta \bar{\gamma}^{-2} \mathcal{S}_{-} \mathcal{S}_{+} \nonumber  \\ 
    & \hspace{2cm} + \delta\left(- \frac{2}{3} \bar{\beta}_{-} \zeta \bar{\gamma}^{-1} \mathcal{S}_{-}^2 -  \frac{2}{3} \bar{\beta}_{+} \zeta \bar{\gamma}^{-1} \mathcal{S}_{-} \mathcal{S}_{+}\right)  -  \frac{7}{6} \zeta \mathcal{S}_{-}^2 \nu\Bigg]\Bigg\} \,,\\
    \left. \Delta J\right|_{\mathrm{bremsstrahlung}}  = - \frac{\alpha \tilde{G} m^2 \overline{\varepsilon} \nu^2 }{c \overline{\jmath}^{1/2}} & \Bigg\{\frac{8}{3} \zeta \mathcal{S}_{-}^2 + \overline{\varepsilon} \left(\frac{16}{5} + \frac{4}{3} \bar{\gamma} + \frac{17}{3} \zeta \mathcal{S}_{-}^2 + \frac{8}{3} \zeta \bar{\gamma} \mathcal{S}_{-}^2 -  \frac{7}{3} \zeta \mathcal{S}_{-}^2 \nu\right)\Bigg\} \,,
\end{align}\end{subequations}\end{widetext}
where the $\Delta E \sim \mathcal{O}(G^{3})$ and $\Delta J \sim \mathcal{O}(G^{2})$ in both the scalar and gravitational sector. These scalings are consistent with results from scattering computations in the PM framework, see~\cite{bini2021radiative,Barack:2023oqp}, where the leading contributions to $\Delta E$ and $\Delta J$ are identified as 3PM and 2PM effects, respectively.

In the Supplemental Material~\cite{Suppl_Mat}, we provide not only the bremsstrahlung limit, but also the subleading corrections in the PM expansion, up to 7PM order at the corresponding reference PN accuracy. As expected, we find a vanishing contribution to the angular momentum loss at 7PM order in both the scalar and gravitational sectors. While the leading gravitational contribution at 7PM is indeed zero, computations performed in~\cite{bini2021radiative} [see Eqs.~(E8)–(E9)] proved that the subleading terms are nonvanishing. However, in the scalar sector, both the leading and subleading contributions vanish at this PM order. Therefore, at the specific PN accuracy considered here, 7PM corrections to the angular momentum loss are vanishing.

\section{Conclusion}
\label{sec:conclusion}

In this work, we have computed the QK parameters for scattering events in massless ST theories, extending previous results that were limited to systems in quasielliptic orbits~\cite{T24_QK, Trestini:2024mfs}. This generalization allowed us to evaluate the energy and angular momentum losses at Newtonian order, namely, 1PN beyond the leading $-1$PN dipolar radiation of the scalar field. With the QK parametrization and the fluxes in hand, we were able to compute the
the leading- and subleading-order dissipative contributions to the QK parameters, which arise  both from the scalar and gravitational sectors. As a result, we completed the analytical description of the scattering angle in this class of theories up to 2.5PN order, including both conservative and dissipative contributions. We also checked that our scattering angle is consistent with the scatter-to-bound prescription. Additionally, we derived explicit expressions for two relevant limiting cases of the dynamics—namely, the parabolic and bremsstrahlung regimes. The latter, in particular, may be of interest to the community investigating scattering events in GR using effective and quantum field theory techniques.
As a byproduct of our work, we revisited the result at 3PN for the impact parameter $b$ in GR, by pointing out inconsistencies in the literature. 

From our perspective, the main challenge in extending this analysis to higher PN orders lies in the appearance of nonlocal-in-time effects in the scalar sector, which first arise at 3PN order. While, in principle, it would be possible to compute these contributions in the dissipative sector, as demonstrated in~\cite{Trestini:2024mfs}, such calculations fall beyond the scope of the present work. We plan to address these contributions in future studies.

\acknowledgments
D.U. thanks Donato Bini for useful conversations and clarifications.
D.U. also acknowledges support from Salvatore Capozziello and thanks Matteo Luca Ruggiero and Lorenzo Fatibene for the hospitality at the Mathematics Department at University of Turin, where most of these calculations where performed.
D.T. thanks Tamanna Jain and Thibault Damour for interesting discussions. A.C. would like to thank Bala Iyer for encouragement and some useful comments on the project during his visit to ICTS, Saswat Tanay for useful discussions, and Huan~Yang for critical comments. A.C. also expresses gratitude to the LUX (Observatoire de Paris) for its warm hospitality during the initial stages of this work.

D.T. acknowledges support from the ERC Consolidator/UKRI Frontier Research Grant GWModels (selected by the ERC and funded by UKRI [grant number EP/Y008251/1]). A.C. is funded by a postdoctoral fellowship (grant number 202504) from the Department of Astronomy of Tsinghua University.  A.C. would also like to acknowledge financial support of IMSc, Chennai for the Institute postdoctoral fellowship from DAE, Government of India, during his postdoctoral tenure, where most of this work was done.

\appendix

\section{Details of the quasi-Keplerian construction at 2PN order}\label{app_A}
\label{app:QK_details}

Unlike in the bound case, there is no natural definition for $\bar{P}$ and $\bar{K}$, so we will instead first solve~\eqref{eq:2PN_motion_radial_angular} and then define $\bar{P}$ and $\bar{K}$ so as to simplify the final QK parametrization. First, note that the polynomial $\mathcal{R}(s)$ has, just like in the bound case, only two real roots, which are nonvanishing in the $c\rightarrow \infty$, which we denote $s_- < 0 < s_p$. Thus, $s_-$ is nonphysical, whereas $r_+ = 1/s_+$ is the radius of closest approach of the encounter. These roots are obtained at leading order  by factorizing the $c\rightarrow \infty$ limit of $\mathcal{R}(s)$, which is of degree 2. Then, PN corrections to these roots are obtained iteratively by requiring that they cancel $\mathcal{R}(s)$, ignoring higher-order PN corrections. One can then factorize the polynomial as 
\begin{align}
    \mathcal{R}(s) = (s_+-s)(s-s_-) \tilde{\mathcal{R}}(s) \,.
\end{align}
The QK parameters \mbox{$\bar{a}_r = - (s_+ + s_-)/(2s_+ s_-) > 0$} and \mbox{$\bar{e}_r = (s_+-s_-)/(s_++s_-) > 1$}  are then trivial to obtain, where we recall $s_{\pm}=1/r_{\pm}$.
Following~\cite{Cho:2018upo}, it is now very useful to introduce the auxiliary variable
\begin{equation}\label{def:vbarprime}
    \vbar' = 2 \arctan\left( \sqrt{\dfrac{\bar{e}_r+1}{\bar {e}_r -1}} \tanh{\dfrac{\ubar}{2}} \right)\,,
\end{equation}
which differs from $\vbar$ defined in \eqref{def:vbar} by the choice of eccentricity. Note that at this stage, $\bar{e}_\phi$ and $\bar{e}_t$ are undetermined, only $\bar{e}_r$ is known.

To determine the Kepler equation, we first integrate \eqref{eq:2PN_motion_radial} to find 
\begin{equation}
    |t - \,t_0| = \int_{s}^{s_+} \!\frac{\mathrm{d} s}{s^2\sqrt{\mathcal{R}(s)}} = \int^{s_+}_{s}\!\!\ ds \, \dfrac{\overline{P}(s)}{s^2 \sqrt{(s_+-s)(s-s_-)}},
\end{equation}
where $t_0$ is the time of closest approach and where $\overline{P}(s)$ arises from the PN expansion of $1/\sqrt{\tilde{R}\smash{(s)}}$. 

We have now reduced the problem  to computing master integrals of the form
\begin{equation}\label{Eq:generic_integral_form}
    \bar{\mathcal{J}}_i = \int^{s_+}_s ds \dfrac{s^{i-2}}{\sqrt{(s_+-s)(s-s_-)}}\,.
\end{equation}
We find that they explicitly are given by
\begin{subequations}
\begin{align}
     \bar{\mathcal{J}}_0 & = \frac{\sinh (\ubar) \left(s_+ - s_- \right) - \ubar \left(s_- + s_+\right)}{2 \left(-s_- s_+\right){}^{3/2}}, \\
     \bar{\mathcal{J}}_1 & = \frac{\ubar}{\sqrt{-s_- s_+}}, \\
     \bar{\mathcal{J}}_2 & = \vbar' , \\
     \bar{\mathcal{J}}_3 & = \frac{1}{2} \sin \left(\vbar'\right) \left(s_+ - s_-\right)+\frac{1}{2} \left(s_+ + s_-\right) \vbar' ,\\ \nonumber
     \bar{\mathcal{J}}_4 & = \frac{1}{16} \sin \left(2 \vbar'\right) \left(s_+ - s_-\right)^2+\frac{1}{2} \sin \left(\vbar'\right) \left(s_+^2-s_-^2\right) \\
    & +\frac{1}{8} \left(3 s_+^2+2 s_+ s_- +3 s_-^2\right) \vbar' , \\ \nonumber
     \bar{\mathcal{J}}_5 & = \frac{1}{96} \sin \left(3 \vbar'\right) \left(s_+ - s_-\right){}^3 \\ \nonumber
    &+\frac{3}{32} \sin \left(2 \vbar'\right) \left(s_+ - s_-\right)^2
   \left(s_+ + s_-\right) \\ \nonumber
   & + \frac{3}{32} \sin \left(\vbar'\right) \left(s_+ - s_-\right) \left(5 s_+^2+6 s_+ s_- + 5 s_-^2\right) \\
   & + \frac{1}{16} \left(s_++s_-\right)
   \left(5 s_+^2 - 2 s_+ s_-  +5 s_-^2\right) \vbar' \,.
\end{align}   
\end{subequations}

Replacing $s_+$ and~$s_-$ by their 2PN-accurate expressions, we find  an intermediate form for the Kepler equation,
\begin{equation}\label{Eq:Intermediate_Kepler_Eq1}
   t-t_0 = - \bar{\kappa}_0\,\ubar + \bar{\kappa}_1 \sinh(\ubar)  + \bar{\kappa}_2\, \vbar' + \bar{\kappa}_3\, \sin(\vbar') \,.
\end{equation}

Similarly, the computation of the angular equation starts by integrating Eq.~\eqref{eq:2PN_motion_angular}  to find
\begin{equation} \label{seq:time_integral}   \left|\phi - \phi_0\right| =  \int_{s}^{s_+} \frac{\mathrm{d} s \,\mathcal{S}(s)}{s^2\sqrt{\mathcal{R}(s)}} = \int_{s}^{s_+} \frac{\mathrm{d} s\, \overline{Q}(s)}{s^2 \sqrt{(s_+-s)(s-s_-)} } \,,  \end{equation}
where $\phi_0$ is the reference angle at $t_0$ and where $\overline{Q}(s)$ arises from the PN expansion of $\mathcal{S}(s)/\sqrt{\tilde{R}\smash{(s)}}$. The master integrals of \eqref{Eq:generic_integral_form} apply, and we find an intermediate form of the angular equation to be
\begin{equation}\label{eq:Intermediate_angEq}
    \phi-\phi_0 = \bar{\lambda}_0 \vbar' + \bar{\lambda}_1\sin(\vbar') + \bar{\lambda}_2\sin(2\vbar') + \bar{\lambda}_3\sin(3\vbar') \,. 
\end{equation}

We now have to express the Kepler and angular equations in terms of $\vbar$. First, we parametrize the 2PN-accurate relation between~$\bar{e}_r$ and~$\bar{e}_\phi$~with the ansatz
\begin{equation}
    \bar{e}_r = \bar{e}_\phi\left(1+\dfrac{\alpha}{c^2} +\dfrac{\beta}{c^4}\right),
\end{equation}
and PN-expand Eq. \eqref{def:vbarprime}. We then find that $\alpha$ and $\beta$ are uniquely determined by the condition that the term proportional to $\sin(\vbar)$ in the 2PN-accurate angular equation must vanish.
 
Thus, we are left with the following parametrization:
\begin{subequations}\begin{align}
    t-t_0 &=  - \bar{\kappa}_0'\, \ubar + \bar{\kappa}_1 \sinh(\ubar) + 
    \bar{\kappa}'_2 \, \vbar + \bar{\kappa}'_3 \, \sin(\vbar), \label{eq:Kepler_unbound_intermediate} \\
    \phi-\phi_0 &= \bar{\lambda}_0' \vbar + \bar{\lambda}'_2\,\sin(2\vbar) + \bar{\lambda}'_3\,\sin(3\vbar), \label{eq:angular_unbound_intermediate}
\end{align}\end{subequations}
where we have introduced new primed constants. Recalling that there is no natural definition of $\bar{n}$ and $\bar{K}$ in the unbound case, we finalize the Kepler and angular equations by requiring that the lower-order PN terms are factored in an analogous way as the bound case. We thus introduce
\begin{align}
    \bar{n} &= \frac{1}{\bar{\kappa}_0'} & \text{and} &&
    \bar{K} &= \bar{\lambda}_0' \,
\end{align}
and obtain the final parametrization
\begin{subequations}    
\begin{align}
    r &= \bar{a}_r(\bar{e}_r \cosh(\ubar) - 1)\,, \\
   \bar{n}(t-t_0) &=  -  \ubar  + \bar{e}_t \sinh(\ubar) + 
    \bar{g}_{t} \, \vbar +  \bar{f}_{t} \, \sin(\vbar)\,, \\
    \frac{\phi-\phi_0}{\bar{K}} &=  \vbar +  \bar{f}_{\phi} \,\sin(2\vbar) +  \bar{g}_{\phi} \,\sin(3\vbar) \,. 
\end{align}\end{subequations}
 
\begin{widetext}
\section{Expressions for quasi-Keplerian parameters in massless scalar-tensor theories}
 \label{app:QK_parameters}

The expressions of the conservative QK parameters for ST theories read (see also the Supplemental Material~\cite{Suppl_Mat}):

\begin{subequations}\label{eq:QK_param_ST}
\begin{align}
    \bar{n} &= \frac{c^3 \overline{\varepsilon}^{3/2}}{\tilde{G} \alpha m}  \left\{1 + \overline{\varepsilon} \left(\frac{15}{8} + \bar{\gamma} -  \frac{1}{8} \nu\right) + \overline{\varepsilon}^2 \left(\frac{555}{128} + \frac{33}{8} \bar{\gamma} + \bar{\gamma}^2 + \frac{15}{64} \nu + \frac{1}{8} \bar{\gamma} \nu + \frac{11}{128} \nu^2\right)\right\} \,,\\*
    \bar{K} &= 1 + \frac{\overline{\varepsilon}}{\overline{\jmath}}\Big(3 -  \bar{\beta}_{+} + 2 \bar{\gamma} + \bar{\beta}_{-} \delta\Big)
    \nonumber \\
    &\hspace{0.65cm} 
    + \frac{\overline{\varepsilon}^2}{\overline{\jmath}^2} \Bigg\{\frac{105}{4} + \frac{3}{2} \bar{\beta}_{-}^2 - 15 \bar{\beta}_{+} + \frac{3}{2} \bar{\beta}_{+}^2 -  \frac{5}{4} \bar{\delta}_{+} + \frac{139}{4} \bar{\gamma} - 12 \bar{\beta}_{+} \bar{\gamma} + \frac{187}{16} \bar{\gamma}^2 - 2 \bar{\chi}_{+} \nonumber\\
    & \hspace{1.8cm}+ \delta \left(15 \bar{\beta}_{-}  - 3 \bar{\beta}_{-} \bar{\beta}_{+}  -  \frac{5}{4} \bar{\delta}_{-} + 12 \bar{\beta}_{-} \bar{\gamma}  + 2 \bar{\chi}_{-} \right) \nonumber\\
    & \hspace{1.8cm}+ \nu \bigg(- \frac{15}{2} - 6 \bar{\beta}_{-}^2 + \frac{21}{2} \bar{\beta}_{+} - 2 \bar{\delta}_{+} - 8 \bar{\gamma}   -  \frac{1}{2} \bar{\gamma}^2 + 24 \bar{\beta}_{-}^2 \bar{\gamma}^{-1} - 24 \bar{\beta}_{+}^2 \bar{\gamma}^{-1} + 4 \bar{\chi}_{+} -  \frac{3}{2} \bar{\beta}_{-} \delta\bigg)  \nonumber \\
    & \hspace{1.8cm} + \overline{\jmath}\bigg[\frac{15}{4} -  \frac{1}{4} \bar{\delta}_{+} + \frac{15}{4} \bar{\gamma} + \frac{15}{16} \bar{\gamma}^2 -  \frac{1}{4} \bar{\delta}_{-} \delta + \nu \left(- \frac{3}{2} + \frac{1}{2} \bar{\beta}_{+} -  \bar{\gamma} -  \frac{1}{2} \bar{\beta}_{-} \delta\right) \bigg]\Bigg\}\,,\\
     \bar{a}_r & = \frac{ \tilde{G}\alpha  m}{c^2 \overline{\varepsilon}}\Bigg( 1 + \overline{\varepsilon} \left(\frac{7}{4} + \bar{\gamma} -  \frac{1}{4} \nu\right) \nonumber \\
     &\hspace{1cm} + \overline{\varepsilon}^2 \Bigg\{\frac{1}{16} + \frac{1}{16} \nu^2 \nonumber + \frac{1}{\overline{\jmath}}\Bigg[4 - 2 \bar{\beta}_{+} -  \frac{2}{3} \bar{\delta}_{+} + \frac{16}{3} \bar{\gamma} - 2 \bar{\beta}_{+} \bar{\gamma} + \frac{11}{6} \bar{\gamma}^2 -  \frac{2}{3} \bar{\chi}_{+}+\delta\Big(2 \bar{\beta}_{-} -  \frac{2}{3} \bar{\delta}_{-} + 2 \bar{\beta}_{-} \bar{\gamma} + \frac{2}{3} \bar{\chi}_{-}\Big)   \nonumber \\
     & \hspace{4cm} + \nu \bigg(-7 + 5 \bar{\beta}_{+} -  \frac{2}{3} \bar{\delta}_{+} -  \frac{17}{3} \bar{\gamma} -  \frac{1}{6} \bar{\gamma}^2 + 8 \bar{\beta}_{-}^2 \bar{\gamma}^{-1} - 8 \bar{\beta}_{+}^2 \bar{\gamma}^{-1} + \frac{4}{3} \bar{\chi}_{+} -  \bar{\beta}_{-} \delta\bigg) \Bigg]\Bigg\}\Bigg) \,,\\ 
    \bar{e}_r & = \sqrt{1 + \overline{\jmath}} + \frac{\overline{\varepsilon}}{\sqrt{1 + \overline{\jmath}}}\left\{- 3 + \bar{\beta}_{+} - 2 \bar{\gamma} -  \bar{\beta}_{-} \delta + \overline{\jmath} \left(- \frac{15}{8} -  \bar{\gamma} + \frac{5}{8} \nu\right) + \frac{1}{2} \nu\right\} \nonumber \\
    & + \frac{\overline{\varepsilon}^2}{(1 + \overline{\jmath})^{3/2}}\Bigg\{ - \frac{35}{4} -  \frac{1}{2} \bar{\beta}_{-}^2 + \frac{21}{4} \bar{\beta}_{+} -  \frac{1}{2} \bar{\beta}_{+}^2 + \frac{5}{2} \bar{\delta}_{+} - 12 \bar{\gamma} + 6 \bar{\beta}_{+} \bar{\gamma} -  \frac{35}{8} \bar{\gamma}^2 + 2 \bar{\chi}_{+} \nonumber \\
    & \hspace{2.2cm} + \delta \left(- \frac{21}{4} \bar{\beta}_{-} + \bar{\beta}_{-} \bar{\beta}_{+} + \frac{5}{2} \bar{\delta}_{-} - 6 \bar{\beta}_{-} \bar{\gamma} - 2 \bar{\chi}_{-}\right)  \nonumber \\
    &\hspace{2.2cm} + \nu\bigg(\frac{99}{4} + 2 \bar{\beta}_{-}^2 -  \frac{65}{4} \bar{\beta}_{+} + 2 \bar{\delta}_{+} + \frac{39}{2} \bar{\gamma} + \frac{1}{2} \bar{\gamma}^2
    - 24 \bar{\beta}_{-}^2 \bar{\gamma}^{-1} + 24 \bar{\beta}_{+}^2 \bar{\gamma}^{-1} - 4 \bar{\chi}_{+} + \frac{13}{4} \bar{\beta}_{-} \delta\bigg)  \nonumber \\
    & \hspace{2.2cm} +  \frac{1}{\overline{\jmath}}\bigg[-8 + 4 \bar{\beta}_{+} +  \frac{4}{3} \bar{\delta}_{+} - \frac{32}{3} \bar{\gamma} + 4 \bar{\beta}_{+} \bar{\gamma} - \frac{11}{3} \bar{\gamma}^2 +  \frac{4}{3} \bar{\chi}_{+} + \delta \left(-4 \bar{\beta}_{-}+ \frac{4}{3} \bar{\delta}_{-} - 4 \bar{\beta}_{-} \bar{\gamma} -  \frac{4}{3} \bar{\chi}_{-}\right)  \nonumber \\
    & \hspace{3cm} + \nu \left(14 - 10 \bar{\beta}_{+} + \frac{4}{3} \bar{\delta}_{+} + \frac{34}{3} \bar{\gamma} + \frac{1}{3} \bar{\gamma}^2 - 16 \bar{\beta}_{-}^2 \bar{\gamma}^{-1} + 16 \bar{\beta}_{+}^2 \bar{\gamma}^{-1} -  \frac{8}{3} \bar{\chi}_{+} + 2 \bar{\beta}_{-} \delta\right) \bigg] \nonumber \\
    & \hspace{2.2cm} + \overline{\jmath} \bigg[\frac{25}{8} + \frac{1}{8} \bar{\beta}_{+} + \frac{7}{6} \bar{\delta}_{+} + \frac{41}{12} \bar{\gamma} + \bar{\beta}_{+} \bar{\gamma} + \frac{19}{24} \bar{\gamma}^2 + \frac{2}{3} \bar{\chi}_{+} +  \delta\left(- \frac{1}{8} \bar{\beta}_{-} + \frac{7}{6} \bar{\delta}_{-} -  \bar{\beta}_{-} \bar{\gamma} -  \frac{2}{3} \bar{\chi}_{-}\right) \nonumber \\
    & \hspace{3cm}  +  \nu\left(\frac{37}{4} -  \frac{51}{8} \bar{\beta}_{+} + \frac{2}{3} \bar{\delta}_{+} + \frac{89}{12} \bar{\gamma} + \frac{1}{6} \bar{\gamma}^2 - 8 \bar{\beta}_{-}^2 \bar{\gamma}^{-1} + 8 \bar{\beta}_{+}^2 \bar{\gamma}^{-1} -  \frac{4}{3} \bar{\chi}_{+} + \frac{11}{8} \bar{\beta}_{-} \delta\right) + \frac{1}{16} \nu^2\bigg]\nonumber \\
    & \hspace{2.2cm}+ \overline{\jmath}^2 \left[\frac{415}{128} + \frac{29}{8} \bar{\gamma} + \bar{\gamma}^2 - \nu\left( \frac{105}{64} +  \frac{7}{8} \bar{\gamma}\right)  + \frac{7}{128} \nu^2\right]\Bigg\}\,, \\ 
   \bar{e}_t & = \sqrt{1 + \overline{\jmath}} + \frac{\overline{\varepsilon}}{\sqrt{1 + \overline{\jmath}}} \left[1 + \bar{\beta}_{+} -  \bar{\beta}_{-} \delta + \overline{\jmath} \left(\frac{17}{8} + \bar{\gamma} -  \frac{7}{8} \nu\right) -  \nu\right] \nonumber \\
   & + \frac{\overline{\varepsilon}^2}{(1 + \overline{\jmath})^{3/2}} \Bigg\{ \frac{21}{4} \bar{\beta}_{+}- \frac{15}{4} -  \frac{1}{2} \bar{\beta}_{-}^2  -  \frac{1}{2} \bar{\beta}_{+}^2 + \frac{7}{6} \bar{\delta}_{+} -  \frac{41}{6} \bar{\gamma} + 4 \bar{\beta}_{+} \bar{\gamma} -  \frac{65}{24} \bar{\gamma}^2 + \frac{2}{3} \bar{\chi}_{+} \nonumber \\
   & \hspace{2.2cm}+\delta \left( -\frac{21}{4} \bar{\beta}_{-} + \bar{\beta}_{-} \bar{\beta}_{+} + \frac{7}{6} \bar{\delta}_{-} - 4 \bar{\beta}_{-} \bar{\gamma} -  \frac{2}{3} \bar{\chi}_{-}\right) + \frac{3}{4} \nu^2 \nonumber\\
   &\hspace{2.2cm} + \nu \bigg(\frac{25}{2} + 2 \bar{\beta}_{-}^2 -  \frac{31}{4} \bar{\beta}_{+} + \frac{2}{3} \bar{\delta}_{+} + \frac{29}{3} \bar{\gamma} + \frac{1}{6} \bar{\gamma}^2 - 8 \bar{\beta}_{-}^2 \bar{\gamma}^{-1} + 8 \bar{\beta}_{+}^2 \bar{\gamma}^{-1} -  \frac{4}{3} \bar{\chi}_{+} + \frac{11}{4} \bar{\beta}_{-} \delta\bigg)  \nonumber \\
   & \hspace{2.2cm} +  \frac{1}{\overline{\jmath}}\bigg[-4 + 2 \bar{\beta}_{+} +  \frac{2}{3} \bar{\delta}_{+} - \frac{16}{3} \bar{\gamma} + 2 \bar{\beta}_{+} \bar{\gamma} - \frac{11}{6} \bar{\gamma}^2 +  \frac{2}{3} \bar{\chi}_{+} +  \delta\left(-2 \bar{\beta}_{-} + \frac{2}{3} \bar{\delta}_{-} - 2 \bar{\beta}_{-} \bar{\gamma} -  \frac{2}{3} \bar{\chi}_{-}\right)  \nonumber \\
   & \hspace{3cm} +  \nu\left(7 - 5 \bar{\beta}_{+} + \frac{2}{3} \bar{\delta}_{+} + \frac{17}{3} \bar{\gamma} + \frac{1}{6} \bar{\gamma}^2 - 8 \bar{\beta}_{-}^2 \bar{\gamma}^{-1} + 8 \bar{\beta}_{+}^2 \bar{\gamma}^{-1} -  \frac{4}{3} \bar{\chi}_{+} + \bar{\beta}_{-} \delta\right) \bigg] \nonumber \\
   & \hspace{2.2cm} + \overline{\jmath} \bigg[\frac{45}{8} + \frac{17}{8} \bar{\beta}_{+} + \frac{1}{2} \bar{\delta}_{+} + 4 \bar{\gamma} + \bar{\beta}_{+} \bar{\gamma} + \frac{5}{8} \bar{\gamma}^2 + \delta\left(- \frac{17}{8} \bar{\beta}_{-} + \frac{1}{2} \bar{\delta}_{-} -  \bar{\beta}_{-} \bar{\gamma}\right)  \nonumber \\
   & \hspace{2.7cm} +  \nu\left(\frac{73}{16} -  \frac{23}{8} \bar{\beta}_{+} + \frac{7}{2} \bar{\gamma} + \frac{15}{8} \bar{\beta}_{-} \delta\right)  + \frac{11}{8} \nu^2 \bigg] \nonumber\\
   & \hspace{2.2cm} + \overline{\jmath}^2 \left[\frac{607}{128} + \frac{35}{8} \bar{\gamma} + \bar{\gamma}^2 - \nu\left( \frac{69}{64} + \frac{5}{8} \bar{\gamma}\right)  + \frac{79}{128} \nu^2\right]\Bigg\}\,,\\ 
    \bar{e}_\phi & = \sqrt{1 + \overline{\jmath}} + \frac{\overline{\varepsilon}}{\sqrt{1 + \overline{\jmath}}} \left[-3 + \bar{\beta}_{+} - 2 \bar{\gamma} -  \bar{\beta}_{-} \delta - \overline{\jmath} \left(\frac{15}{8} +  \bar{\gamma} + \frac{1}{8} \nu\right)\right] \nonumber \\
    & + \frac{\overline{\varepsilon}^2}{(1 + \overline{\jmath})^{3/2}} \Bigg\{ \frac{37}{4} \bar{\beta}_{+}- \frac{75}{4} -  \frac{1}{2} \bar{\beta}_{-}^2  -  \frac{1}{2} \bar{\beta}_{+}^2 + \frac{11}{6} \bar{\delta}_{+} -  \frac{74}{3} \bar{\gamma} + 10 \bar{\beta}_{+} \bar{\gamma} -  \frac{205}{24} \bar{\gamma}^2 + \frac{10}{3} \bar{\chi}_{+}  \nonumber \\
    & \hspace{2.2cm} + \delta \left(- \frac{37}{4} \bar{\beta}_{-} + \bar{\beta}_{-} \bar{\beta}_{+} + \frac{11}{6} \bar{\delta}_{-} - 10 \bar{\beta}_{-} \bar{\gamma} -  \frac{10}{3} \bar{\chi}_{-}\right) + \frac{33}{32} \nu^2 \nonumber\\
    & \hspace{2.2cm} + \nu\bigg(\frac{125}{32} + 2 \bar{\beta}_{-}^2 -  \frac{59}{4} \bar{\beta}_{+} + \frac{10}{3} \bar{\delta}_{+} + \frac{47}{6} \bar{\gamma}  + \frac{5}{6} \bar{\gamma}^2 - 40 \bar{\beta}_{-}^2 \bar{\gamma}^{-1} + 40 \bar{\beta}_{+}^2 \bar{\gamma}^{-1} -  \frac{20}{3} \bar{\chi}_{+} + \frac{7}{4} \bar{\beta}_{-} \delta\bigg)   \nonumber \\
    & \hspace{2.2cm}  +  \frac{1}{\overline{\jmath}}\bigg[-13 + 6 \bar{\beta}_{+} +  \bar{\delta}_{+} - 17 \bar{\gamma} + 6 \bar{\beta}_{+} \bar{\gamma} - \frac{23}{4} \bar{\gamma}^2 + 2 \bar{\chi}_{+} +  \delta\Big(-6 \bar{\beta}_{-} + \bar{\delta}_{-} - 6 \bar{\beta}_{-} \bar{\gamma} - 2 \bar{\chi}_{-}\Big)  \nonumber \\
    & \hspace{3cm} + \nu \bigg(\frac{91}{32} - 9 \bar{\beta}_{+} + 2 \bar{\delta}_{+} + 5 \bar{\gamma} + \frac{1}{2} \bar{\gamma}^2 - 24 \bar{\beta}_{-}^2 \bar{\gamma}^{-1} + 24 \bar{\beta}_{+}^2 \bar{\gamma}^{-1} - 4 \bar{\chi}_{+} + \bar{\beta}_{-} \delta\bigg)  +  \frac{15}{32} \nu^2\bigg] \nonumber \\
    & \hspace{2.2cm} + \overline{\jmath} \bigg[ \frac{17}{8} \bar{\beta}_{+}- \frac{15}{8}  + \frac{5}{6} \bar{\delta}_{+} -  \frac{35}{12} \bar{\gamma} + 3 \bar{\beta}_{+} \bar{\gamma} -  \frac{31}{24} \bar{\gamma}^2 + \frac{4}{3} \bar{\chi}_{+} +  \delta \left(- \frac{17}{8} \bar{\beta}_{-} + \frac{5}{6} \bar{\delta}_{-} - 3 \bar{\beta}_{-} \bar{\gamma} - \frac{4}{3} \bar{\chi}_{-}\right) \nonumber \\
    & \hspace{2.6cm} +  \nu\left(\frac{15}{32} -  \frac{47}{8} \bar{\beta}_{+} + \frac{4}{3} \bar{\delta}_{+} + \frac{31}{12} \bar{\gamma} + \frac{1}{3} \bar{\gamma}^2 - 16 \bar{\beta}_{-}^2 \bar{\gamma}^{-1} + 16 \bar{\beta}_{+}^2 \bar{\gamma}^{-1} -  \frac{8}{3} \bar{\chi}_{+} + \frac{7}{8} \bar{\beta}_{-} \delta\right) + \frac{21}{32} \nu^2\bigg] \nonumber\\
    & \hspace{2.2cm} + \overline{\jmath}^2 \bigg[\frac{415}{128} + \frac{29}{8} \bar{\gamma} + \bar{\gamma}^2 +  \nu \left(- \frac{47}{64} -  \frac{3}{8} \bar{\gamma}\right) + \frac{11}{128} \nu^2\bigg] 
    \Bigg\}\,,\\
     \bar{f}_{t} & = \frac{\overline{\varepsilon}^2  \sqrt{1 + \overline{\jmath}} \,\nu\,(15 + 8 \bar{\gamma} -  \nu)}{8\sqrt{\overline{\jmath}}} \,,\\
     \bar{f}_{\phi} & = \frac{\overline{\varepsilon}^2 (1 + \overline{\jmath})}{8 \overline{\jmath}^2}\Big[1 + \bar{\delta}_{+} + \bar{\gamma} + \frac{1}{4} \bar{\gamma}^2 + \bar{\delta}_{-} \delta + \nu\Big(19 - 6 \bar{\beta}_{+} + 12 \bar{\gamma} + 2 \bar{\beta}_{-} \delta\Big)  - 3 \nu^2\Big]\,, \\
     \bar{g}_{t} & = \frac{\overline{\varepsilon}^2 }{8 \sqrt{\overline{\jmath}}}\Big[60 - 4 \bar{\delta}_{+} + 60 \bar{\gamma} + 15 \bar{\gamma}^2 - 4 \bar{\delta}_{-} \delta + \nu\Big(-24 + 8 \bar{\beta}_{+} - 16 \bar{\gamma} - 8 \bar{\beta}_{-} \delta \Big) \Big]\,, \\
     \bar{g}_{\phi} & =  \frac{\overline{\varepsilon}^2(1 + \overline{\jmath})^{3/2} \nu (1 - 3 \nu)}{32 \overline{\jmath}^2}\,, \\
     b & = \frac{ \tilde{G} \alpha  m \sqrt{\overline{\jmath}}}{c^2 \overline{\varepsilon}}\left\{1 + \overline{\varepsilon} \left( -\frac{1}{8}+ \frac{3}{8} \nu\right) + \overline{\varepsilon}^2 \left(\frac{3}{128} -  \frac{5}{64} \nu -  \frac{5}{128} \nu^2\right)\right\} \,, \label{subeq:b_ST}\\
     \frac{\chi}{2} =& 
     \arctan\left(\frac{1}{\sqrt{\overline{\jmath}}}\right)\Bigg(1 + \frac{\overline{\varepsilon}}{\overline{\jmath}}\left(3 -  \bar{\beta}_{+} + 2 \bar{\gamma} + \bar{\beta}_{-} \delta\right)   \nonumber \\
     &\hspace{1.5cm} + \frac{\overline{\varepsilon}^2}{\overline{\jmath}} \Bigg\{
     \frac{15}{4} -  \frac{1}{4} \bar{\delta}_{+} + \frac{15}{4} \bar{\gamma} + \frac{15}{16} \bar{\gamma}^2 -  \frac{1}{4} \bar{\delta}_{-} \delta +  \nu  \left(- \frac{3}{2} + \frac{1}{2} \bar{\beta}_{+} -  \bar{\gamma} -  \frac{1}{2} \bar{\beta}_{-} \delta\right)\nonumber \\
     &\hspace{2cm} + 
     \frac{1}{\overline{\jmath}}\Bigg[\frac{105}{4} + \frac{3}{2} \bar{\beta}_{-}^2 - 15 \bar{\beta}_{+} + \frac{3}{2} \bar{\beta}_{+}^2 -  \frac{5}{4} \bar{\delta}_{+} + \frac{139}{4} \bar{\gamma} - 12 \bar{\beta}_{+} \bar{\gamma} + \frac{187}{16} \bar{\gamma}^2 - 2 \bar{\chi}_{+} \nonumber \\
     &\hspace{2.5cm} + \delta\left(15 \bar{\beta}_{-} - 3 \bar{\beta}_{-} \bar{\beta}_{+} -  \frac{5}{4} \bar{\delta}_{-} + 12 \bar{\beta}_{-} \bar{\gamma} + 2 \bar{\chi}_{-}\right)  \nonumber \\
     & \hspace{2.5cm} + \nu \left( - \frac{15}{2} - 6 \bar{\beta}_{-}^2 + \frac{21}{2} \bar{\beta}_{+} - 2 \bar{\delta}_{+} - 8 \bar{\gamma} -  \frac{1}{2} \bar{\gamma}^2 + 24 \bar{\beta}_{-}^2 \bar{\gamma}^{-1} - 24 \bar{\beta}_{+}^2 \bar{\gamma}^{-1} + 4 \bar{\chi}_{+} -  \frac{3}{2} \bar{\beta}_{-} \delta\right) \Bigg]
     \Bigg\}\Bigg) \nonumber  \\
     &    +\frac{\overline{\varepsilon}}{\overline{\jmath}} \Bigg\{\pi\left[\frac{3}{2} -  \frac{1}{2} \bar{\beta}_{+} + \bar{\gamma}+ \frac{1}{2} \bar{\beta}_{-}  \delta\right]   + \frac{\sqrt{\overline{\jmath}}}{1 + \overline{\jmath}} \left(3 -  \bar{\beta}_{+} + 2 \bar{\gamma} + \bar{\beta}_{-} \delta\right)   + \frac{\overline{\jmath}^{3/2}}{1 + \overline{\jmath}} \left(\frac{15}{8} + \bar{\gamma} -  \frac{1}{8} \nu\right)\Bigg\} \nonumber \\
     & + \frac{\overline{\varepsilon}^2}{\overline{\jmath}^2 (1 + \overline{\jmath})^2} \Bigg\{ \pi\Bigg[\frac{105}{8} + \frac{3}{4} \bar{\beta}_{-}^2 -  \frac{15}{2} \bar{\beta}_{+} + \frac{3}{4} \bar{\beta}_{+}^2 -  \frac{5}{8} \bar{\delta}_{+} + \frac{139}{8} \bar{\gamma} - 6 \bar{\beta}_{+} \bar{\gamma} + \frac{187}{32} \bar{\gamma}^2 -  \bar{\chi}_{+}\nonumber \\
     & \hspace{2.8cm}+  \delta\left(\frac{15}{2} \bar{\beta}_{-} -  \frac{3}{2} \bar{\beta}_{-} \bar{\beta}_{+} -  \frac{5}{8} \bar{\delta}_{-} + 6 \bar{\beta}_{-} \bar{\gamma} + \bar{\chi}_{-}\right)  \nonumber \\
     & \hspace{2.8cm}+ \nu\bigg(-\frac{15}{4} - 3 \bar{\beta}_{-}^2 + \frac{21}{4} \bar{\beta}_{+} -  \bar{\delta}_{+} - 4 \bar{\gamma} -  \frac{1}{4} \bar{\gamma}^2 + 12 \bar{\beta}_{-}^2 \bar{\gamma}^{-1} - 12 \bar{\beta}_{+}^2 \bar{\gamma}^{-1} + 2 \bar{\chi}_{+}  -  \frac{3}{4} \bar{\beta}_{-} \delta \bigg)  \Bigg]\nonumber \\
     & \hspace{1.5cm} + \sqrt{\overline{\jmath}}\Bigg[\frac{105}{4} + \frac{3}{2} \bar{\beta}_{-}^2 - 15 \bar{\beta}_{+} + \frac{3}{2} \bar{\beta}_{+}^2 -  \frac{5}{4} \bar{\delta}_{+} + \frac{139}{4} \bar{\gamma} - 12 \bar{\beta}_{+} \bar{\gamma} + \frac{187}{16} \bar{\gamma}^2 - 2 \bar{\chi}_{+} \nonumber \\
     & \hspace{2.5cm} + \delta\left(15 \bar{\beta}_{-} - 3 \bar{\beta}_{-} \bar{\beta}_{+} -  \frac{5}{4} \bar{\delta}_{-} + 12 \bar{\beta}_{-} \bar{\gamma} + 2 \bar{\chi}_{-}\right)  \nonumber \\
     & \hspace{2.5cm}   +  \nu\bigg(- \frac{15}{2} - 6 \bar{\beta}_{-}^2 + \frac{21}{2} \bar{\beta}_{+} - 2 \bar{\delta}_{+} - 8 \bar{\gamma} -  \frac{1}{2} \bar{\gamma}^2 + 24 \bar{\beta}_{-}^2 \bar{\gamma}^{-1} - 24 \bar{\beta}_{+}^2 \bar{\gamma}^{-1} + 4 \bar{\chi}_{+} -  \frac{3}{2} \bar{\beta}_{-} \delta\bigg) \Bigg] \nonumber \\
    & \hspace{1.5cm} + \overline{\jmath}\, \pi \Bigg[\frac{225}{8} + \frac{3}{2} \bar{\beta}_{-}^2 - 15 \bar{\beta}_{+} + \frac{3}{2} \bar{\beta}_{+}^2 -  \frac{11}{8} \bar{\delta}_{+} + \frac{293}{8} \bar{\gamma} - 12 \bar{\beta}_{+} \bar{\gamma} + \frac{389}{32} \bar{\gamma}^2 - 2 \bar{\chi}_{+} \nonumber \\
     & \hspace{2.5cm} + \delta\left(15 \bar{\beta}_{-} - 3 \bar{\beta}_{-} \bar{\beta}_{+} -  \frac{11}{8} \bar{\delta}_{-} + 12 \bar{\beta}_{-} \bar{\gamma} + 2 \bar{\chi}_{-}\right) \nonumber \\
     & \hspace{2.5cm} + \nu\bigg(- \frac{33}{4} - 6 \bar{\beta}_{-}^2 + \frac{43}{4} \bar{\beta}_{+} - 2 \bar{\delta}_{+} -  \frac{17}{2} \bar{\gamma} -  \frac{1}{2} \bar{\gamma}^2 + 24 \bar{\beta}_{-}^2 \bar{\gamma}^{-1} - 24 \bar{\beta}_{+}^2 \bar{\gamma}^{-1} + 4 \bar{\chi}_{+} - \frac{7}{4} \bar{\beta}_{-}  \delta   \bigg) \Bigg] \nonumber \\
     & \hspace{1.5cm} + \overline{\jmath}^{3/2} \Bigg[\frac{95}{2} + \frac{5}{2} \bar{\beta}_{-}^2 - 25 \bar{\beta}_{+} + \frac{5}{2} \bar{\beta}_{+}^2 -  \frac{7}{3} \bar{\delta}_{+} + \frac{185}{3} \bar{\gamma} - 20 \bar{\beta}_{+} \bar{\gamma} + \frac{245}{12} \bar{\gamma}^2 -  \frac{10}{3} \bar{\chi}_{+} \nonumber \\
     & \hspace{2.5cm} + \delta\left(25 \bar{\beta}_{-} - 5 \bar{\beta}_{-} \bar{\beta}_{+} -  \frac{7}{3} \bar{\delta}_{-} + 20 \bar{\beta}_{-} \bar{\gamma} + \frac{10}{3} \bar{\chi}_{-}\right) \nonumber\\
     & \hspace{2.5cm} + \nu\bigg(-14 -10 \bar{\beta}_{-}^2 + 18 \bar{\beta}_{+} -  \frac{10}{3} \bar{\delta}_{+} -  \frac{43}{3} \bar{\gamma} -  \frac{5}{6} \bar{\gamma}^2 + 40 \bar{\beta}_{-}^2 \bar{\gamma}^{-1} - 40 \bar{\beta}_{+}^2 \bar{\gamma}^{-1} + \frac{20}{3} \bar{\chi}_{+} - 3 \bar{\beta}_{-} \delta\bigg) \Bigg] \nonumber \\
     & \hspace{1.5cm} + \overline{\jmath}^2\pi\Bigg[\frac{135}{8} + \frac{3}{4} \bar{\beta}_{-}^2 -  \frac{15}{2} \bar{\beta}_{+} + \frac{3}{4} \bar{\beta}_{+}^2 -  \frac{7}{8} \bar{\delta}_{+} + \frac{169}{8} \bar{\gamma} - 6 \bar{\beta}_{+} \bar{\gamma} + \frac{217}{32} \bar{\gamma}^2 -  \bar{\chi}_{+}\nonumber \\
     & \hspace{2.5cm} + \delta\left(\frac{15}{2} \bar{\beta}_{-} -  \frac{3}{2} \bar{\beta}_{-} \bar{\beta}_{+} -  \frac{7}{8} \bar{\delta}_{-} + 6 \bar{\beta}_{-} \bar{\gamma} + \bar{\chi}_{-}\right)  \nonumber \\
     & \hspace{2.5cm}  +  \nu\bigg(-  \frac{21}{4} - 3 \bar{\beta}_{-}^2 + \frac{23}{4} \bar{\beta}_{+} -  \bar{\delta}_{+} - 5 \bar{\gamma} -  \frac{1}{4} \bar{\gamma}^2 + 12 \bar{\beta}_{-}^2 \bar{\gamma}^{-1} - 12 \bar{\beta}_{+}^2 \bar{\gamma}^{-1} + 2 \bar{\chi}_{+} - \frac{5}{4} \bar{\beta}_{-}  \delta  \bigg)\Bigg] \nonumber \\
     & \hspace{1.5cm} + \overline{\jmath}^{5/2} \Bigg[\frac{2593}{128} -  \frac{31}{4} \bar{\beta}_{+} -  \frac{13}{12} \bar{\delta}_{+} + \frac{595}{24} \bar{\gamma} - 6 \bar{\beta}_{+} \bar{\gamma} + \frac{371}{48} \bar{\gamma}^2 -  \frac{4}{3} \bar{\chi}_{+} \nonumber \\
     & \hspace{2.5cm} + \delta \left(\frac{31}{4} \bar{\beta}_{-} -  \frac{13}{12} \bar{\delta}_{-} + 6 \bar{\beta}_{-} \bar{\gamma} + \frac{4}{3} \bar{\chi}_{-}\right) + \frac{1}{128} \nu^2 \nonumber \\
     & \hspace{2.5cm} + \nu  \left(- \frac{419}{64} + \frac{31}{4} \bar{\beta}_{+} -  \frac{4}{3} \bar{\delta}_{+} -  \frac{155}{24} \bar{\gamma} -  \frac{1}{3} \bar{\gamma}^2 + 16 \bar{\beta}_{-}^2 \bar{\gamma}^{-1} - 16 \bar{\beta}_{+}^2 \bar{\gamma}^{-1} + \frac{8}{3} \bar{\chi}_{+} -  \frac{7}{4} \bar{\beta}_{-} \delta\right) \Bigg] \nonumber \\
     & \hspace{1.5cm} + \overline{\jmath}^3 \pi\Bigg[\frac{15}{8} -  \frac{1}{8} \bar{\delta}_{+} + \frac{15}{8} \bar{\gamma} + \frac{15}{32} \bar{\gamma}^2 -  \frac{1}{8} \bar{\delta}_{-}   \delta + \nu\bigg( -\frac{3}{4} + \frac{1}{4} \bar{\beta}_{+} -  \frac{1}{2} \bar{\gamma}  -  \frac{1}{4} \bar{\beta}_{-}  \delta \bigg) \Bigg] \nonumber \\
     & \hspace{1.5cm} + \overline{\jmath}^{7/2} \Bigg[\frac{35}{128} + \frac{1}{8} \bar{\gamma} +  \nu\left(\frac{15}{64} + \frac{1}{8} \bar{\gamma}\right) + \frac{3}{128} \nu^2\Bigg]\Bigg\} \label{scattering_angle_ST_cons} \, .
\end{align}\end{subequations}
\newpage

\section{Expressions for dissipative corrections to the impact parameter, asymptotic velocity and time eccentricity} \label{app:diss_quantities}

Due to dissipation, the QK parameters evolve throughout the scattering process. The difference between the incoming and outgoing values of $b$, $v_\infty$, and $\bar{e}_t$ reads (see also the Supplemental Material~\cite{Suppl_Mat})

\begin{subequations}
    \begin{align}
        \Delta b & =  \frac{\alpha \tilde{G} m^2 \overline{\varepsilon}^{1/2} \nu^2}{2 c^2} \Bigg\{4 \zeta \mathcal{S}_{-}^2 \left[ \frac{2}{\overline{\jmath}^{3/2}} - \frac{4}{3 \overline{\jmath}^{1/2}} + \left( \frac{2}{\overline{\jmath}^2} - \frac{2}{3 \overline{\jmath}}\right) \left(\pi -  \arctan\sqrt{\overline{\jmath}})\right)\right] \nonumber \\*
        & + \frac{\overline{\varepsilon}}{1 + \overline{\jmath}} \Bigg[\overline{\jmath}^{1/2} \left(-\frac{32}{5} - \frac{8}{3} \bar{\gamma} - \frac{32}{3} \zeta \mathcal{S}_{-}^2 - \frac{16}{3} \zeta \bar{\gamma} \mathcal{S}_{-}^2 +  \frac{8}{3} \zeta \mathcal{S}_{-}^2 \nu\right) \nonumber \\
        &  \hspace{1.2cm} + \frac{1}{\overline{\jmath}^{1/2}}\Bigg(\frac{244}{45} + \frac{416}{9} \bar{\beta}_{+} -  \frac{49}{9} \bar{\gamma} -  \frac{208}{3} \bar{\beta}_{+}^2 \bar{\gamma}^{-1} + \frac{37}{9} \zeta \mathcal{S}_{-}^2 + \frac{112}{9} \bar{\beta}_{+} \zeta \mathcal{S}_{-}^2 -  \frac{88}{9} \zeta \bar{\gamma} \mathcal{S}_{-}^2  \nonumber \\
        & \hspace{2.5cm}-  \frac{680}{9} \bar{\beta}_{+} \zeta \bar{\gamma}^{-1} \mathcal{S}_{-}^2 + \frac{416}{3} \bar{\beta}_{-}^2 \zeta \bar{\gamma}^{-2} \mathcal{S}_{-}^2 + \frac{416}{3} \bar{\beta}_{+}^2 \zeta \bar{\gamma}^{-2} \mathcal{S}_{-}^2 + \frac{832}{3} \bar{\beta}_{-} \bar{\beta}_{+} \zeta \bar{\gamma}^{-2} \mathcal{S}_{-} \mathcal{S}_{+}   \nonumber \\
        & \hspace{2.5cm} -  \frac{680}{9} \bar{\beta}_{-} \zeta \bar{\gamma}^{-1} \mathcal{S}_{-} \mathcal{S}_{+}   -  \frac{71}{9} \zeta \mathcal{S}_{-}^2 \nu+  \delta \left(-\frac{112}{9} \bar{\beta}_{-} \zeta \mathcal{S}_{-}^2 +  \frac{56}{9} \bar{\beta}_{-} \zeta \bar{\gamma}^{-1} \mathcal{S}_{-}^2 +  \frac{56}{9} \bar{\beta}_{+} \zeta \bar{\gamma}^{-1} \mathcal{S}_{-} \mathcal{S}_{+}\right)  \Bigg) \nonumber \\
        & \hspace{1.2cm} +  \frac{1}{\overline{\jmath}^{5/2}}\Bigg(\frac{340}{3} + \frac{160}{3} \bar{\beta}_{+} + \frac{115}{3} \bar{\gamma} - 80 \bar{\beta}_{+}^2 \bar{\gamma}^{-1} + \frac{380}{3} \zeta \mathcal{S}_{-}^2 -  \frac{280}{3} \bar{\beta}_{+} \zeta \mathcal{S}_{-}^2 + \frac{280}{3} \zeta \bar{\gamma} \mathcal{S}_{-}^2 -  \frac{40}{3} \bar{\beta}_{+} \zeta \bar{\gamma}^{-1} \mathcal{S}_{-}^2 \nonumber \\
        & \hspace{2.5cm}+ 160 \bar{\beta}_{-}^2 \zeta \bar{\gamma}^{-2} \mathcal{S}_{-}^2 + 160 \bar{\beta}_{+}^2 \zeta \bar{\gamma}^{-2} \mathcal{S}_{-}^2 -  \frac{40}{3} \bar{\beta}_{-} \zeta \bar{\gamma}^{-1} \mathcal{S}_{-} \mathcal{S}_{+} + 320 \bar{\beta}_{-} \bar{\beta}_{+} \zeta \bar{\gamma}^{-2} \mathcal{S}_{-} \mathcal{S}_{+} +  \frac{100}{3} \zeta \mathcal{S}_{-}^2 \nu \nonumber \\
        & \hspace{2.5cm}+ \delta \left( \frac{280}{3} \bar{\beta}_{-} \zeta \mathcal{S}_{-}^2 - \frac{200}{3} \bar{\beta}_{-} \zeta \bar{\gamma}^{-1} \mathcal{S}_{-}^2 - \frac{200}{3} \bar{\beta}_{+} \zeta \bar{\gamma}^{-1} \mathcal{S}_{-} \mathcal{S}_{+}\right)  \Bigg) \nonumber \\
        & \hspace{1.2cm} +  \frac{1}{\overline{\jmath}^{3/2}}\Bigg(\frac{5632}{45} + \frac{896}{9} \bar{\beta}_{+} + \frac{320}{9} \bar{\gamma} -  \frac{448}{3} \bar{\beta}_{+}^2 \bar{\gamma}^{-1} + \frac{1165}{9} \zeta \mathcal{S}_{-}^2 -  \frac{632}{9} \bar{\beta}_{+} \zeta \mathcal{S}_{-}^2 + \frac{704}{9} \zeta \bar{\gamma} \mathcal{S}_{-}^2  \nonumber \\
        & \hspace{2.5cm}-  \frac{800}{9} \bar{\beta}_{+} \zeta \bar{\gamma}^{-1} \mathcal{S}_{-}^2+ \frac{896}{3} \bar{\beta}_{-}^2 \zeta \bar{\gamma}^{-2} \mathcal{S}_{-}^2 + \frac{896}{3} \bar{\beta}_{+}^2 \zeta \bar{\gamma}^{-2} \mathcal{S}_{-}^2  + \frac{1792}{3} \bar{\beta}_{-} \bar{\beta}_{+} \zeta \bar{\gamma}^{-2} \mathcal{S}_{-} \mathcal{S}_{+} -  \frac{407}{9} \zeta \mathcal{S}_{-}^2 \nu \nonumber \\
        & \hspace{2.5cm} -  \frac{800}{9} \bar{\beta}_{-} \zeta \bar{\gamma}^{-1} \mathcal{S}_{-} \mathcal{S}_{+}  +  \delta\left( \frac{632}{9} \bar{\beta}_{-} \zeta \mathcal{S}_{-}^2 - \frac{544}{9} \bar{\beta}_{-} \zeta \bar{\gamma}^{-1} \mathcal{S}_{-}^2 - \frac{544}{9} \bar{\beta}_{+} \zeta \bar{\gamma}^{-1} \mathcal{S}_{-} \mathcal{S}_{+}\right)  \Bigg)\Bigg] \nonumber \\
        & \hspace{1.2cm} + \frac{\overline{\varepsilon}}{1 + \overline{\jmath}} \left(\pi -  \arctan\sqrt{\overline{\jmath}}\right)\Bigg[-\frac{188}{15} +  \frac{32}{3} \bar{\beta}_{+} - 7 \bar{\gamma} - 16 \bar{\beta}_{+}^2 \bar{\gamma}^{-1} - 9 \zeta \mathcal{S}_{-}^2 +  \frac{16}{3} \bar{\beta}_{+} \zeta \mathcal{S}_{-}^2 - 8 \zeta \bar{\gamma} \mathcal{S}_{-}^2 \nonumber \\
        & \hspace{2.2cm}- 24 \bar{\beta}_{+} \zeta \bar{\gamma}^{-1} \mathcal{S}_{-}^2 + 32 \bar{\beta}_{-}^2 \zeta \bar{\gamma}^{-2} \mathcal{S}_{-}^2 + 32 \bar{\beta}_{+}^2 \zeta \bar{\gamma}^{-2} \mathcal{S}_{-}^2 - 24 \bar{\beta}_{-} \zeta \bar{\gamma}^{-1} \mathcal{S}_{-} \mathcal{S}_{+} + 64 \bar{\beta}_{-} \bar{\beta}_{+} \zeta \bar{\gamma}^{-2} \mathcal{S}_{-} \mathcal{S}_{+} \nonumber \\
        & \hspace{2.2cm}+ \delta\left(-\frac{16}{3} \bar{\beta}_{-} \zeta \mathcal{S}_{-}^2 + 8 \bar{\beta}_{-} \zeta \bar{\gamma}^{-1} \mathcal{S}_{-}^2 + 8 \bar{\beta}_{+} \zeta \bar{\gamma}^{-1} \mathcal{S}_{-} \mathcal{S}_{+}\right)  + 3 \zeta \mathcal{S}_{-}^2 \nu \nonumber \\
        & \hspace{1.2cm} +  \frac{1}{\overline{\jmath}}\Bigg(\frac{556}{15} + \frac{224}{3} \bar{\beta}_{+} + 3 \bar{\gamma} - 112 \bar{\beta}_{+}^2 \bar{\gamma}^{-1} + 36 \zeta \mathcal{S}_{-}^2 -  \frac{8}{3} \bar{\beta}_{+} \zeta \mathcal{S}_{-}^2 + 8 \zeta \bar{\gamma} \mathcal{S}_{-}^2 - 104 \bar{\beta}_{+} \zeta \bar{\gamma}^{-1} \mathcal{S}_{-}^2  \nonumber \\
        & \hspace{2.0cm} + 224 \bar{\beta}_{-}^2 \zeta \bar{\gamma}^{-2} \mathcal{S}_{-}^2+ 224 \bar{\beta}_{+}^2 \zeta \bar{\gamma}^{-2} \mathcal{S}_{-}^2 - 104 \bar{\beta}_{-} \zeta \bar{\gamma}^{-1} \mathcal{S}_{-} \mathcal{S}_{+} + 448 \bar{\beta}_{-} \bar{\beta}_{+} \zeta \bar{\gamma}^{-2} \mathcal{S}_{-} \mathcal{S}_{+} - 20 \zeta \mathcal{S}_{-}^2 \nu \nonumber \\
        & \hspace{2.0cm}+ \delta\left( \frac{8}{3} \bar{\beta}_{-} \zeta \mathcal{S}_{-}^2 - 8 \bar{\beta}_{-} \zeta \bar{\gamma}^{-1} \mathcal{S}_{-}^2 - 8 \bar{\beta}_{+} \zeta \bar{\gamma}^{-1} \mathcal{S}_{-} \mathcal{S}_{+}\right)  \Bigg)\nonumber \\
        & \hspace{1.2cm} +  \frac{1}{\overline{\jmath}^3}\Bigg(\frac{340}{3} + \frac{160}{3} \bar{\beta}_{+} + \frac{115}{3} \bar{\gamma} - 80 \bar{\beta}_{+}^2 \bar{\gamma}^{-1} + \frac{380}{3} \zeta \mathcal{S}_{-}^2 -  \frac{280}{3} \bar{\beta}_{+} \zeta \mathcal{S}_{-}^2 + \frac{280}{3} \zeta \bar{\gamma} \mathcal{S}_{-}^2 -  \frac{40}{3} \bar{\beta}_{+} \zeta \bar{\gamma}^{-1} \mathcal{S}_{-}^2 \nonumber \\
        & \hspace{2.0cm}+ 160 \bar{\beta}_{-}^2 \zeta \bar{\gamma}^{-2} \mathcal{S}_{-}^2 + 160 \bar{\beta}_{+}^2 \zeta \bar{\gamma}^{-2} \mathcal{S}_{-}^2 -  \frac{40}{3} \bar{\beta}_{-} \zeta \bar{\gamma}^{-1} \mathcal{S}_{-} \mathcal{S}_{+} + 320 \bar{\beta}_{-} \bar{\beta}_{+} \zeta \bar{\gamma}^{-2} \mathcal{S}_{-} \mathcal{S}_{+}  -  \frac{100}{3} \zeta \mathcal{S}_{-}^2 \nu \nonumber \\
        & \hspace{2.0cm}+  \delta\left( \frac{280}{3} \bar{\beta}_{-} \zeta \mathcal{S}_{-}^2 - \frac{200}{3} \bar{\beta}_{-} \zeta \bar{\gamma}^{-1} \mathcal{S}_{-}^2 - \frac{200}{3} \bar{\beta}_{+} \zeta \bar{\gamma}^{-1} \mathcal{S}_{-} \mathcal{S}_{+}\right) \Bigg) \nonumber \\
        & \hspace{1.2cm} +  \frac{1}{\overline{\jmath}^2}\Bigg(\frac{2444}{15} + \frac{352}{3} \bar{\beta}_{+} + \frac{145}{3} \bar{\gamma} - 176 \bar{\beta}_{+}^2 \bar{\gamma}^{-1} + \frac{515}{3} \zeta \mathcal{S}_{-}^2 -  \frac{304}{3} \bar{\beta}_{+} \zeta \mathcal{S}_{-}^2 + \frac{328}{3} \zeta \bar{\gamma} \mathcal{S}_{-}^2 -  \frac{280}{3} \bar{\beta}_{+} \zeta \bar{\gamma}^{-1} \mathcal{S}_{-}^2 \nonumber \\
        & \hspace{2.0cm}+ 352 \bar{\beta}_{-}^2 \zeta \bar{\gamma}^{-2} \mathcal{S}_{-}^2 + 352 \bar{\beta}_{+}^2 \zeta \bar{\gamma}^{-2} \mathcal{S}_{-}^2 -  \frac{280}{3} \bar{\beta}_{-} \zeta \bar{\gamma}^{-1} \mathcal{S}_{-} \mathcal{S}_{+} + 704 \bar{\beta}_{-} \bar{\beta}_{+} \zeta \bar{\gamma}^{-2} \mathcal{S}_{-} \mathcal{S}_{+} \nonumber \\
        & \hspace{2.0cm}+ \delta\left( \frac{304}{3} \bar{\beta}_{-} \zeta \mathcal{S}_{-}^2 - \frac{248}{3} \bar{\beta}_{-} \zeta \bar{\gamma}^{-1} \mathcal{S}_{-}^2 - \frac{248}{3} \bar{\beta}_{+} \zeta \bar{\gamma}^{-1} \mathcal{S}_{-} \mathcal{S}_{+}\right)  -  \frac{169}{3} \zeta \mathcal{S}_{-}^2 \nu\Bigg)\Bigg]\Bigg\}\,
        ,\\
        \Delta v_\infty & = - c m \overline{\varepsilon}^2 \nu^2 \Bigg\{\dfrac{4 \zeta \mathcal{S}_{-}^2}{3 \overline{\jmath}^{5/2}}  \left(3 \overline{\jmath}^{1/2} + \left(3 + \overline{\jmath}\right) \bigl(\pi -  \arctan\sqrt{\overline{\jmath}}\bigr)\right)  \nonumber \\*
        & + \frac{\overline{\varepsilon}}{1 + \overline{\jmath}} \Bigg[\frac{1}{\overline{\jmath}^3}\Bigg(\frac{170}{3} + \frac{80}{3} \bar{\beta}_{+} + \frac{115}{6} \bar{\gamma} - 40 \bar{\beta}_{+}^2 \bar{\gamma}^{-1} + \frac{190}{3} \zeta \mathcal{S}_{-}^2 -  \frac{140}{3} \bar{\beta}_{+} \zeta \mathcal{S}_{-}^2 + \frac{140}{3} \zeta \bar{\gamma} \mathcal{S}_{-}^2 -  \frac{20}{3} \bar{\beta}_{+} \zeta \bar{\gamma}^{-1} \mathcal{S}_{-}^2 \nonumber \\
        & \hspace{2cm}+ 80 \bar{\beta}_{-}^2 \zeta \bar{\gamma}^{-2} \mathcal{S}_{-}^2 + 80 \bar{\beta}_{+}^2 \zeta \bar{\gamma}^{-2} \mathcal{S}_{-}^2 -  \frac{20}{3} \bar{\beta}_{-} \zeta \bar{\gamma}^{-1} \mathcal{S}_{-} \mathcal{S}_{+} + 160 \bar{\beta}_{-} \bar{\beta}_{+} \zeta \bar{\gamma}^{-2} \mathcal{S}_{-} \mathcal{S}_{+} -  \frac{50}{3} \zeta \mathcal{S}_{-}^2 \nu \nonumber \\
        & \hspace{2cm}+ \delta\left(\frac{140}{3} \bar{\beta}_{-} \zeta \mathcal{S}_{-}^2 -  \frac{100}{3} \bar{\beta}_{-} \zeta \bar{\gamma}^{-1} \mathcal{S}_{-}^2 -  \frac{100}{3} \bar{\beta}_{+} \zeta \bar{\gamma}^{-1} \mathcal{S}_{-} \mathcal{S}_{+}\right)  \Bigg) \nonumber \\
        & \hspace{1.4cm} + \frac{1}{\overline{\jmath}^2}\Bigg(\frac{3896}{45} + \frac{448}{9} \bar{\beta}_{+} + \frac{250}{9} \bar{\gamma} -  \frac{224}{3} \bar{\beta}_{+}^2 \bar{\gamma}^{-1} + \frac{1435}{18} \zeta \mathcal{S}_{-}^2 -  \frac{460}{9} \bar{\beta}_{+} \zeta \mathcal{S}_{-}^2 + \frac{496}{9} \zeta \bar{\gamma} \mathcal{S}_{-}^2  \nonumber \\
        & \hspace{2.4cm} + \frac{448}{3} \bar{\beta}_{-}^2 \zeta \bar{\gamma}^{-2} \mathcal{S}_{-}^2 + \frac{448}{3} \bar{\beta}_{+}^2 \zeta \bar{\gamma}^{-2} \mathcal{S}_{-}^2 -  \frac{256}{9} \bar{\beta}_{-} \zeta \bar{\gamma}^{-1} \mathcal{S}_{-} \mathcal{S}_{+} + \frac{896}{3} \bar{\beta}_{-} \bar{\beta}_{+} \zeta \bar{\gamma}^{-2} \mathcal{S}_{-} \mathcal{S}_{+}  \nonumber \\
        & \hspace{2.4cm} -  \frac{281}{18} \zeta \mathcal{S}_{-}^2 \nu -  \frac{256}{9} \bar{\beta}_{+} \zeta \bar{\gamma}^{-1} \mathcal{S}_{-}^2 + \delta\left(\frac{460}{9} \bar{\beta}_{-} \zeta \mathcal{S}_{-}^2 -  \frac{416}{9} \bar{\beta}_{-} \zeta \bar{\gamma}^{-1} \mathcal{S}_{-}^2 -  \frac{416}{9} \bar{\beta}_{+} \zeta \bar{\gamma}^{-1} \mathcal{S}_{-} \mathcal{S}_{+}\right)  \Bigg)  \nonumber \\
        & \hspace{1.2cm} +  \frac{1}{\overline{\jmath}}\Bigg(\frac{1346}{45} + \frac{208}{9} \bar{\beta}_{+} + \frac{155}{18} \bar{\gamma} -  \frac{104}{3} \bar{\beta}_{+}^2 \bar{\gamma}^{-1} + \frac{349}{18} \zeta \mathcal{S}_{-}^2 -  \frac{64}{9} \bar{\beta}_{+} \zeta \mathcal{S}_{-}^2 + \frac{100}{9} \zeta \bar{\gamma} \mathcal{S}_{-}^2 -  \frac{196}{9} \bar{\beta}_{+} \zeta \bar{\gamma}^{-1} \mathcal{S}_{-}^2 \nonumber \\
        & \hspace{2.2cm} + \frac{208}{3} \bar{\beta}_{-}^2 \zeta \bar{\gamma}^{-2} \mathcal{S}_{-}^2 + \frac{208}{3} \bar{\beta}_{+}^2 \zeta \bar{\gamma}^{-2} \mathcal{S}_{-}^2 -  \frac{196}{9} \bar{\beta}_{-} \zeta \bar{\gamma}^{-1} \mathcal{S}_{-} \mathcal{S}_{+} + \frac{416}{3} \bar{\beta}_{-} \bar{\beta}_{+} \zeta \bar{\gamma}^{-2} \mathcal{S}_{-} \mathcal{S}_{+} + \frac{25}{18} \zeta \mathcal{S}_{-}^2 \nu \nonumber \\
        & \hspace{2.2cm} + \delta\left(\frac{64}{9} \bar{\beta}_{-} \zeta \mathcal{S}_{-}^2 -  \frac{116}{9} \bar{\beta}_{-} \zeta \bar{\gamma}^{-1} \mathcal{S}_{-}^2 -  \frac{116}{9} \bar{\beta}_{+} \zeta \bar{\gamma}^{-1} \mathcal{S}_{-} \mathcal{S}_{+}\right)  \Bigg) \Bigg] \nonumber \\
        &  + \frac{\overline{\varepsilon}}{1 + \overline{\jmath}} \left(\pi -  \arctan\sqrt{\overline{\jmath}}\right)\Bigg[ \frac{1}{\overline{\jmath}^{5/2}}\Bigg(\frac{1582}{15} + \frac{176}{3} \bar{\beta}_{+} + \frac{205}{6} \bar{\gamma} - 88 \bar{\beta}_{+}^2 \bar{\gamma}^{-1} + \frac{605}{6} \zeta \mathcal{S}_{-}^2 -  \frac{200}{3} \bar{\beta}_{+} \zeta \mathcal{S}_{-}^2 + \frac{212}{3} \zeta \bar{\gamma} \mathcal{S}_{-}^2 \nonumber \\
        & \hspace{2.2cm} -  \frac{92}{3} \bar{\beta}_{+} \zeta \bar{\gamma}^{-1} \mathcal{S}_{-}^2 + 176 \bar{\beta}_{-}^2 \zeta \bar{\gamma}^{-2} \mathcal{S}_{-}^2 + 176 \bar{\beta}_{+}^2 \zeta \bar{\gamma}^{-2} \mathcal{S}_{-}^2 -  \frac{92}{3} \bar{\beta}_{-} \zeta \bar{\gamma}^{-1} \mathcal{S}_{-} \mathcal{S}_{+} + 352 \bar{\beta}_{-} \bar{\beta}_{+} \zeta \bar{\gamma}^{-2} \mathcal{S}_{-} \mathcal{S}_{+} \nonumber \\
        & \hspace{2.2cm} + \delta\left(\frac{200}{3} \bar{\beta}_{-} \zeta \mathcal{S}_{-}^2 -  \frac{172}{3} \bar{\beta}_{-} \zeta \bar{\gamma}^{-1} \mathcal{S}_{-}^2 -  \frac{172}{3} \bar{\beta}_{+} \zeta \bar{\gamma}^{-1} \mathcal{S}_{-} \mathcal{S}_{+}\right)  -  \frac{127}{6} \zeta \mathcal{S}_{-}^2 \nu\Bigg) \nonumber \\
        & \hspace{1.2cm} + \frac{1}{\overline{\jmath}^{7/2}}\Bigg(\frac{170}{3} + \frac{80}{3} \bar{\beta}_{+} + \frac{115}{6} \bar{\gamma} - 40 \bar{\beta}_{+}^2 \bar{\gamma}^{-1} + \frac{190}{3} \zeta \mathcal{S}_{-}^2 -  \frac{140}{3} \bar{\beta}_{+} \zeta \mathcal{S}_{-}^2 + \frac{140}{3} \zeta \bar{\gamma} \mathcal{S}_{-}^2 -  \frac{20}{3} \bar{\beta}_{+} \zeta \bar{\gamma}^{-1} \mathcal{S}_{-}^2 \nonumber \\
        & \hspace{2.2cm} + 80 \bar{\beta}_{-}^2 \zeta \bar{\gamma}^{-2} \mathcal{S}_{-}^2 + 80 \bar{\beta}_{+}^2 \zeta \bar{\gamma}^{-2} \mathcal{S}_{-}^2 -  \frac{20}{3} \bar{\beta}_{-} \zeta \bar{\gamma}^{-1} \mathcal{S}_{-} \mathcal{S}_{+} + 160 \bar{\beta}_{-} \bar{\beta}_{+} \zeta \bar{\gamma}^{-2} \mathcal{S}_{-} \mathcal{S}_{+} -  \frac{50}{3} \zeta \mathcal{S}_{-}^2 \nu\nonumber \\
        & \hspace{2.2cm} + \delta\left(\frac{140}{3} \bar{\beta}_{-} \zeta \mathcal{S}_{-}^2 -  \frac{100}{3} \bar{\beta}_{-} \zeta \bar{\gamma}^{-1} \mathcal{S}_{-}^2 -  \frac{100}{3} \bar{\beta}_{+} \zeta \bar{\gamma}^{-1} \mathcal{S}_{-} \mathcal{S}_{+}\right)  \Bigg) \nonumber \\
        & \hspace{1.2cm} + \frac{1}{\overline{\jmath}^{3/2}}\Bigg(\frac{806}{15} + \frac{112}{3} \bar{\beta}_{+} + \frac{97}{6} \bar{\gamma} - 56 \bar{\beta}_{+}^2 \bar{\gamma}^{-1} + \frac{121}{3} \zeta \mathcal{S}_{-}^2 - 20 \bar{\beta}_{+} \zeta \mathcal{S}_{-}^2 + \frac{76}{3} \zeta \bar{\gamma} \mathcal{S}_{-}^2 -  \frac{92}{3} \bar{\beta}_{+} \zeta \bar{\gamma}^{-1} \mathcal{S}_{-}^2 \nonumber \\
        & \hspace{2.2cm} + 112 \bar{\beta}_{-}^2 \zeta \bar{\gamma}^{-2} \mathcal{S}_{-}^2 + 112 \bar{\beta}_{+}^2 \zeta \bar{\gamma}^{-2} \mathcal{S}_{-}^2 -  \frac{92}{3} \bar{\beta}_{-} \zeta \bar{\gamma}^{-1} \mathcal{S}_{-} \mathcal{S}_{+} + 224 \bar{\beta}_{-} \bar{\beta}_{+} \zeta \bar{\gamma}^{-2} \mathcal{S}_{-} \mathcal{S}_{+} -  \frac{7}{3} \zeta \mathcal{S}_{-}^2 \nu \nonumber \\
        & \hspace{2.2cm} + \delta\left(20 \bar{\beta}_{-} \zeta \mathcal{S}_{-}^2 -  \frac{76}{3} \bar{\beta}_{-} \zeta \bar{\gamma}^{-1} \mathcal{S}_{-}^2 -  \frac{76}{3} \bar{\beta}_{+} \zeta \bar{\gamma}^{-1} \mathcal{S}_{-} \mathcal{S}_{+}\right)  \Bigg) \nonumber \\
        & \hspace{1.2cm} + \frac{1}{\overline{\jmath}^{1/2}}\Bigg(\frac{74}{15} + \frac{16}{3} \bar{\beta}_{+} + \frac{7}{6} \bar{\gamma} - 8 \bar{\beta}_{+}^2 \bar{\gamma}^{-1} + \frac{17}{6} \zeta \mathcal{S}_{-}^2 + \frac{4}{3} \zeta \bar{\gamma} \mathcal{S}_{-}^2 -  \frac{20}{3} \bar{\beta}_{+} \zeta \bar{\gamma}^{-1} \mathcal{S}_{-}^2 + 16 \bar{\beta}_{-}^2 \zeta \bar{\gamma}^{-2} \mathcal{S}_{-}^2  \nonumber \\
        & \hspace{2.2cm} + \frac{13}{6} \zeta \mathcal{S}_{-}^2 \nu + 16 \bar{\beta}_{+}^2 \zeta \bar{\gamma}^{-2} \mathcal{S}_{-}^2 -  \frac{20}{3} \bar{\beta}_{-} \zeta \bar{\gamma}^{-1} \mathcal{S}_{-} \mathcal{S}_{+} + 32 \bar{\beta}_{-} \bar{\beta}_{+} \zeta \bar{\gamma}^{-2} \mathcal{S}_{-} \mathcal{S}_{+} \nonumber \\
        & \hspace{2.2cm} + \delta\left(- \frac{4}{3} \bar{\beta}_{-} \zeta \bar{\gamma}^{-1} \mathcal{S}_{-}^2 -  \frac{4}{3} \bar{\beta}_{+} \zeta \bar{\gamma}^{-1} \mathcal{S}_{-} \mathcal{S}_{+}\right)  \Bigg) \Bigg] \Bigg\} \,,\\
        \Delta \bar{e}_t & = - m \overline{\varepsilon}^{3/2} \nu^2 \Bigg\{\frac{4 \zeta \mathcal{S}_{-}^2}{(1 + \overline{\jmath})^{1/2}\overline{\jmath}^{3/2}} \left[\overline{\jmath}^{1/2}+\frac{2}{3}\overline{\jmath}^{3/2} + \left(1 + \overline{\jmath}\right) \left(\pi -  \arctan\sqrt{\overline{\jmath}}\right)\right] \nonumber \\*
        &  + \frac{\overline{\varepsilon}}{(1 + \overline{\jmath})^{3/2}}\Bigg[ \frac{514}{9} + \frac{208}{9} \bar{\beta}_{+} + \frac{359}{18} \bar{\gamma} -  \frac{104}{3} \bar{\beta}_{+}^2 \bar{\gamma}^{-1} + \frac{1453}{18} \zeta \mathcal{S}_{-}^2 -  \frac{208}{9} \bar{\beta}_{+} \zeta \mathcal{S}_{-}^2 + \frac{400}{9} \zeta \bar{\gamma} \mathcal{S}_{-}^2 -  \frac{52}{9} \bar{\beta}_{+} \zeta \bar{\gamma}^{-1} \mathcal{S}_{-}^2  \nonumber \\
        & \hspace{2cm} + \frac{208}{3} \bar{\beta}_{-}^2 \zeta \bar{\gamma}^{-2} \mathcal{S}_{-}^2 + \frac{208}{3} \bar{\beta}_{+}^2 \zeta \bar{\gamma}^{-2} \mathcal{S}_{-}^2 -  \frac{52}{9} \bar{\beta}_{-} \zeta \bar{\gamma}^{-1} \mathcal{S}_{-} \mathcal{S}_{+} + \frac{416}{3} \bar{\beta}_{-} \bar{\beta}_{+} \zeta \bar{\gamma}^{-2} \mathcal{S}_{-} \mathcal{S}_{+} -  \frac{635}{18} \zeta \mathcal{S}_{-}^2 \nu \nonumber \\
        & \hspace{2cm} + \delta\left(\frac{208}{9} \bar{\beta}_{-} \zeta \mathcal{S}_{-}^2 -  \frac{260}{9} \bar{\beta}_{-} \zeta \bar{\gamma}^{-1} \mathcal{S}_{-}^2 -  \frac{260}{9} \bar{\beta}_{+} \zeta \bar{\gamma}^{-1} \mathcal{S}_{-} \mathcal{S}_{+}\right)  \nonumber \\
        & \hspace{2cm}  + \frac{1}{\overline{\jmath}}\Bigg(\frac{4976}{45} + \frac{448}{9} \bar{\beta}_{+} + \frac{340}{9} \bar{\gamma} -  \frac{224}{3} \bar{\beta}_{+}^2 \bar{\gamma}^{-1} + \frac{1280}{9} \zeta \mathcal{S}_{-}^2 -  \frac{568}{9} \bar{\beta}_{+} \zeta \mathcal{S}_{-}^2 + \frac{784}{9} \zeta \bar{\gamma} \mathcal{S}_{-}^2  \nonumber \\
        & \hspace{2.7cm} -  \frac{112}{9} \bar{\beta}_{+} \zeta \bar{\gamma}^{-1} \mathcal{S}_{-}^2  + \frac{448}{3} \bar{\beta}_{+}^2 \zeta \bar{\gamma}^{-2} \mathcal{S}_{-}^2 -  \frac{112}{9} \bar{\beta}_{-} \zeta \bar{\gamma}^{-1} \mathcal{S}_{-} \mathcal{S}_{+} + \frac{896}{3} \bar{\beta}_{-} \bar{\beta}_{+} \zeta \bar{\gamma}^{-2} \mathcal{S}_{-} \mathcal{S}_{+}  \nonumber \\
        & \hspace{2.7cm} + \frac{448}{3} \bar{\beta}_{-}^2 \zeta \bar{\gamma}^{-2} \mathcal{S}_{-}^2  -  \frac{496}{9} \zeta \mathcal{S}_{-}^2 \nu + \delta\left(\frac{568}{9} \bar{\beta}_{-} \zeta \mathcal{S}_{-}^2 -  \frac{560}{9} \bar{\beta}_{-} \zeta \bar{\gamma}^{-1} \mathcal{S}_{-}^2 -  \frac{560}{9} \bar{\beta}_{+} \zeta \bar{\gamma}^{-1} \mathcal{S}_{-} \mathcal{S}_{+}\right)  \Bigg) \nonumber \\
        & \hspace{2cm} + \frac{1}{\overline{\jmath}^2}\Bigg(\frac{170}{3} + \frac{80}{3} \bar{\beta}_{+} + \frac{115}{6} \bar{\gamma} - 40 \bar{\beta}_{+}^2 \bar{\gamma}^{-1} + \frac{214}{3} \zeta \mathcal{S}_{-}^2 -  \frac{116}{3} \bar{\beta}_{+} \zeta \mathcal{S}_{-}^2 + \frac{140}{3} \zeta \bar{\gamma} \mathcal{S}_{-}^2 -  \frac{20}{3} \bar{\beta}_{+} \zeta \bar{\gamma}^{-1} \mathcal{S}_{-}^2 \nonumber \\
        & \hspace{2.7cm} + 80 \bar{\beta}_{-}^2 \zeta \bar{\gamma}^{-2} \mathcal{S}_{-}^2 + 80 \bar{\beta}_{+}^2 \zeta \bar{\gamma}^{-2} \mathcal{S}_{-}^2 -  \frac{20}{3} \bar{\beta}_{-} \zeta \bar{\gamma}^{-1} \mathcal{S}_{-} \mathcal{S}_{+} + 160 \bar{\beta}_{-} \bar{\beta}_{+} \zeta \bar{\gamma}^{-2} \mathcal{S}_{-} \mathcal{S}_{+}  \nonumber \\
        & \hspace{2.7cm} -  \frac{74}{3} \zeta \mathcal{S}_{-}^2 \nu + \delta\left(\frac{116}{3} \bar{\beta}_{-} \zeta \mathcal{S}_{-}^2 -  \frac{100}{3} \bar{\beta}_{-} \zeta \bar{\gamma}^{-1} \mathcal{S}_{-}^2 -  \frac{100}{3} \bar{\beta}_{+} \zeta \bar{\gamma}^{-1} \mathcal{S}_{-} \mathcal{S}_{+}\right)  \Bigg) \nonumber \\
        & \hspace{2cm} + \overline{\jmath} \left(\frac{16}{5} + \frac{4}{3} \bar{\gamma} + \frac{34}{3} \zeta \mathcal{S}_{-}^2 + \frac{16}{3} \zeta \bar{\gamma} \mathcal{S}_{-}^2 -  \frac{14}{3} \zeta \mathcal{S}_{-}^2 \nu\right)\Bigg]  \nonumber \\
        &  + \frac{\overline{\varepsilon}}{(1 + \overline{\jmath})^{3/2}}\left(\pi -  \arctan\sqrt{\overline{\jmath}}\right)\Bigg[\frac{1}{\overline{\jmath}^{3/2}}\Bigg(\frac{1942}{15} + \frac{176}{3} \bar{\beta}_{+} + \frac{265}{6} \bar{\gamma} - 88 \bar{\beta}_{+}^2 \bar{\gamma}^{-1} + 166 \zeta \mathcal{S}_{-}^2 - 76 \bar{\beta}_{+} \zeta \mathcal{S}_{-}^2 + \frac{308}{3} \zeta \bar{\gamma} \mathcal{S}_{-}^2  \nonumber \\
        & \hspace{2.7cm} -  \frac{44}{3} \bar{\beta}_{+} \zeta \bar{\gamma}^{-1} \mathcal{S}_{-}^2 + 176 \bar{\beta}_{-}^2 \zeta \bar{\gamma}^{-2} \mathcal{S}_{-}^2 + 176 \bar{\beta}_{+}^2 \zeta \bar{\gamma}^{-2} \mathcal{S}_{-}^2 -  \frac{44}{3} \bar{\beta}_{-} \zeta \bar{\gamma}^{-1} \mathcal{S}_{-} \mathcal{S}_{+}-  \frac{190}{3} \zeta \mathcal{S}_{-}^2\nu  \nonumber \\
        & \hspace{2.7cm} + 352 \bar{\beta}_{-} \bar{\beta}_{+} \zeta \bar{\gamma}^{-2} \mathcal{S}_{-} \mathcal{S}_{+}  \nu + \delta\left(76 \bar{\beta}_{-} \zeta \mathcal{S}_{-}^2 -  \frac{220}{3} \bar{\beta}_{-} \zeta \bar{\gamma}^{-1} \mathcal{S}_{-}^2 -  \frac{220}{3} \bar{\beta}_{+} \zeta \bar{\gamma}^{-1} \mathcal{S}_{-} \mathcal{S}_{+}\right)   \Bigg) \nonumber  \\
        & \hspace{1.8cm} + \frac{1}{\overline{\jmath}^{1/2}}\Bigg(\frac{1334}{15} + \frac{112}{3} \bar{\beta}_{+} + \frac{185}{6} \bar{\gamma} - 56 \bar{\beta}_{+}^2 \bar{\gamma}^{-1} + \frac{245}{2} \zeta \mathcal{S}_{-}^2 - 40 \bar{\beta}_{+} \zeta \mathcal{S}_{-}^2 + \frac{208}{3} \zeta \bar{\gamma} \mathcal{S}_{-}^2  \nonumber \\
        & \hspace{2.7cm} + 112 \bar{\beta}_{-}^2 \zeta \bar{\gamma}^{-2} \mathcal{S}_{-}^2 + 112 \bar{\beta}_{+}^2 \zeta \bar{\gamma}^{-2} \mathcal{S}_{-}^2 -  \frac{28}{3} \bar{\beta}_{-} \zeta \bar{\gamma}^{-1} \mathcal{S}_{-} \mathcal{S}_{+} + 224 \bar{\beta}_{-} \bar{\beta}_{+} \zeta \bar{\gamma}^{-2} \mathcal{S}_{-} \mathcal{S}_{+}  \nonumber \\
        & \hspace{2.7cm} -  \frac{313}{6} \zeta \mathcal{S}_{-}^2 \nu -  \frac{28}{3} \bar{\beta}_{+} \zeta \bar{\gamma}^{-1} \mathcal{S}_{-}^2+ \delta\left(40 \bar{\beta}_{-} \zeta \mathcal{S}_{-}^2 -  \frac{140}{3} \bar{\beta}_{-} \zeta \bar{\gamma}^{-1} \mathcal{S}_{-}^2 -  \frac{140}{3} \bar{\beta}_{+} \zeta \bar{\gamma}^{-1} \mathcal{S}_{-} \mathcal{S}_{+}\right)  \Bigg) \nonumber \\
        & \hspace{1.8cm} + \frac{1}{\overline{\jmath}^{5/2}}\Bigg(\frac{170}{3} + \frac{80}{3} \bar{\beta}_{+} + \frac{115}{6} \bar{\gamma} - 40 \bar{\beta}_{+}^2 \bar{\gamma}^{-1} + \frac{214}{3} \zeta \mathcal{S}_{-}^2 -  \frac{116}{3} \bar{\beta}_{+} \zeta \mathcal{S}_{-}^2 + \frac{140}{3} \zeta \bar{\gamma} \mathcal{S}_{-}^2  \nonumber \\
        & \hspace{2.7cm} -  \frac{20}{3} \bar{\beta}_{+} \zeta \bar{\gamma}^{-1} \mathcal{S}_{-}^2 + 80 \bar{\beta}_{-}^2 \zeta \bar{\gamma}^{-2} \mathcal{S}_{-}^2 + 80 \bar{\beta}_{+}^2 \zeta \bar{\gamma}^{-2} \mathcal{S}_{-}^2 -  \frac{20}{3} \bar{\beta}_{-} \zeta \bar{\gamma}^{-1} \mathcal{S}_{-} \mathcal{S}_{+}  -  \frac{74}{3} \zeta \mathcal{S}_{-}^2 \nu \nonumber \\
        & \hspace{2.7cm} + 160 \bar{\beta}_{-} \bar{\beta}_{+} \zeta \bar{\gamma}^{-2} \mathcal{S}_{-} \mathcal{S}_{+} + \delta\left(\frac{116}{3} \bar{\beta}_{-} \zeta \mathcal{S}_{-}^2 -  \frac{100}{3} \bar{\beta}_{-} \zeta \bar{\gamma}^{-1} \mathcal{S}_{-}^2 -  \frac{100}{3} \bar{\beta}_{+} \zeta \bar{\gamma}^{-1} \mathcal{S}_{-} \mathcal{S}_{+}\right)  \Bigg) \nonumber \\
        & \hspace{1.8cm} + \overline{\jmath}^{1/2} \Bigg(\frac{242}{15} + \frac{16}{3} \bar{\beta}_{+} + \frac{35}{6} \bar{\gamma} - 8 \bar{\beta}_{+}^2 \bar{\gamma}^{-1} + \frac{167}{6} \zeta \mathcal{S}_{-}^2 -  \frac{8}{3} \bar{\beta}_{+} \zeta \mathcal{S}_{-}^2 + \frac{40}{3} \zeta \bar{\gamma} \mathcal{S}_{-}^2 -  \frac{4}{3} \bar{\beta}_{+} \zeta \bar{\gamma}^{-1} \mathcal{S}_{-}^2 \nonumber \\
        & \hspace{2.7cm} + 16 \bar{\beta}_{-}^2 \zeta \bar{\gamma}^{-2} \mathcal{S}_{-}^2 + 16 \bar{\beta}_{+}^2 \zeta \bar{\gamma}^{-2} \mathcal{S}_{-}^2 -  \frac{4}{3} \bar{\beta}_{-} \zeta \bar{\gamma}^{-1} \mathcal{S}_{-} \mathcal{S}_{+} + 32 \bar{\beta}_{-} \bar{\beta}_{+} \zeta \bar{\gamma}^{-2} \mathcal{S}_{-} \mathcal{S}_{+} \nonumber \\
        & \hspace{2.7cm} +  \delta\left(\frac{8}{3} \bar{\beta}_{-} \zeta \mathcal{S}_{-}^2 -  \frac{20}{3} \bar{\beta}_{-} \zeta \bar{\gamma}^{-1} \mathcal{S}_{-}^2 -  \frac{20}{3} \bar{\beta}_{+} \zeta \bar{\gamma}^{-1} \mathcal{S}_{-} \mathcal{S}_{+}\right) -  \frac{27}{2} \zeta \mathcal{S}_{-}^2 \nu\Bigg)\Bigg]\Bigg\}\,.
    \end{align}
\end{subequations}
\end{widetext}

\section{The impact parameter at 3PN in general relativity} \label{app:b_GR}

In GR, the impact parameter at 3PN reads:  
\begin{align} \label{eq:b_GR}
    b_\mathrm{GR}   =&\  \frac{G m \sqrt{\overline{\jmath}}}{\bar{\varepsilon }} \Bigg\{1+\left(-\frac{1}{8} +\frac{3}{8}  \nu \right) \bar{\varepsilon }\nonumber \\*
    &  +\bar{\varepsilon }^2\left(\frac{3}{128} -\frac{5}{64}  \nu -\frac{5}{128}  \nu
   ^2\right)\nonumber \\*
   & +\bar{\varepsilon }^3\left(-\frac{5}{1024}+\frac{21}{1024} \nu -\frac{7}{1024} \nu ^2-\frac{7}{1024} \nu ^3\right)
    \Bigg\},\nonumber \\*
\end{align}
where, here, the reduced energy and angular momentum are defined as in GR, namely, 
\begin{subequations}
\begin{align}
    \bar{\varepsilon} &= \frac{2 E}{\mu c^2}   &\text{and}&&
    \overline{\jmath} &= \frac{2 J^2 E}{\mu^3 (G m)^2}\,.
\end{align}
\end{subequations}
Our result is also in agreement at 1PN with Eq.~(40) of~\cite{Junker:1992kle} and Eq.~(2.18) of~\cite{DeVittori:2014psa}, but it disagrees already at 2PN with the 3PN-accurate result of Eq.~(3.8)~of~\cite{Cho:2018upo}. 

The agreement with~\cite{Junker:1992kle} was immediate, but the agreement with~\cite{DeVittori:2014psa} was obtained after a typo was corrected in that work: an extra minus sign was needed in front of the 1PN correction. The result obtained by~\cite{Cho:2018upo} also had a missing minus sign at 1PN, but starting at 2PN, we found an unresolvable disagreement with the result claimed there. This disagreement should propagate to~\cite{Cho:2021onr}, which uses this result, see their Eq. (32).

\bibliography{references}

\end{document}